\documentclass[acmlarge]{acmart}
\AtBeginDocument{%
  }

\setcopyright{acmlicensed}
\acmJournal{IMWUT}
\acmYear{2025} \acmVolume{9} \acmNumber{2} \acmArticle{63}
\acmMonth{6}\acmDOI{10.1145/3729499}

\acmSubmissionID{8707}

\usepackage{hyperref}
\usepackage{hyperxmp}

\usepackage{threeparttable}
\def\sysname{\textsc{TAPOR}\xspace}
\usepackage{xcolor}
\usepackage{xurl}
\usepackage{booktabs}
\usepackage{tcolorbox}
\usepackage{fp}
\usepackage{romannum}
\usepackage[ruled]{algorithm2e}
\usepackage{algpseudocode}
\usepackage{enumerate}
\usepackage[english]{babel}
\usepackage{blindtext}
\usepackage[caption=false,font=footnotesize,subrefformat=parens]{subfig}
\usepackage{amsmath}

\usepackage{amssymb}
\usepackage{xspace}
\usepackage{multirow}
\usepackage{threeparttable}
\usepackage{float}
\usepackage{flafter}
\usepackage{comment}
\usepackage[skip=12pt]{caption}
\usepackage{graphicx}
\usepackage{soul}
\usepackage{color}
\usepackage{subfig}
\usepackage{bbding}
\usepackage{pifont}

\ifodd 0
\newcommand{\rev}[1]{{\color{blue}#1}} 
\newcommand{\com}[1]{\textbf{\color{red}(COMMENT: #1)}} 
\newcommand{\todo}[1]{\textbf{{\color{orange}(TODO: #1)}}}
\else
\newcommand{\rev}[1]{#1}
\newcommand{\com}[1]{}
\newcommand{\todo}[1]{}
\fi

\usepackage{tikz}

\def\fig{Fig.\xspace}
\def\eqn{Eq.\xspace}

\def\tab{Tab.\xspace}

\def\ie{{\textit{i.e.}\xspace}} 
\def\eg{{\textit{e.g.}\xspace}}
 
\def\etal{{\textit{et al.}\xspace}}
\def\etc{{\textit{etc.}\xspace}}

\newcommand{\head}[1]{{\noindent \textbf{#1:}}}

\graphicspath{{./figures/}}
\graphicspath{{./pdf/}}
\DeclareGraphicsExtensions{.pdf,.jpeg,.png}

\usepackage{fp}

\newlength\maxlentime

\def\headertime{New}
\begin{document}

\pagenumbering{arabic}
\title{\sysname: 3D Hand Pose Reconstruction with Fully Passive Thermal Sensing for Around-Device Interactions}


\author{Xie Zhang}
\affiliation{%
  \institution{The University of Hong Kong}
  \city{Hong Kong SAR}
  \country{Hong Kong SAR, China}}
\email{zhangxie@connect.hku.hk}
\orcid{0000-0003-4103-6256}

\author{Chengxiao Li}
\affiliation{%
  \institution{The University of Hong Kong}
  \city{Hong Kong SAR}
  \country{Hong Kong SAR, China}}
\email{chengxiaoli@connect.hku.hk}
\orcid{0009-0002-0371-428X}

\author{Chenshu Wu}
\authornote{Corresponding author.}
\affiliation{%
  \institution{The University of Hong Kong}
  \city{Hong Kong SAR}
  \country{Hong Kong SAR, China}}
\email{chenshu@cs.hku.hk}
\orcid{0000-0002-9700-4627}





\renewcommand{\shortauthors}{Zhang et al.}

\begin{abstract}
This paper presents the design and implementation of \sysname, a privacy-preserving, non-contact, and fully passive sensing system for accurate and robust 3D hand pose reconstruction for around-device interaction using a single low-cost thermal array sensor.
Thermal sensing using inexpensive and miniature thermal arrays emerges with an excellent utility-privacy balance, offering an imaging resolution significantly lower than cameras but far superior to RF signals like radar or WiFi. 
The design of \sysname, however, is challenging, mainly because the captured temperature maps are low-resolution and textureless.
To overcome the challenges, we investigate thermo-depth and thermo-pose properties, proposing a novel physics-inspired neural network that learns effective 3D spatial representations of potential hand poses.
We then formulate the 3D pose reconstruction problem as a distinct retrieval task, enabling accurate hand pose determination from the input temperature map.
To deploy \sysname on IoT devices, we introduce an effective heterogeneous knowledge distillation method, reducing computation by 377$\times$.
\sysname is fully implemented and tested in real-world scenarios, showing remarkable performance, supported by four gesture control and finger tracking case studies.
We envision \sysname to be a ubiquitous interface for around-device control and have open-sourced it at 
\href{https://github.com/aiot-lab/TAPOR}{\color{blue}https://github.com/aiot-lab/TAPOR}.
\end{abstract}




\maketitle


\begin{figure}[t]
    \includegraphics[width=0.96\textwidth]{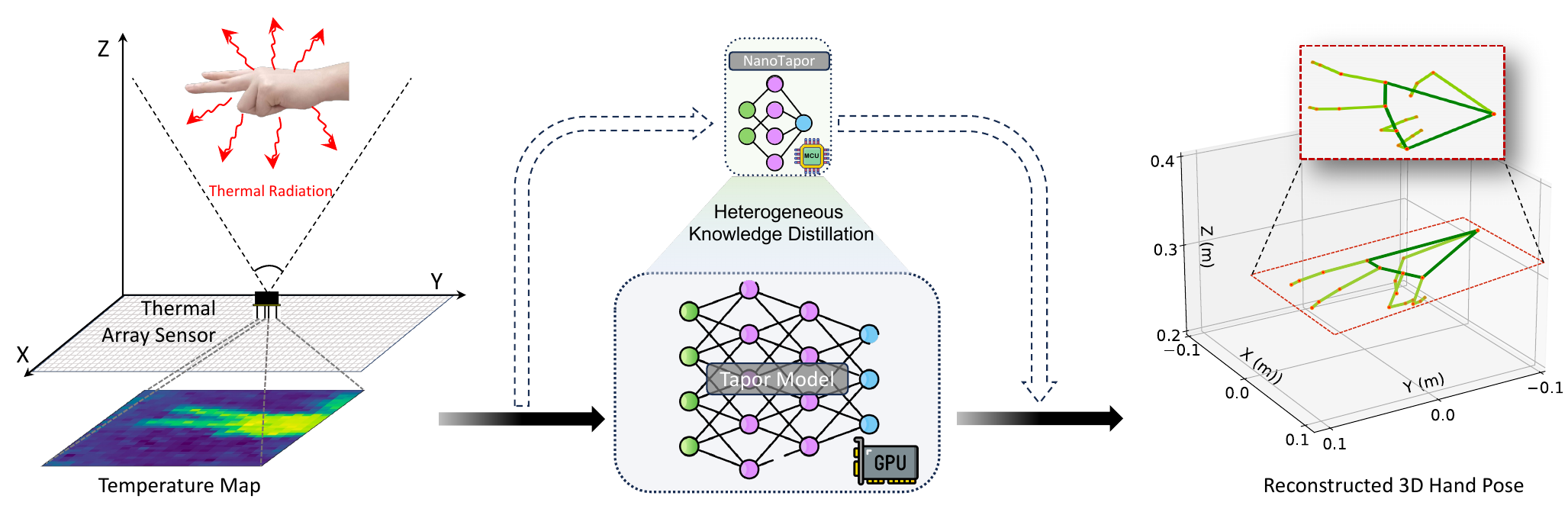}
    \vspace{-0.8\baselineskip}
    \caption{Overview of \sysname. \rm{\sysname reconstructs the 3D hand pose from a temperature map generated by a single low-cost thermal array sensor through a physics-inspired neural network model. \rev{Leveraging the distilled NanoTapor model, the system operates in real-time on IoT devices with negligible performance degradation, making it suitable for a wide range of around-device interaction applications.}}}
\label{fig:sys_overview}
\end{figure}

\section{Introduction}
\label{sec:intro}

\rev{As the Internet of Things (IoT) devices and edge computing continue to proliferate globally, the need for seamless and efficient interaction has been growing rapidly. 
Around-device gesture interaction \cite{nandakumarFingerIOUsingActive2016, lienSoliUbiquitousGesture2016} has become an integral part of modern interactive systems, providing an increasingly essential interface for applications such as smart homes, autonomous vehicles, virtual and augmented reality environments, human-robot interaction, and wearables/hearables, \etc. 
Around-device gesture interaction, focusing on close-range sensing, enables users to control many devices in the same environment using a consistent set of touchless gestures. 
This capability is particularly crucial for resource-constrained IoT devices, which often lack traditional input mechanisms such as touchscreens or keyboards, thereby driving the rapid expansion of the gesture control market, which is projected to exceed \$57 billion by 2033 \cite{futuremarketinsightsGestureControlMarket}.}

\rev{Despite its potential, enabling ubiquitous around-device gesture interaction is fraught with challenges, including constraints related to privacy, accuracy, computational complexity, cost, and power consumption. 
Significant efforts have been undertaken to address these challenges, with notable contributions from both industry and academia. 
Commercial solutions such as Google Soli \cite{lienSoliUbiquitousGesture2016} and Vtouch \cite{Holo_Button}, alongside research initiatives \cite{yu2019rfid,xiao2021onefi,truong2018capband,nguyen2019handsense}, have made strides in advancing gesture recognition technologies. 
However, many existing approaches are limited to recognizing a predefined set of gestures, which inherently restricts their flexibility in supporting user-defined gestures.}
Recently, 3D hand pose reconstruction using various technologies has gained significant attention as a key enabler for more flexible and scalable gesture control. 
Despite notable advances, existing solutions face several limitations in the context of ubiquitous around-device gesture interaction.
For example, wearable systems create adherence issues as users are required to wear specific devices, in addition to an implicit gap in integration into various IoT \cite{chenMagXWearableUntethered2021,zhangRoFin3DHand2023}. 
Camera-based approaches raise significant privacy concerns and are usually computation-intensive \cite{zhaoIfSightedPeople2023,akterPrivacyConsiderationsVisually2020,rezaei2023trihorn}. 
Wireless sensing solutions using Wi-Fi \cite{Liu_Yu_Wang_Guo_Li_Yi_Zhang_2024, Gao_Li_Xie_Yi_Wang_Wu_Zhang_2022} or millimeter-wave \cite{Li_Zhang_Chen_Wan_Zhang_Hu_Sun_Chen_2023, lienSoliUbiquitousGesture2016} signals preserve privacy, yet suffer from deficient domain generalization \cite{zhang2021wifi}, and cumbersome hardware setup (\eg, a pair of separately deployed Wi-Fi transceivers) \cite{jiConstruct3DHand2023}. 
Similarly, acoustic methods face comparable issues and are additionally vulnerable to interference from environmental noise \cite{li2022experience, nandakumarFingerIOUsingActive2016}.

\begin{table}[tp]
\caption{Contactless hand pose reconstruction techniques.}
\centering
\label{tab:modality_comparison}

\begingroup

\renewcommand{\arraystretch}{1.3} 

\begin{tabular}{cccccc}
\hline
\textbf{Sensing Modality} & \textbf{Privacy Protection} & \textbf{Sensing Mode}  & \textbf{Sample Device}    & \textbf{Power}             & \textbf{Cost}  \\ \hline
RF Devices                & Strong                      & Active                 & IWR1843 Radar             & \textgreater{}2 W          & $\sim$299\$         \\
Acoustic Devices          & Weak                        & Active                 & miniDSP UMA-8-SP          & \textgreater{}20 W         & $\sim$155\$         \\
RGB/IR Cameras            & Very Weak                   & Quasi-passive          & Basler IMX334               & \textgreater{}2.7 W        & $\sim$199\$         \\
\textbf{Thermal Array}    & \textbf{Medium}             & \textbf{Fully Passive} & \textbf{Melexis MLX90640} & \textbf{\textless{}0.08 W} & \textbf{$\sim$20\$} \\ \hline
\end{tabular}

\endgroup

\end{table}

In this paper, we present the design, implementation, and deployment of \sysname, a \textbf{T}hermal \textbf{A}rray-based 3D hand \textbf{Po}se \textbf{R}econstruction system that is privacy-preserving, fully passive, accurate, and cost-effective. 
\sysname explores and exploits thermal sensing for 3D hand pose reconstruction using a low-cost thermal array. 
As shown in \fig\ref{fig:sys_overview}, a thermal array is a miniature sensor that captures thermal radiation from a user's hand. 
Compared with the aforementioned modalities, a thermal array produces frames with magnitude lower imaging resolution (\eg, $32\times24$ thermal pixels) than RGB/depth/thermal cameras, while offering significantly higher resolution than RF or acoustic signals, striking an excellent balance between utility and privacy. 
As further detailed in \tab\ref{tab:modality_comparison}, since the sensing signal, \ie, thermal radiation, is emitted directly by the target (\ie, the hand), \sysname is a \textit{fully passive sensing} system that requires no external signal sources. 
It can operate in darkness, unlike vision-based methods, and consumes less power compared to mmWave radars and WiFi devices.
Built upon thermal array-based sensing, \sysname achieves accurate and robust 3D hand pose reconstruction, offering a scalable and flexible solution for hand gesture control. 
Notably, \sysname requires only a single tiny thermal array, making it a low-cost and efficient solution that can be easily integrated into various IoT devices.

However, 3D hand pose reconstruction from a temperature map produced by the thermal sensors using a lightweight model poses significant challenges. 
First, thermal arrays are fully passive sensors, capturing only the thermal radiation emitted by the target without transmitting any signals.
Therefore, depth/position estimation from the captured temperature readings is exceedingly challenging due to the lack of a reliable model correlating thermal radiation with distance. 
The difficulty is exacerbated by the use of a single, low-resolution thermal sensor.
Hence, the problem necessitates a novel solution different from existing methods designed for depth or RGB cameras, which either use dedicated infrared lights for ranging \cite{zanuttigh2016time} or rely on high-resolution textured features \cite{yang2024depth}. 
Second, key point detection from the low-resolution, textureless temperature maps is far from straightforward. 
Despite extensive research on low-resolution images or textureless data (\eg, depth cameras or lidar point clouds) separately, low-resolution coupling with the lack of texture complicates key point identification for various hand poses. 
Last, it is non-trivial to build an efficient end-to-end system for real-time deployment on IoT devices with highly constrained resources (\eg, ESP32 with an MCU). 

\sysname presents two unique designs to overcome these challenges and deliver a practical 3D hand pose reconstruction system deployable on IoT. 
First, we propose a novel physics-inspired neural network that formulates the 3D hand pose reconstruction as a retrieval problem, enabling accurate and robust 3D hand pose estimation from a single temperature map. 
Based on an in-depth understanding of thermal data from the principles of thermography, we identify a distinct opportunity to learn from the thermo-depth relationship and the 2D thermo-pose context in conjunction with hand structure knowledge. 
Our design incorporates (i) a thermo-space encoder that integrates the thermo-depth property and 2D thermal imaging context to construct a 3D spatial representation for hand poses; (ii) a key point (KP) feature extractor that generates 2D KP features by incorporating common knowledge of hand structure and temporal coherence; and (iii) a position retrieval decoder that retrieves the 3D hand pose by matching the KP features (constraints) to the 3D spatial representation (solution space). 

Second, we propose a heterogeneous knowledge distillation approach to transfer the physics-inspired model into an extremely lightweight network for IoT deployment. 
Specifically, rather than involving attention blocks, we constrain the model structure to purely convolutional neural networks (CNNs), which are efficiently supported by edge devices.
We also perform feature dimension reduction to minimize computational overhead.
To transfer knowledge between models of different structures and feature dimensions, we propose heterogeneous knowledge distillation with a learnable adapter to align structures and feature dimensions, significantly reducing the computation and successfully delivering a tiny version of \sysname, named NanoTapor, for real-time on-device deployment for IoT.

We implement \sysname as an end-to-end prototype using a Melexis MLX90640BAB thermal array ($32\times24$ pixels, \$20) and conduct experiments in diverse real-world settings. 
We collect a dataset from 10 users across 5 environments, yielding over 60,000 hand pose samples under varied lighting, hand coverings, distances, and ambient temperatures. 
Our results show that \sysname achieves a mean joint position error of 2.26 cm, sufficient for many around-device gesture interaction applications. 
Notably, NanoTapor reduces computational overhead by $377\times$ compared to the original Tapor model, with only a slight increase in mean error to 3.26 cm, supporting real-time IoT operations.
To further highlight \sysname's IoT applicability, we deploy it on an ESP32-S3 device and conduct four case studies: wearable AI pin control, online meeting gestures, bathroom appliance control, and in-car gestures, all demonstrating promising performance.
We foresee \sysname as a ubiquitous interface for an interactive Internet of Thermal Things.

In summary, this paper contributes the following: (1) \sysname, a privacy-preserving, non-contact, passive system for robust 3D hand pose reconstruction in IoT using a single, low-cost thermal sensor; (2) a physics-inspired neural network utilizing thermo-depth relationship and 2D thermal context with hand structure knowledge for 3D pose reconstruction from temperature maps; (3) a heterogeneous knowledge distillation approach, reducing computational load by 377× to enable NanoTapor for real-time IoT deployment; (4) a complete implementation and extensive experiments with real-world case studies highlighting exciting applications.

The rest of this paper first presents a primer on thermal array sensing (\S\ref{sec:primer}), then introduces \sysname's design (\S\ref{sec:method}), followed by system implementation (\S\ref{sec:impl}) and experimental evaluation and case studies (\S\ref{sec:exp}). 
We discuss future work in \S\ref{sec:discussion}, review related work in \S\ref{sec:related_works}, and conclude the paper in \S\ref{sec:conclusion}.

\begin{figure}[t]
    \subfloat[Temperature map vs. images.]{%
    \label{subfig:modality_compare_examples}
      \includegraphics[width=0.6\textwidth]{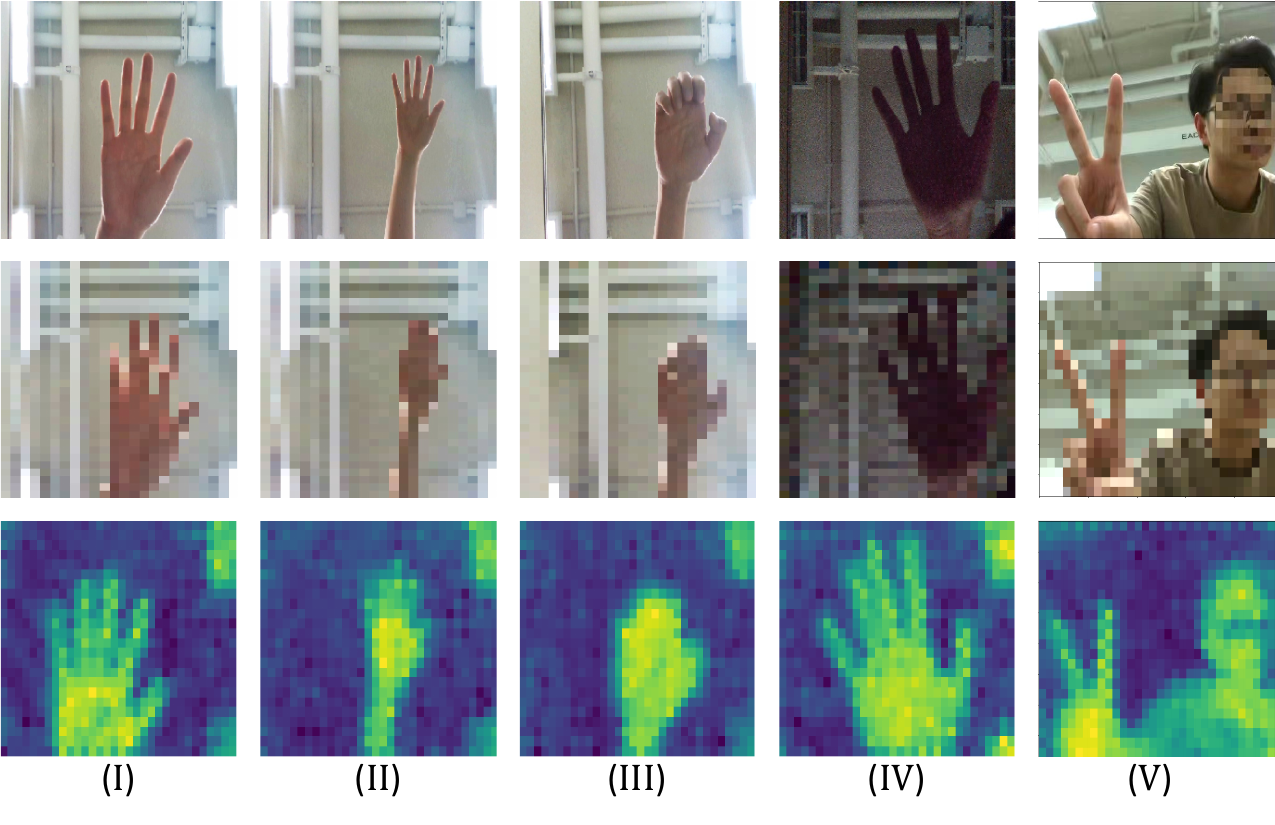}
	  }
    \subfloat[Mediapipe results.]{%
    \label{subfig:modality_compare_results}
      \includegraphics[width=0.25\textwidth]{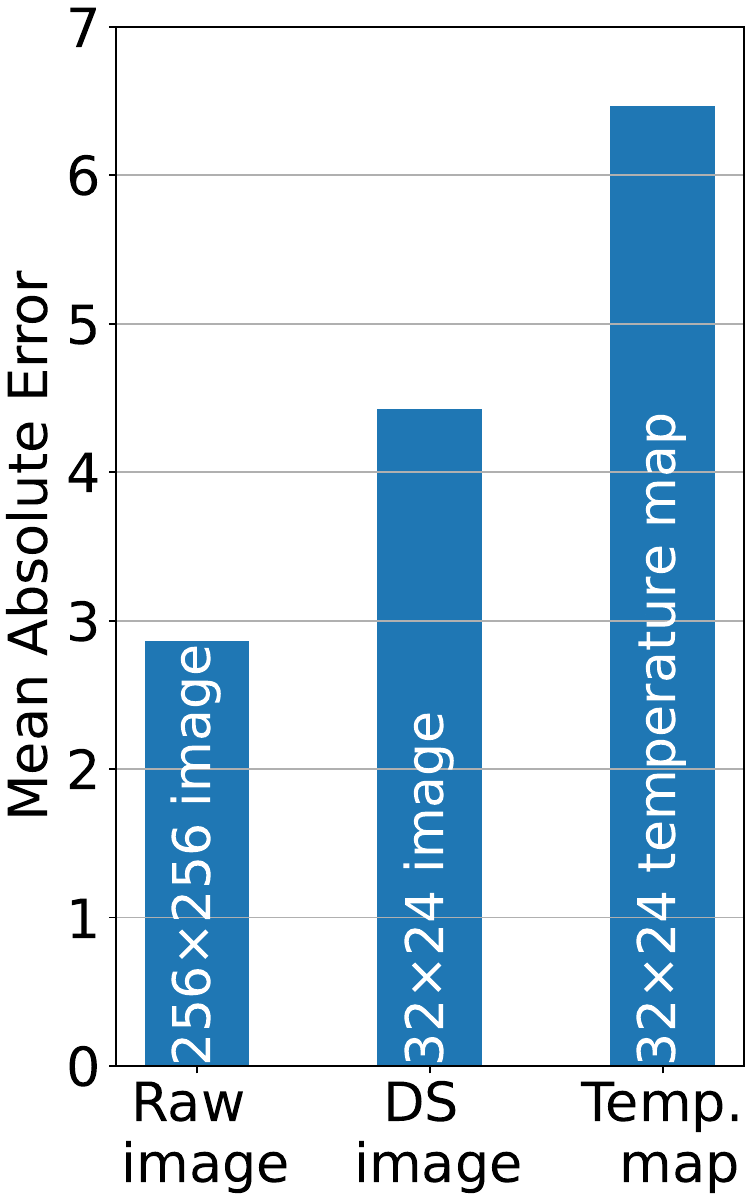}
	  }
    \caption{Modality comparison. {\rm (a) Example samples highlight distinct properties of the temperature maps: (i) single-channel, (ii) low-resolution, (iii) textureless, (iv) illumination immunity, and (v) privacy-preserving. (b) Performance comparison of Mediapipe on original RGB images (raw image), downsampled (DS) images, and temperature maps. }}
	\label{fig:modality_compare}
\end{figure}

\section{Primer on Thermal Array Sensing}
\label{sec:primer}
We first present a primer on thermal array sensors, with a preliminary experiment (\S\ref{ssec:investigative_exp}) to understand the distinct properties of temperature maps and an in-depth understanding of thermography  (\S\ref{ssec:key_insights}) to gain insights for \sysname design.

\subsection{Thermal Array}
\label{ssec:thermal_array_sensor}

A thermal array is a small sensor comprising a miniature array of thermal detectors arranged in a grid pattern. Each detector captures the thermal radiant power $\Phi$ emitted by objects within its field of view (FOV). The sensor converts the captured thermal radiation into an estimated surface temperature for each target area, resulting in a temperature map as illustrated in \fig\ref{subfig:modality_compare_examples}. 
Designed in a compact size (approximately 12 mm in height and 9.5 $\times$ 9.5 mm in footprint), thermal arrays usually have a resolution at least two orders lower than mainstream thermal cameras and are correspondingly significantly cheaper. 
For example, \sysname utilizes the Melexis MLX90640BAB thermal array sensor, which features a resolution of 32$\times$24 and costs about US\$20, with a reported temperature accuracy of $\pm 1.5^{\circ}C$, a FOV of $35^{\circ} \times 55^{\circ}$, and a frame rate up to 64 Hz. 
Thanks to the low price and compact size, thermal array sensors emerge as an attractive sensing modality, especially for IoT devices. 

\subsection{Investigative Experiment}
\label{ssec:investigative_exp}

\rev{Computer vision techniques for hand pose reconstruction from RGB images have reached a high level of maturity in recent years.
Consequently, a natural approach is to leverage existing RGB-based models, such as Mediapipe \cite{zhangMediaPipeHandsOndevice2020}, and apply them directly to temperature maps. 
However, this strategy proves to be suboptimal, as temperature maps exhibit distinct characteristics that differ significantly from RGB images. 
These unique properties, which are explored in detail in the following investigation, hinder the direct applicability of existing RGB-based models and necessitate specialized solutions tailored to the thermal domain.}

\begin{figure}[t]
\subfloat[Temperature map.]{%
    \label{subfig:hand_kp}
      \includegraphics[width=0.33\textwidth]{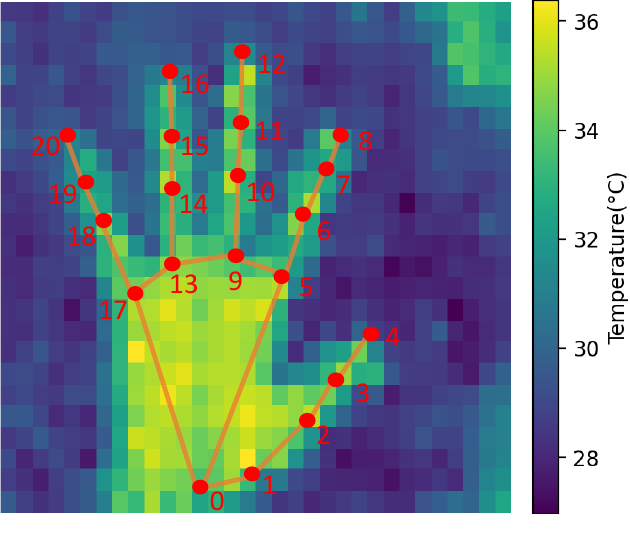}}
\subfloat[Detection schematic.]{%
    \label{subfig:diagran}
      \includegraphics[width=0.33\textwidth]{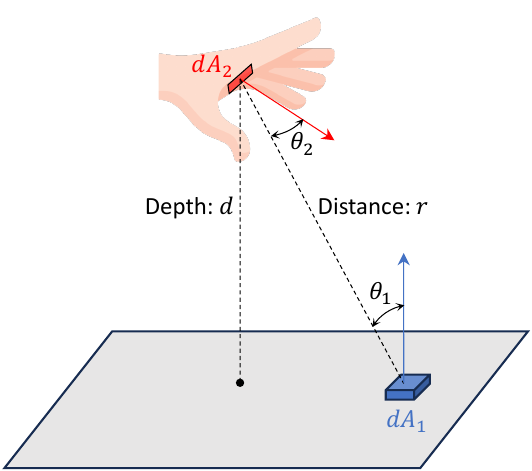}
	  }
   \vspace{-0.6\baselineskip}
    \caption{Temperature map and thermal radiation detection diagram. {\rm 
    (a) Temperature map of an outstretched hand, with red dots marking key points.
    (b) The thermal radiation detection schematic of the thermal detector.}}
\label{fig:theory_example}
\end{figure}

Our preliminary study involves five participants, four males and one female, who performed various hand poses in front of a thermal array sensor and an RGB camera to collect thermal and RGB sample pairs, as shown in \fig\ref{subfig:modality_compare_examples}. In total, over 1000 samples were collected from each participant. 
To examine the effectiveness of existing RGB-based models, we employ Mediapipe \cite{zhangMediaPipeHandsOndevice2020}, a well-known and widely used model for 2D hand pose estimation. 
The hand pose is represented by 21 key points, as depicted in \fig\ref{subfig:hand_kp}. 
For a comprehensive evaluation, we retrain and test the MediaPipe model under three input conditions: 1) the original RGB images (256 $\times$ 256), 2) the downsampled RGB images (32 $\times$ 24), and 3) the thermal array temperature maps (32 $\times$ 24). All models were trained and evaluated using the same 7:1:2 train/val/test dataset split.
As illustrated in \fig\ref{subfig:modality_compare_results}, the results show that, Mediapipe performs considerably worse on the thermal array data than on RGB images, with a performance drop of up to 100\%. 
This suggests that existing RGB-based pose estimation models have difficulty adapting to thermal array data, highlighting the need for a novel model design. 

Based on an in-depth investigation of the temperature maps reported by the thermal array sensor, we identify several distinct properties that lead to the unsatisfactory performance of existing RGB-based models, which are designed for very different RGB images. 
As illustrated in \fig\ref{subfig:modality_compare_examples}, the temperature maps exhibit the following unique characteristics: 
1) \textbf{Low resolution}: Thermal arrays typically generate a temperature map of 32$\times$24 (or even fewer) pixels. While the low resolution inherently preserves privacy, most prior CV models assume a minimum resolution of 128$\times$128 pixels \cite{deng2009imagenet,cordts2016cityscapes,lin2014microsoft}. 
2) \textbf{Textureless}: As opposed to visual images, the temperature maps lack texture information, making it more difficult to identify overlapped objects (\eg, crossed fingers as in \fig\ref{subfig:modality_compare_examples}(iii)). 
3) \textbf{Single channel}: The temperature map possesses only one channel characterizing the captured thermal radiation from the target, unlike the multi-channel RGB images. 
The combination of these properties makes temperature maps a unique data type, posing significant challenges for 3D hand pose reconstruction using thermal arrays, a novel problem that requires a new solution.

Despite these challenges, thermal arrays offer many advantageous features for practical hand pose reconstruction, as they are \textbf{low-cost}, \textbf{privacy-preserving}, and, importantly, \textbf{illumination-immune}. 
This motivates us to overcome these challenges and design \sysname for 3D hand pose reconstruction using a thermal array sensor. 

\subsection{Thermo-Pose Modeling}
\label{ssec:key_insights}

We investigate the principles of thermography theory \cite{mollmannInfraredThermalImaging2018} to see if it is possible to derive depth and position directly from the temperature readings, an essential component for 3D hand pose reconstruction. 

\head{Thermo-Depth Relation}
Thermography theory explains how a thermal detector measures the radiant power emitted from a target, which is then converted to temperature readings. 
As shown in \fig\ref{subfig:diagran}, the measured radiant power $d\Phi$ of the thermal radiance from a unit surface area $dA_2$ to a unit detector area $dA_1$ is expressed as {\cite{budzierThermalInfraredSensors2011a}
\begin{equation}
    \Phi(d) = \epsilon e^{- \gamma d/ \cos{\theta_1}} L \cdot A_1\int_{A_2}  \frac{\cos \theta_1 \cos \theta_2}{ {(r/ \cos{\theta_1})}^2} dA_2,
    \label{eq:total_radiant_power}
\end{equation}
where $L$ is the radiance from the source, $r$ is the distance between surfaces $dA_2$ and $dA_1$, $\theta_1$ and $\theta_2$ are angles between the normal vectors of the surfaces and the radiation direction. $\gamma$ is the attenuation coefficient, and $\epsilon$ denotes the emissivity of the target, \ie, human hand.

Ideally, \eqn\eqref{eq:total_radiant_power} should provide an opportunity to derive depth from the measured temperature $T$, given that $T$ can be calculated from and is proportional to the captured radiant power, \ie, $T \propto \sqrt[4]{\Phi}$. 
In practice, however, this intuitive approach does not work. The main reason is that the source radiance $L$ and the effective radiant area $A_2$ heavily depend on the target temperature, surface coverings, environmental heat interference, etc, all time-varying and unpredictable. 
Additionally, measurement noise in low-cost thermal sensors adds further complexity.
Specifically, as shown in \fig\ref{subfig:distance_observation}, we make the following observation: 

\head{\textit{Observation I}} \textit{The captured thermal radiant power has a non-linear and negative correlation with hand depth.} \\
This consistent, monotonous thermo-depth relationship, when combined with the imaging resolution, presents an opportunity to address the problem via data-driven learning (\S\ref{sssection:thermo-sapce-encoder}).

\begin{figure}[t]
    \subfloat[Temperature vs. depth.]{%
        \label{subfig:distance_observation}
          \includegraphics[width=0.33\textwidth]{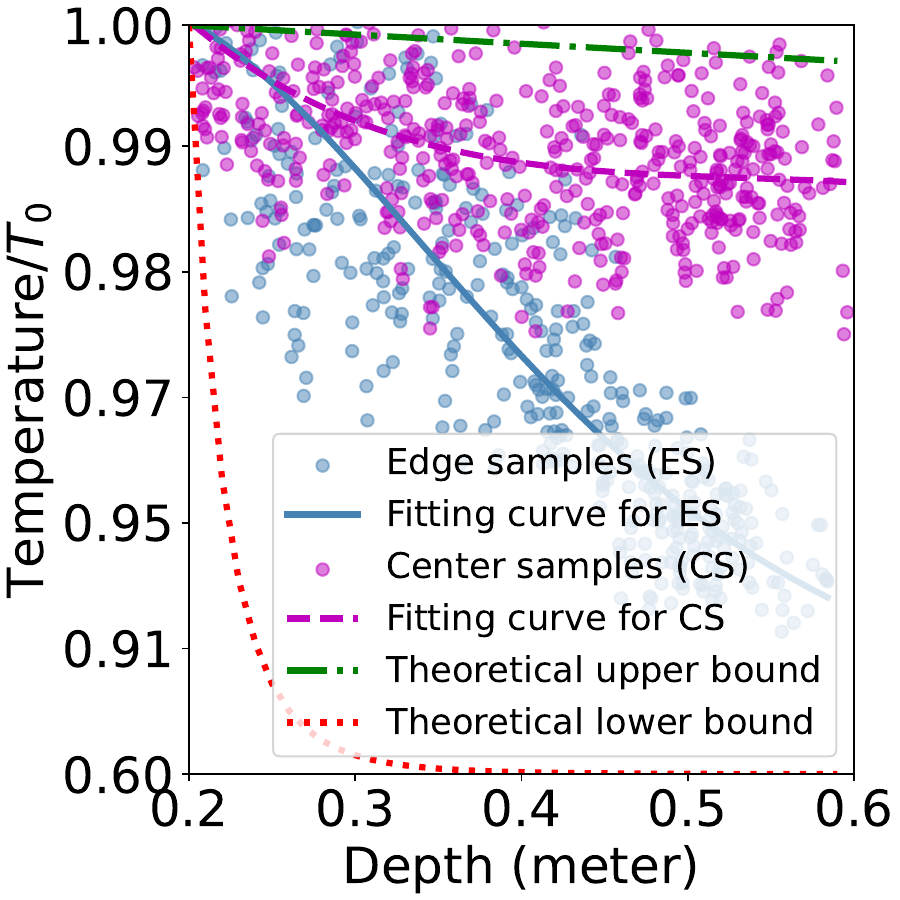}
    	  }
    \subfloat[Radiation area vs. depth.]{%
    \label{subfig:pose_distance}
      \includegraphics[width=0.45\textwidth]{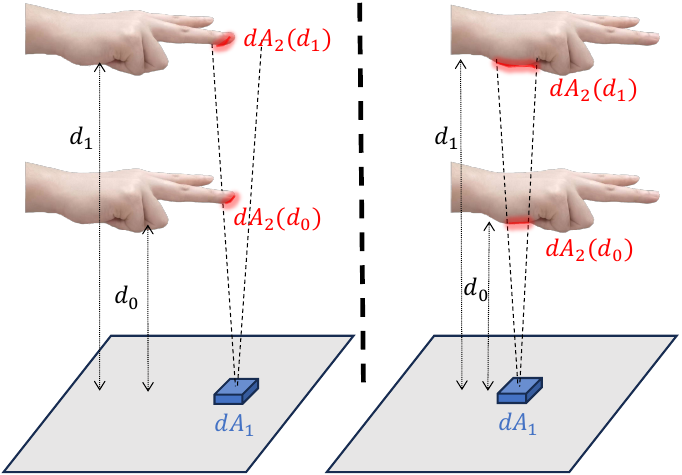}
	  }
    \vspace{-0.6\baselineskip}
    \caption{Thermo-depth analysis. {\rm (a) The detected temperature of hand key points overall decreases, at a large scale, regarding depths. Yet the decreasing trends vary significantly, depending on the locations of the key points relative to the hand. $T_0$ is the detected temperature at a reference depth of 0.2 meters. This phenomenon is explained in (b): The radiation area $A_2$ increases with larger depth for center key points, compensating for the increased air attenuation. The radiation area, however, remains nearly constant for edge key points, leading to more drastic temperature decreases.}}
	\label{fig:distance_theory}
\end{figure}

\head{Thermo-Pose Relation}
The above analysis models the hand as a single body. 
To further understand the relationship, we dive into the detailed hand pose structure and model its impact on the captured temperature map. 
Recalling \eqn\eqref{eq:total_radiant_power} and the relation of $T \propto \sqrt[4]{\Phi}$, the temperature readings for different key points of the hand should be similar, since they are all at approximately the same depth. 
As shown in \fig\ref{subfig:distance_observation}, however, both the temperature readings and the changing trends with respect to depth vary significantly across different key points. 

To understand the root causes, we can formulate the impacts of depth on different key points as follows. 
As shown in \fig\ref{subfig:pose_distance}, for key points at the edge of the hand, depth changes do not significantly affect the total radiant surface area $A_2$. Thus, according to \eqn\eqref{eq:total_radiant_power}, the captured temperature decreases with respect to $d$ as follows:
\begin{equation}
    \frac{\hat{T}(d_1)}{\hat{T}(d_0)} \propto e^{- \gamma / \cos{\theta_1} \frac{d_1}{d_0}} \cdot \frac{d_0^2}{d_1^2},
    \label{eq:temperature_distance_lb}
\end{equation}
Conversely, since the radiant surface area $A_2$ increases proportionally to $d^2$, the captured temperature of key points in the center of the hand is less impacted by depth and changes differently as:
\begin{equation}
    \frac{\hat{T}(d_1)}{\hat{T}(d_0)} \propto e^{- \gamma / \cos{\theta_1} \frac{d_1}{d_0}}.
    \label{eq:temperature_distance_ub}
\end{equation} 
\fig\ref{subfig:distance_observation} demonstrates both the theoretical bounds and practical measurements of these highly diverse trends, where the theoretical bounds are obtained based on  \cite{mollmannInfraredThermalImaging2018} by setting $d_0 = 0.2 m$, $\theta_1 = 0$, and $\gamma = 0.025$. 
On this basis, we make another key observation as follows: 

\head{\textit{Observation II}} \textit{The captured temperature maps exhibit distinct 2D spatial patterns for different hand poses with different key point structures, even at approximately the same depth.} \\
Therefore, subtle variations corresponding to the hand region in the captured temperature maps will imply changes in hand poses, allowing rich latent information for learning the detailed pose structure (\S\ref{sssection:kp-feature-extracor}). 

In the next section, we explore leveraging these observations in a novel neural network to learn positional information and key point features for 3D hand pose reconstruction. 

\begin{figure*}[t]
    \centering
    \includegraphics[width=0.98\textwidth]{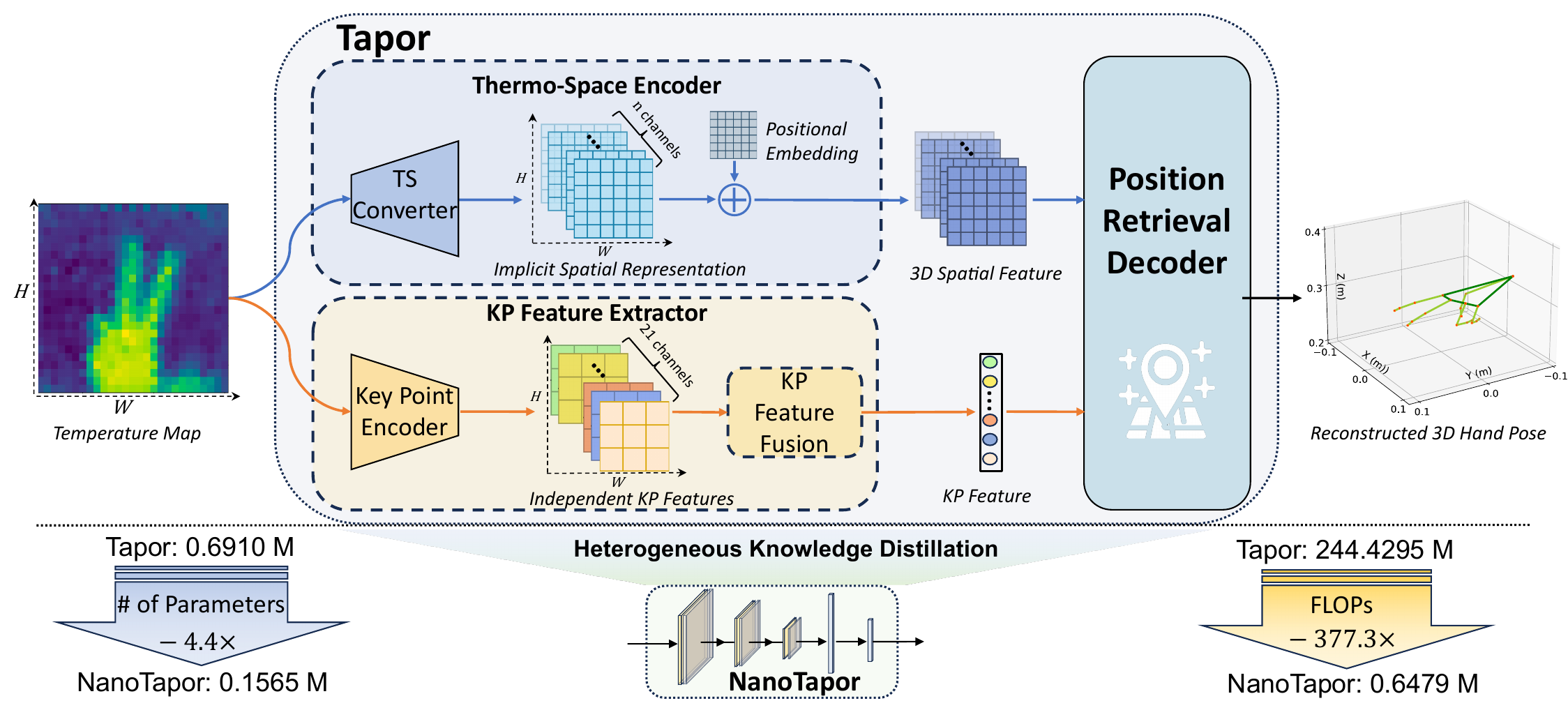}
    \caption{A structure overview of the Tapor and NanoTapor models.}
    \label{fig:model_overview}
\end{figure*}

\section{Physics-inspired Design}
\label{sec:method}

This section presents our physics-inspired neural network design, followed by its ultra-lightweight variant for mobile and IoT devices. 

\subsection{Overview}
\label{ssec:overview}
Tapor model consists of three main components: Thermo-Space Encoder, Key Point (KP) Feature Extractor, and Position Retrieval Decoder, as depicted in \fig\ref{fig:model_overview}.

Like RF ranging, thermal radiation attenuation, as described in Observation I in \S\ref{ssec:key_insights}, can be utilized for range/depth estimation. 
However, direct depth estimation based solely on temperature values results in large errors, \eg, about $\pm 20 cm$ for human bodies \cite{naserHumanDistanceEstimation2021, zhang2024tadar}, which can be even worse for the significantly smaller hands. 
In \sysname, we exploit a complementary opportunity thanks to the 2D imaging capability of the thermal arrays. 
By leveraging the contextual information (\eg, 2D hand shape) in the low-resolution temperature map, we establish a context-aware mapping, called the thermo-space encoder, that converts detected hand temperature into a 3D spatial representation of the hand's depth and all possible poses.
This approach reduces the depth estimation error to around $2 cm$, as demonstrated in our experiments (\S\ref{sec:exp}). 

On this basis, unlike previous works that mostly consider a regression \cite{li3DHumanPose2015,oberwegerHandsDeepDeep2016,zhangMediaPipeHandsOndevice2020} or detection problem \cite{ramakrishnaReconstructing3DHuman2012,tomeLiftingDeepConvolutional2017}, we formulate the 3D hand pose reconstruction as a \textit{retrieval task}. 
To reconstruct the 3D hand pose, we first draw inspiration from Observation II (\S\ref{ssec:key_insights}) and build a Key Point Feature Extractor to extract key point features from the 2D temperature maps with the help of common hand pose knowledge. 
We then design a Position Retrieval Decoder to utilize the key point features as queries to retrieve the 3D position of each key point from the 3D spatial representation.
At a high level, this process can be viewed as finding the unique solution (\ie, the output 3D hand pose) from the possible solution space (\ie, the 3D spatial representation) with constraints (\ie, the key point features). 
By doing so, \sysname reconstructs the 3D hand pose $J \in \mathbf{R}^{21\times 3}$, represented by the 3D physical positions of 21 key points, from the temperature map $T \in \mathbf{R}^{32 \times 24}$.

\begin{figure}[t]
    \includegraphics[width=0.75\textwidth]{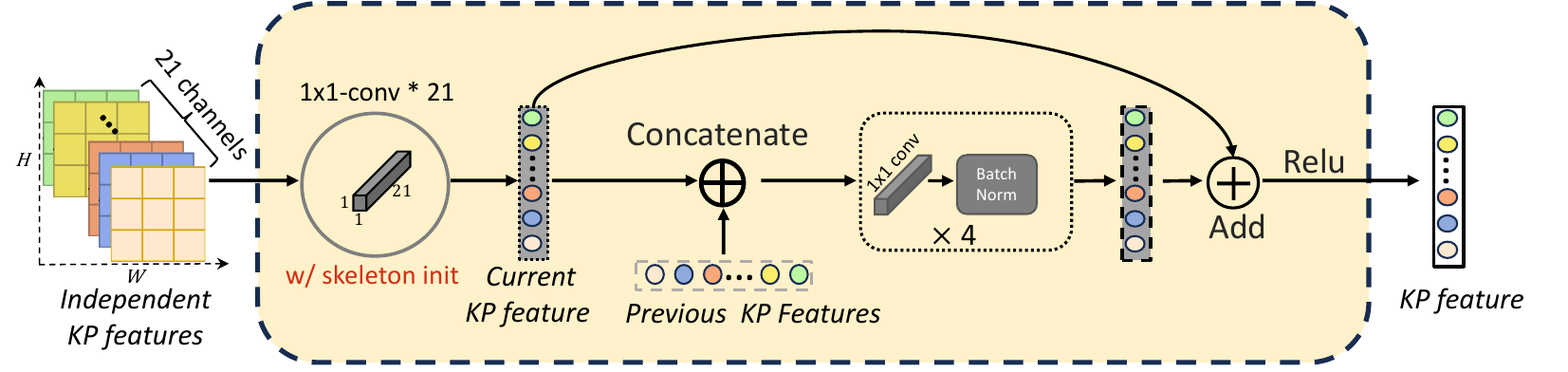}
    \caption{The KP feature fusion module. {\rm 
    This module facilitates information exchange across different key points and includes an optional residual connection for feature fusion from previous inputs.
    }}
\label{fig:kp_feature_fusion}
\end{figure}

\begin{figure}[t]
    \subfloat[The initialized weights.]{%
    \label{subfig:kernel_initial}
      \includegraphics[width=0.25\textwidth]{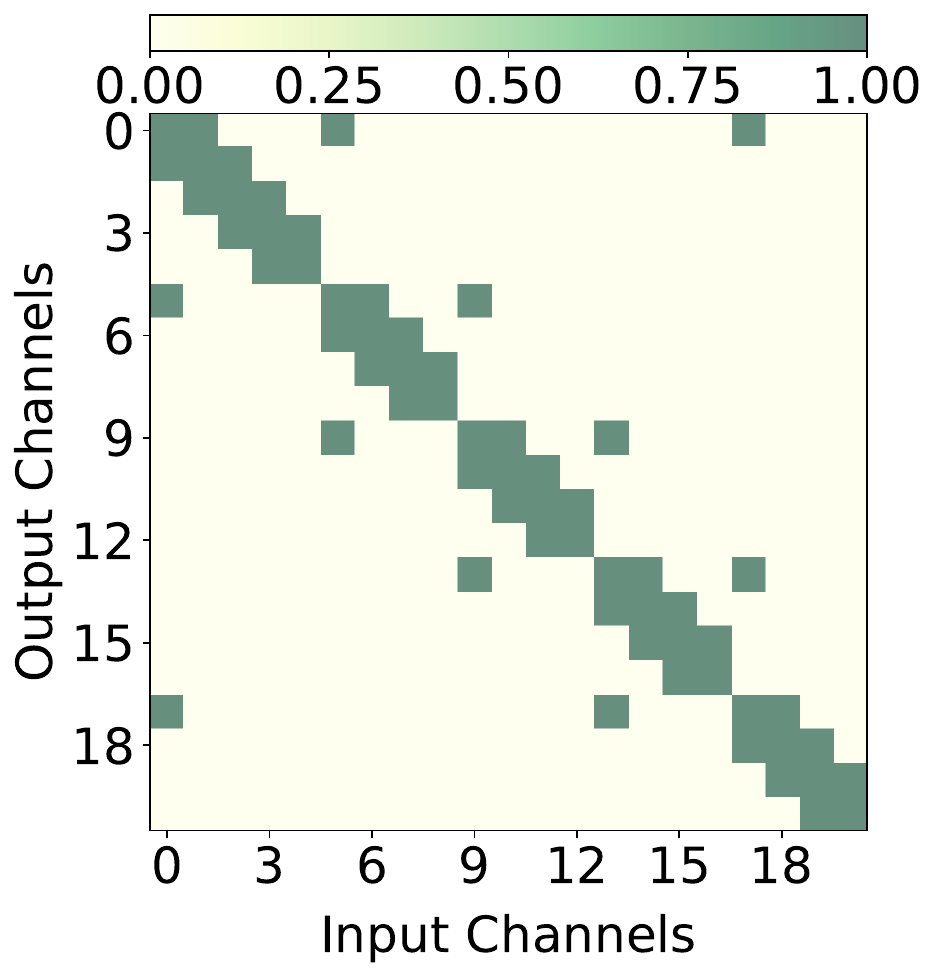}
	  }
    \subfloat[The weights after training.]{%
    \label{subfig:kernel_trained}
      \includegraphics[width=0.25\textwidth]{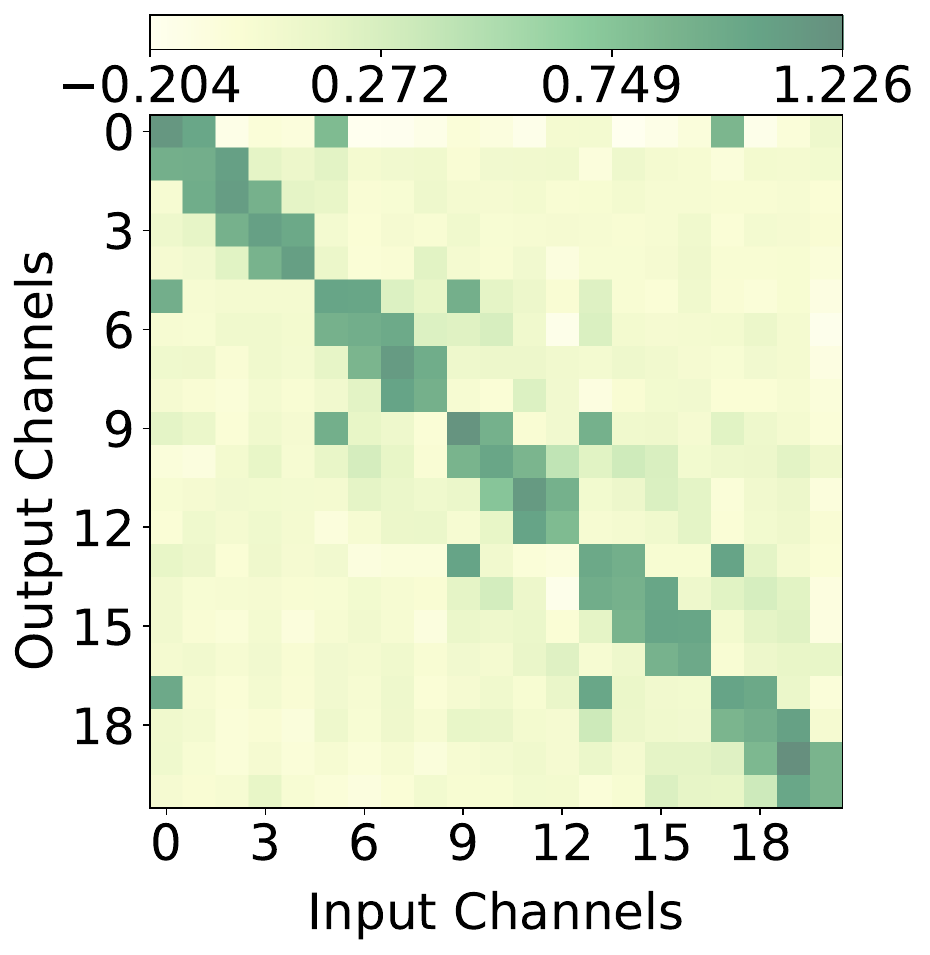}
	  }
    \caption{Skeleton init. {\rm 
    (a) Initial weights of the 1x1 convolutional kernel based on hand structure. (b) The learned weights after training retain hand structure knowledge.
    }}
	\label{fig:kernel_weights}
\end{figure}

\subsection{Tapor Model Design}
\label{ssec:modules}

\subsubsection{Thermo-Space Encoder}
\label{sssection:thermo-sapce-encoder}
The thermo-space encoder is dedicated to constructing a representation of all possible 3D hand poses from the current temperature map.
To achieve this, we introduce a Temperature Space (TS) converter that transforms the temperature values and contextual information into an implicit spatial representation of possible hand locations and poses.
By integrating a positional embedding with the implicit spatial representation, the thermo-space encoder produces a 3D spatial feature that describes the potential solution space of the hand pose. 

The TS converter, represented by $E_{ts}(\cdot)$, extracts information from the input temperature map, $T$. For model efficiency, we employ the inverted residual block from MobileNetV2 \cite{sandlerMobileNetV2InvertedResiduals2018} to construct the converter. However, the convolution operation reduces the $H$ and $W$ dimensions of $T$ as shown in \fig\ref{fig:model_overview}. 
To counter this, we use an upsample layer to scale up the original dimensions along the $H$ and $W$ axes, with an upscaling factor $\alpha$. 
The output of the TS converter is the implicit spatial representation, denoted as $f_{ISR} \in \mathbf{R}^{n \times 15 \times 20}$, where $n=128$ is the number of channels.
The 3D spatial feature is then produced as $f_{spatial} = E_{ts}(T) \oplus P_e$, where $P_e$ is the positional embedding \cite{vaswani2017attention}, $\oplus$ represents element-wise addition, and $f_{spatial} \in \mathbf{R}^{128 \times 15 \times 20}$. 
Thus, the input $32 \times 24$ 2D temperature space is extended to the 3D spatial feature space with dimensions $128 \times 15 \times 20$.

\subsubsection{KP Feature Extractor}
\label{sssection:kp-feature-extracor}
Extracting distinct features for all key points is essential for accurately retrieving their locations from the 3D spatial feature space.
We first introduce a key point encoder to derive the feature of each key point directly from the entire temperature map, capturing both hand shape and temperature information.
To address the challenges of occluded key points and reconstructing a realistic hand pose, we design an efficient KP feature fusion module. 
This module compensates for missing information in occluded key points by leveraging features from other key points and incorporating information from previous inputs.
It further enhances the coherence of the hand structure by incorporating prior knowledge of hand anatomy, enabling the extraction of all key point features without the need to see all fingers.

\begin{figure}[t]
    \includegraphics[width=0.85\textwidth]{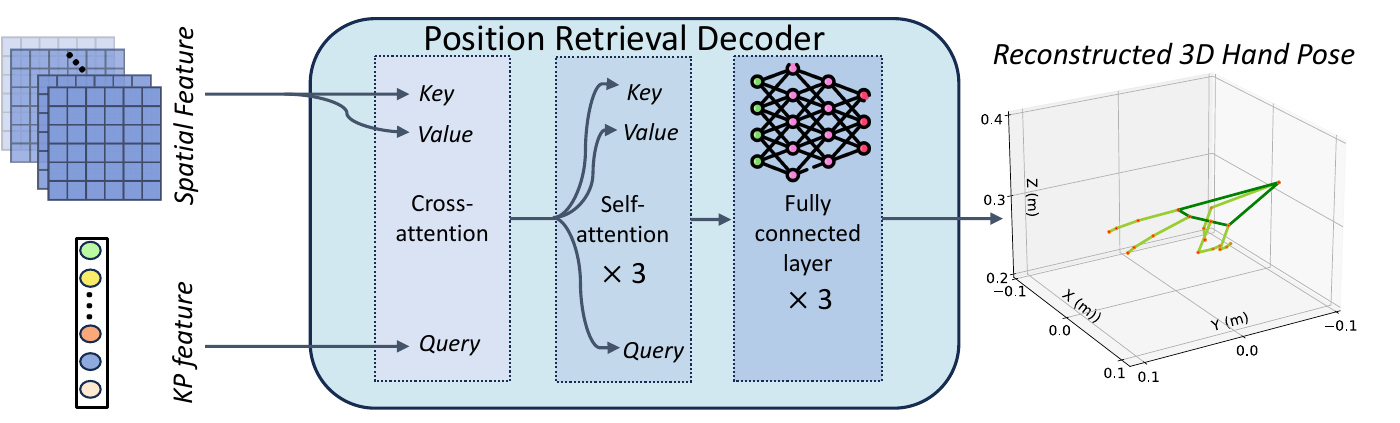}
    \caption{Position retrieval decoder structure. {\rm }}
\label{fig:attention_decoder}
\end{figure}

Specifically, the key point encoder, $E_{kp}(\cdot)$, shares a similar structure with the TS converter but with two key differences: 
(1) it does not need to preserve the $H$ and $W$ dimensions, so it skips the upsample layer, and 
(2) each channel in its output feature maps represents the features of a corresponding key point, with the total number of channels equal to the number of key points, which is 21. 
The encoder produces independent key point features as $f_{ikp} = E_{kp}(T)$, where $f_{ikp} \in \mathbf{R}^{21 \times 6 \times 8}$. In the KP feature fusion module, $E_{kpf}(\cdot)$, as depicted in \fig\ref{fig:kp_feature_fusion}, the independent key point features $f_{ikp}$ are first processed with a $1\times1$ convolution $\kappa$ to enable information exchange.
Additionally, four cascaded convolutional blocks merge the current feature with preceding features, incorporating temporal coherence.
Given the sensor's variable refresh rate, a residual connection is introduced to make this component optional without altering the model's overall structure. 
Eventually, the KP feature extractor outputs the distinct features, $f_{kp} \in \mathbf{R}^{21 \times 48}$, of the 21 key points.

To inject hand structure knowledge, we propose a \textit{skeleton init} method to initialize the convolutional kernel $\kappa$.
As shown in \fig\ref{fig:kernel_weights}, we tailor the kernel weights to the hand's structure by assigning a value of 1 to diagonal elements and elements corresponding to adjacent key points, while other elements are set to 0. 
After training, the learned weights maintain this pattern, effectively preserving the hand's structural information.
The skeleton init technique boosts Tapor's performance without additional computational costs, outperforming random initialization, as evidenced in \S\ref{ssec:ablation_study}.

\subsubsection{Position Retrieval Decoder}
\label{sssection:position-retrieval-decoder}
The position retrieval decoder aims to accurately identify the position of each key point. 
As shown in \fig\ref{fig:attention_decoder}, it features a cross-attention layer, $D_{cross}(\cdot, \cdot)$, which uses the KP feature $f_{kp}$ as a query to identify its corresponding location feature $f_{loc}$ within the 3D spatial feature $f_{spatial}$, \ie, $f_{loc} = D_{cross}(f_{spatial}, f_{kp})$. Then, three self-attention layers, $D_{self}(\cdot)$, fuse the location features of all key points to produce the pose feature, $f_{pose} = D_{self}(f_{loc})$. Finally, three cascaded fully connected layers, $L(\cdot)$, convert the pose feature into 3D physical positions of all hand key points, reconstructing the hand pose $\hat{J} \in \mathbf{R}^{21 \times 3}$. 

To understand whether the model functions as intended, we inspect the model in the following two aspects: 
(1) We assess the model's operation to ascertain whether it concentrates on the hand region (\ie, the context). As illustrated in \fig\ref{subfig:att_in_hand}, the attention map of the cross-attention layer clearly demonstrates the model's focus on the hand region, signifying that the learned model is context-aware. (2) Moreover, to confirm whether the key point features possess distinct semantic meanings (\ie, each feature corresponding to a specific hand key point), we showcase attention maps for various hand parts generated using the relevant key point features, as depicted in \fig\ref{subfig:att_finger}. The attention maps distinctly highlight their target hand regions. Additionally, the attentions to different fingers and the root point consistently emphasize the palm's center, which aligns with our objective to reconstruct an accurate and coherent 3D hand pose.



\begin{figure}[t]
    \subfloat[Attention map of the hand.]{%
    \label{subfig:att_in_hand}
      \includegraphics[width=0.26\textwidth]{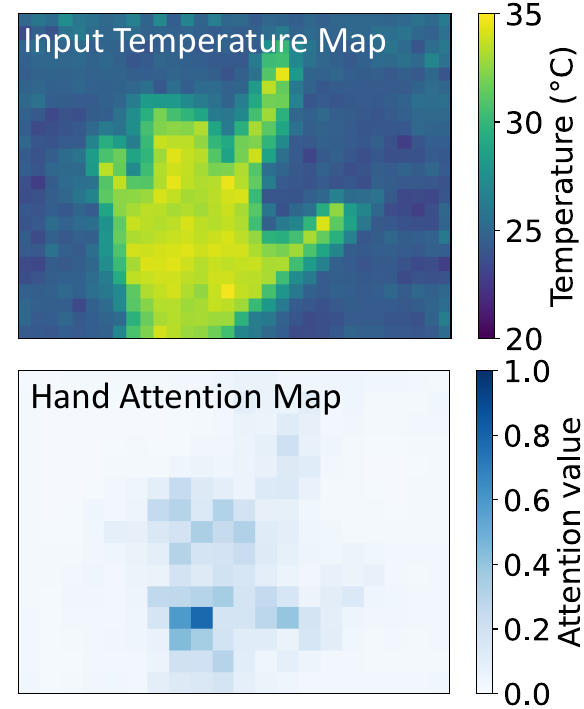}
	  }
    \subfloat[Attention maps of root and fingers.]{%
    \label{subfig:att_finger}
      \includegraphics[width=0.64\textwidth]{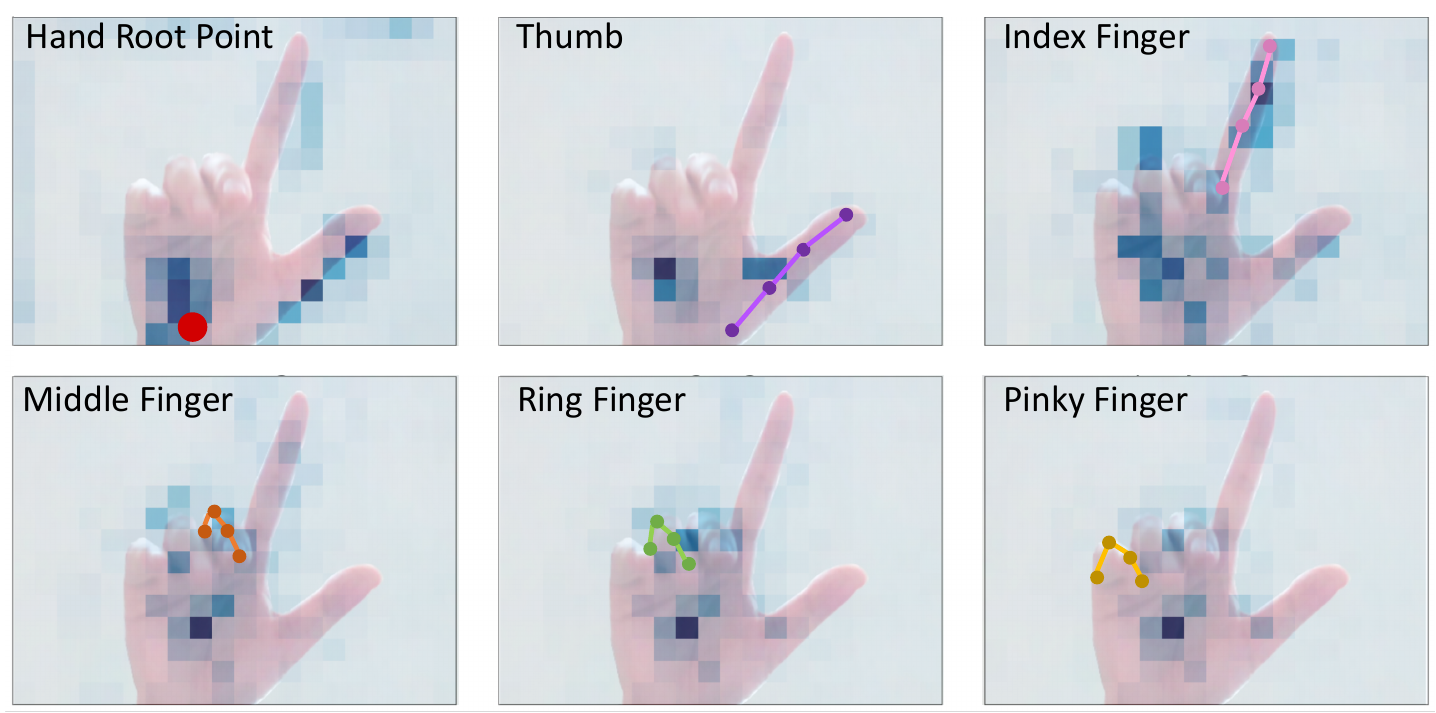}
	  }
    \caption{Attention maps of \sysname. {\rm (a) The attention map of the cross-attention layer highlights the model's correct focus on the hand region, indicating effective context awareness. (b) The attention maps demonstrate that the key point encoder effectively interprets the semantic meanings of the input temperature map. Note that the RGB images are included solely for visualization purposes.}}
	\label{fig:att_maps}
\end{figure}

\head{Training Loss}
\label{ssec:end2end_training}
To train Tapor, we use two loss functions: the mean square error (MSE) loss and the bone length (BL) loss \cite{sun2017compositional}. The MSE loss, $\mathcal{L}_{mse} = \frac{1}{N} \sum_{i=0}^{N}{\| J-\hat{J} \|_{2}^{2}}$, where $N$ is the number of training samples, will cause the reconstructed pose to collapse around the palm's center. To address this, we add the BL loss, $\mathcal{L}_{bl} = \frac{1}{N} \sum_{i=0}^{N}{\| \mathbf{b_i} - \mathbf{\hat{b_i}} \|_{2}^{2}}$, where $\mathbf{b_i}$ and $\mathbf{\hat{b_i}}$ are the bone lengths of the ground truth and reconstructed poses, respectively. 
Thus, the overall loss function is defined as follows for end-to-end training of the model: 
\begin{equation}
\mathcal{L}_{Tapor} = \mathcal{L}_{mse}(\hat{J}, J) + \mathcal{L}_{bl}(\hat{J}, J).
\label{eq:tapor_loss}
\end{equation}

\subsection{NanoTapor: Extending to the IoT}
\label{ssec: lightweight}
Despite Tapor's lightweight design with fewer parameters and floating-point operations per second (FLOPs) than all baseline models, as shown in \S\ref{sec:exp}, it remains too large for deployment on IoT devices like the ESP32 MCU. 
To enable \sysname on IoT devices and ensure real-time operation, we introduce a novel method of \textit{heterogeneous knowledge distillation} to deliver a tiny version named NanoTapor.

Inspired by EdgeSAM \cite{zhou2023edgesam}, NanoTapor is designed as a purely CNN-based architecture, making it well-suited for edge devices.
Specifically, NanoTapor comprises four convolutional layers as the encoder $E_{nano}(\cdot)$ to extract features $f_{n}$ from the input temperature map $T$, followed by a fully-connected layer $D_{nano}(\cdot)$ to transform the features into the final pose, \ie, $\hat{J} = D_{nano}(f_{n})$.

\begin{figure}[t]
    \includegraphics[width=0.68\textwidth]{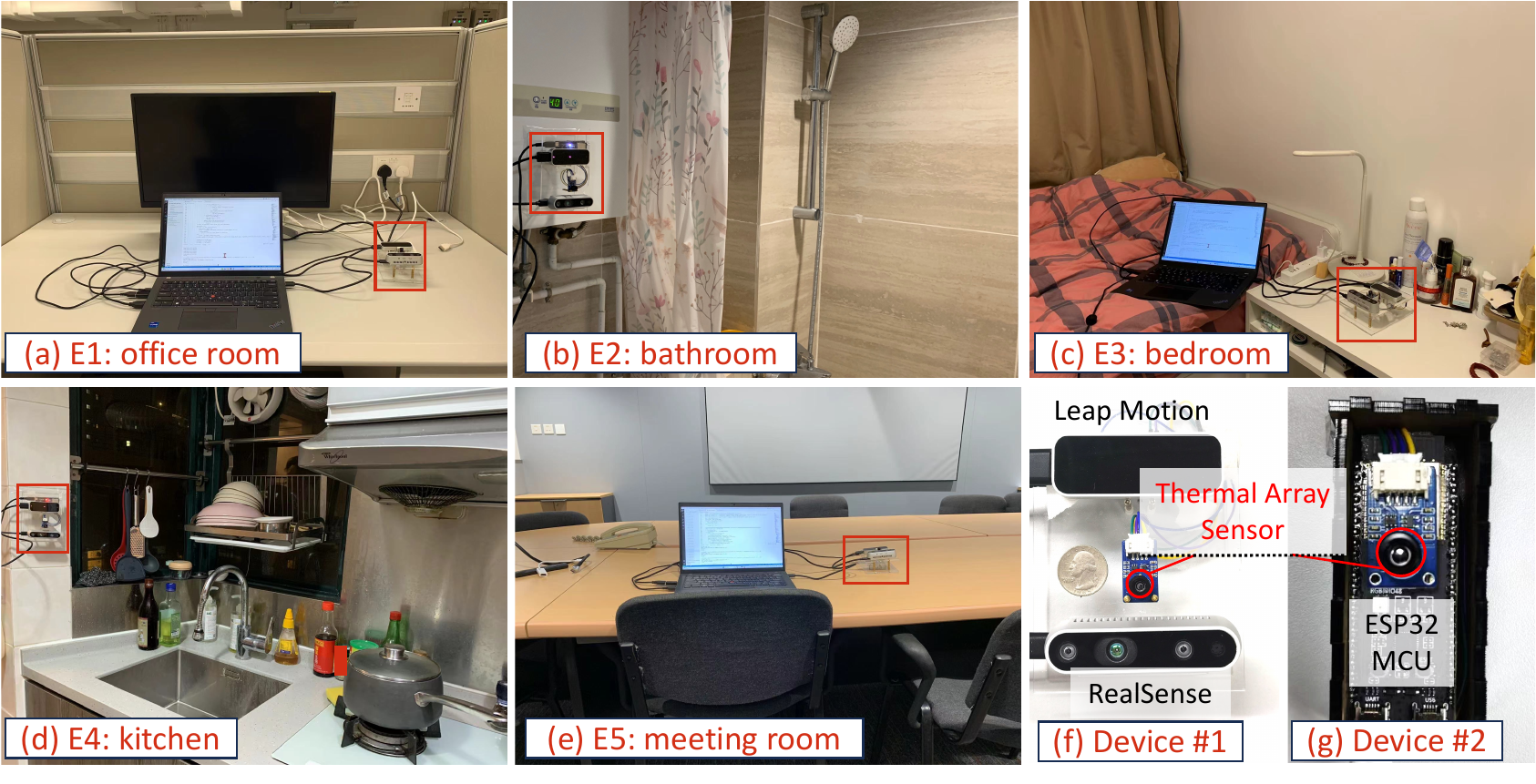}
    \caption{Experimental environments and devices. {\rm We conduct experiments in various settings (a)-(e). (f) Device setup for data collection. (g) IoT prototype on ESP32-S3.}}
\label{fig:data_env}
\end{figure}

To transfer knowledge from Tapor to NanoTapor, we employ knowledge distillation \cite{hinton2015distilling} by adding an extra feature distillation loss to the Tapor loss $\mathcal{L}_{Tapor}$:
\begin{equation}
    \mathcal{L} = \mathcal{L}_{mse}(\hat{J}, J) + \mathcal{L}_{bl}(\hat{J}, J) + \mathcal{L}_{mse}(f_n, f_{Tapor}),
    \label{eq:temp_loss}
\end{equation}
where $f_{Tapor}$ is the feature from the Tapor model used as supervision. 
However, standard knowledge distillation requires matching dimensions for $f_n$ and $f_{Tapor}$, which restricts the room for reducing parameters and computational overhead for NanoTapor and further hinders the ability to operate in real-time on IoT devices. For instance, processing one input temperature map takes more than 1.5 seconds on the ESP32S3 MCU, which is impractical for most interactive applications.

To tackle this issue, a straightforward solution is to reduce the dimension of $f_n$. This necessitates addressing the challenge of transferring knowledge from the teacher model, Tapor, to the student model, NanoTapor, despite their heterogeneous feature dimensions and structures. 
To overcome this challenge, we introduce an adaptor $A: \mathbf{R}^{21 \times 48} \to \mathbf{R}^{21 \times 16}$ to align the Tapor and NanoTapor models. We explore two types of adaptors: a non-learnable dimension reduction algorithm, namely principal component analysis (PCA), and a learnable linear model. Due to NanoTapor's heterogeneous structure, the PCA method fails to adapt and leads to performance degradation, as demonstrated in \S\ref{ssec:ablation_study}. Consequently, we opt for a learnable linear layer as the adaptor $A(\cdot)$, which offers superior performance. Finally, the training loss function for NanoTapor is defined as:
\begin{equation}
    \mathcal{L}_{NanoTapor} = \mathcal{L}_{mse}(\hat{J}, J) + \mathcal{L}_{bl}(\hat{J}, J) + \delta \mathcal{L}_{mse}(f_n, f_{super}),
    \label{eq:NanoTapor_loss}
\end{equation}
where $f_{super} = A(f_{Tapor})$ is the supervision feature generated from the Tapor model, and $\delta$ is set to 0.1 empirically.

\section{Implementation}
\label{sec:impl}
\head{Hardware} We fully implement \sysname using the Melexis MLX90640BAB thermal array sensor, featuring a $32 \times 24$ element array with a $35^{\circ} \times 55^{\circ}$ FOV. The FOV can be expanded to $75^{\circ} \times 110^{\circ}$ with the MLX90640BAA model. The sensor’s refresh rate, configurable between 0.5 Hz and 64 Hz, is set to 10 Hz for our experiments. NanoTapor runs on an ESP32-S3 board with an ESP32S3-WROOM-1 MCU.

\head{Software} We develop the Tapor and NanoTapor models using PyTorch \cite{paszke2019pytorch}. 
The NanoTapor model is then converted to TensorFlow Lite Micro \cite{david2021tensorflow} using the ONNX format \cite{bai2019onnx} for deployment on IoT devices. 
We have implemented an end-to-end system for both Tapor and NanoTapor to run on diverse devices like Jetson, Raspberry PI, and ESP32. 

\head{Training} Tapor and NanoTapor models are trained using the Adam optimizer with a learning rate of 0.0001. 
Tapor is trained with a batch size of 48 for 200 epochs, while NanoTapor uses a batch size of 128 for 1000 epochs.
All firmware, code, data, models, and demo videos are available here: \href{https://github.com/aiot-lab/TAPOR}{\color{blue}\texttt{https://github.com/aiot-lab/TAPOR}}.

\begin{figure*}[t]
    \centering
    \includegraphics[width=0.96\textwidth]{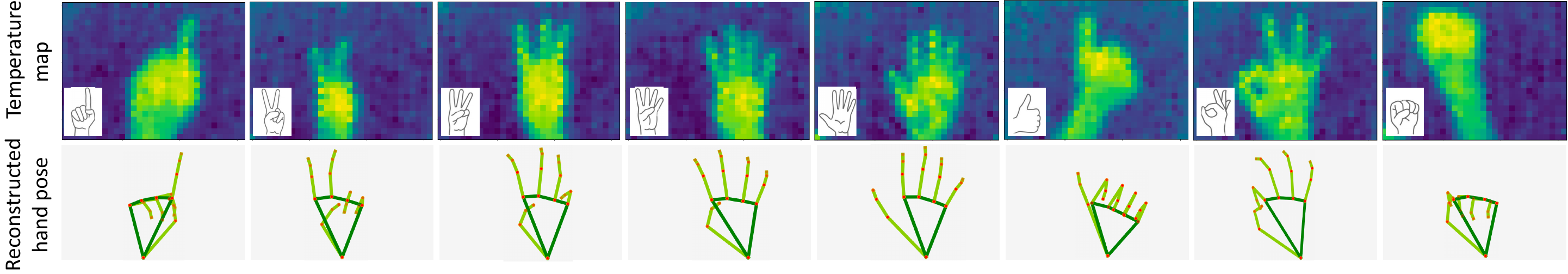}
    \vspace{-0.15in}
    \caption{Reconstructed 3D hand poses by \sysname. \rm{The top row displays the input temperature maps alongside their corresponding gesture icons. The bottom row presents the top-down view of the reconstructed 3D hand poses.}}
    \label{fig:reconstruction_examples}
\end{figure*}

\section{Experiments}
\label{sec:exp}

\subsection{Experimental Setup}
\label{sec:dataset_metrics_baselines}
\head{Dataset} 
To evaluate \sysname, we build a dataset comprising over 60,000 arbitrary hand pose samples (\ie, 32$\times$24 temperature maps). 
As shown in \fig\ref{fig:data_env}, the dataset is gathered in five different environments covering different use cases: an office, a bathroom, a bedroom, a kitchen, and a meeting room. 
Ten participants (8 males, 2 females), aged 21–27, with hand widths from 9.4 to 13.1 cm and lengths from 17.3 to 20.5 cm, contributed to the dataset. All procedures were approved by our institution’s IRB.
During each trial, participants freely moved their hands above the upward-facing sensor, rotating and translating them within the FoV to cover diverse angles and positions, typically with palms facing downward in a natural interaction posture.
This ensures variation in hand orientation and position while remaining within the effective sensing range. 
For ground truth data, we use the Leap Motion to capture 3D hand poses, while the RealSense D435 depth camera is utilized to record depth and RGB images for our investigative experiments in \S\ref{ssec:investigative_exp}. 
To train the Tapor and NanoTapor models, we use a subset of data from six users in the office for training, while the rest is reserved for testing. Specifically, the dataset is split into 26,557 training samples, 3,319 validation samples, and 34,520 testing samples. 

\head{Metrics}  
To assess the performance of \sysname in 3D hand pose reconstruction, we employ the following three metrics: 

{\noindent \textit{(i) Mean Per Joint Position Error (MPJPE)}} \cite{ionescu2013human3,chenModelBased3DHand2021}: MPJPE is the average Euclidean distance between estimated key points $\hat{\mathbf{j}}$ and ground truth $\mathbf{j}$: $\text { MPJPE }=\frac{1}{J}\sum_{i=0}^J\left(\left\|\hat{\mathbf{j}}_i-\mathbf{j}_i\right\|_2\right)$, where $J = 21$ is the number of key points.

{\noindent \textit{(ii) Percentage of Correct Key points (PCK)}} \cite{yangSemiHandSemisupervisedHand2021}: PCK calculates the percentage of correctly estimated joints, with a joint considered correct if its distance from the ground truth is below a specified threshold $a$: $\text {PCK@}a=\frac{1}{J}\sum_{i=1}^J \mathrm{1}{\left(\left\|\hat{\mathbf{j}}_i-\mathbf{j}_i\right\|_2 \leq a\right)}$, where $\mathrm{1}(\cdot)$ is the indicator function.

{\noindent \textit{(iii) Mean Per Joint Relative Position Error (MPJRPE)}}: For many tasks, including gesture recognition, the relative structure of the hand pose is adequate. Therefore, we further propose MPJRPE to measure the average Euclidean distance between estimated and ground truth joint coordinates after aligning their root points (0-th joint): $\text {MPJRPE}=\frac{1}{J}\sum_{i=0}^J\left(\left\|\hat{\mathbf{j}}_i^{'}-\mathbf{j}_i^{'}\right\|_2\right)$
where $\hat{\mathbf{j}}_i^{'} = \hat{\mathbf{j}}_i -\hat{\mathbf{j}}_0 $ and $\mathbf{j}_i^{'} = \mathbf{j}_i -\mathbf{j}_0 $ represent the shifted positions.

Note that lower MPJPE/MPJRPE values indicate better performance, while higher PCK signifies improved accuracy.

\begin{figure}[t]
    \subfloat[Complexity and performance.]{%
    \label{subfig:model_compare}
      \includegraphics[width=0.46\textwidth]{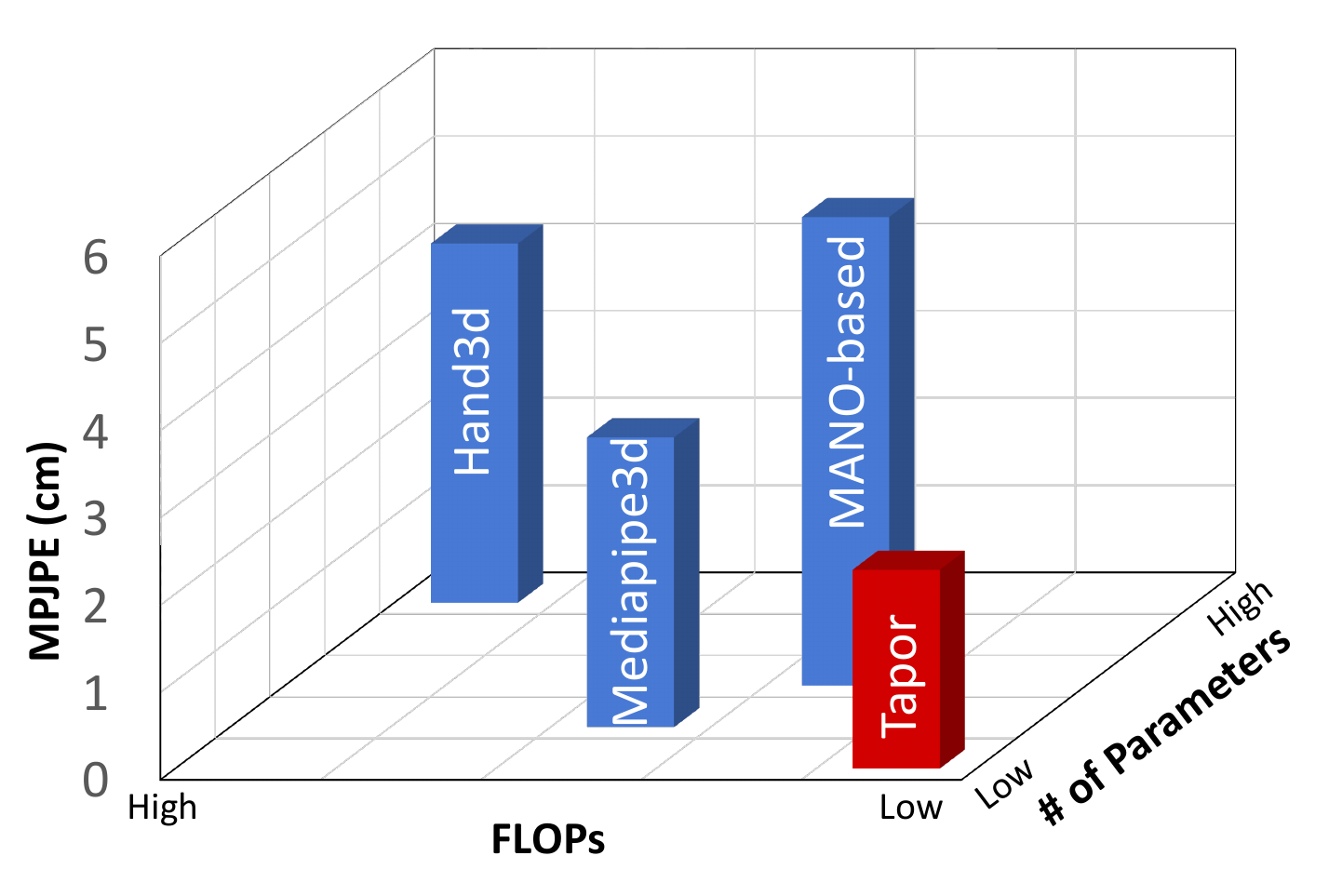}
	  }
    \subfloat[PCK curve.]{%
    \label{subfig:model_pck}
      \includegraphics[width=0.32\textwidth]{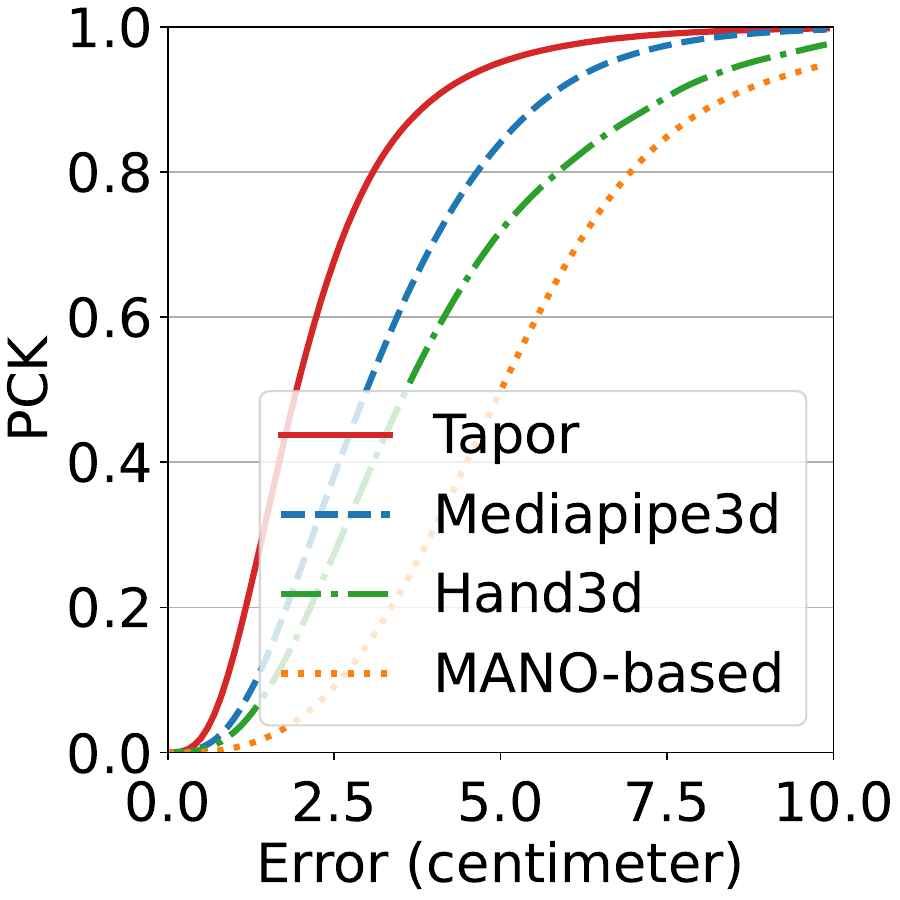}
	  }
   \hfill
   \subfloat[Performance comparison {\rm (\textbf{Best}, \underline{Second best})}.]
   {
   \label{subfig:model_compare_tb}
    
    \begingroup

    \renewcommand{\arraystretch}{1.3} 
    \begin{tabular}{ccccc}
    \hline
    Method         & MPJPE (cm)$\downarrow$         & PCK@4cm$\uparrow$               & FLOPs (M)$\downarrow$           & \# of Params (M)$\downarrow$         \\ \hline
    MANO-based           & 5.3681          & 0.3078            & 277.6867          & 2.2886          \\
    Hand3d             & 4.1137          & 0.5750           & 30103.7655          & 18.9612          \\
    Mediapipe3d      & 3.3196          & 0.7048          & 385.8148          & 2.0134         \\ 
    \textbf{NanoTapor}      & \underline{3.2683}          & \underline{0.7238}         & \textbf{0.6479}         & \textbf{0.1565}       \\ 
    \textbf{Tapor} & \textbf{2.2636} & \textbf{0.9041} & \underline{244.4295} & \underline{0.6910} \\ \hline
    \end{tabular}
    \endgroup
   }
   \vspace{-0.15in}
    \caption{Model comparison. \rm{NanoTapor, with its significantly lower FLOPs, is excluded in (a) for better visualization.}}
\label{fig:model_compare}
\end{figure}

\head{Baselines} 
\sysname is the first 3D hand pose reconstruction system using a single thermal array sensor.
To showcase the advantages of \sysname, we compare it with three mainstream deep learning baselines in our experiments:

{\noindent \textit{1) Mediapipe3d}} \cite{zhangMediaPipeHandsOndevice2020} is a well-known model that directly learns the mapping from RGB images to 3D hand poses. 

{\noindent \textit{2) Hand3d}} \cite{zimmermannLearningEstimate3D2017} employs a different approach by first detecting hand key points in the 2D image and then transforming these 2D key points to 3D physical space positions.

{\noindent \textit{3) MANO-based Model}} \cite{romeroEmbodiedHandsModeling2017} learns to map the input image to a parametric hand model derived from high-resolution 3D hand scans. The predicted parameters are then transformed into 3D key point positions using predefined rules. 

\textit{All baseline models have been adapted to accept temperature maps as input}. We adjust the input channel from 3 to 1 to match the single-channel temperature map and interpolate the map to match the baselines' input dimensions.

\subsection{Overall Performance}
\label{overall_perform}
We first illustrate examples of the reconstructed hand poses, as shown in \fig\ref{fig:reconstruction_examples}. 
As seen, \sysname reconstructs 3D hand poses accurately across a variety of gestures.

\head{Tapor vs. Baselines} As shown in \fig\ref{fig:model_compare}, Tapor outperforms the baseline models in all aspects, demonstrating the superiority of our physics-inspired design. Detailed in \fig\ref{subfig:model_compare_tb}, Tapor achieves an MPJPE of 2.26 cm and a PCK@4 cm of 90.41\%. 
NanoTapor, though slightly less accurate than Tapor, still exceeds all the baseline models, highlighting the effectiveness of the heterogeneous knowledge distillation technique. In terms of computational complexity, both Tapor and NanoTapor are remarkably more efficient than the baselines, with NanoTapor requiring only 0.65 million FLOPs and 0.16 million parameters.

\head{Tapor vs. NanoTapor} NanoTapor significantly reduces the computational cost, achieving 377.3$\times$ fewer FLOPs and 4.4$\times$ fewer parameters than Tapor. 
This efficiency translates to minimized processing latency, as shown in \fig\ref{subfig:cross_device_latency}. On a GPU (A100), Tapor and NanoTapor report 6.63 ms and 0.28 ms latencies, respectively. 
Tapor runs more slowly on embedded devices Jetson Nano and Raspberry Pi 4, with a respective latency of 54.41 ms and 319.06 ms. 
Notably, NanoTapor significantly reduces the latencies by 16$\times$ and 85$\times$ to 3.64 ms and 3.74 ms latencies respectively, making it the only model deployable on resource-constrained IoT. 
Specifically, NanoTapor achieves a latency of 313 ms on ESP32-S3, which can support real-time IoT applications. 
Despite its lightweight design, NanoTapor's performance in hand pose reconstruction is only slightly lower than Tapor's but still sufficient for downstream tasks. In gesture recognition, as shown in \fig\ref{subfig:gesture_reco_compare}, Tapor and NanoTapor achieve F1 scores of 94.9\% and 87.2\%, respectively, for six gestures detailed in \S\ref{ssec:case_study}, indicating effective recognition capabilities. Notably, both models achieve an F1 score of 100\% in the absence of a hand (Gesture \#0), underscoring \sysname's excellent performance in avoiding false alarms, which is crucial for practical gesture control applications.

\begin{figure}[t]
    \begin{minipage}{0.55\textwidth}
    \centering
    \subfloat[Latency across devices.]{%
    \label{subfig:cross_device_latency}
      \includegraphics[width=0.5\textwidth]{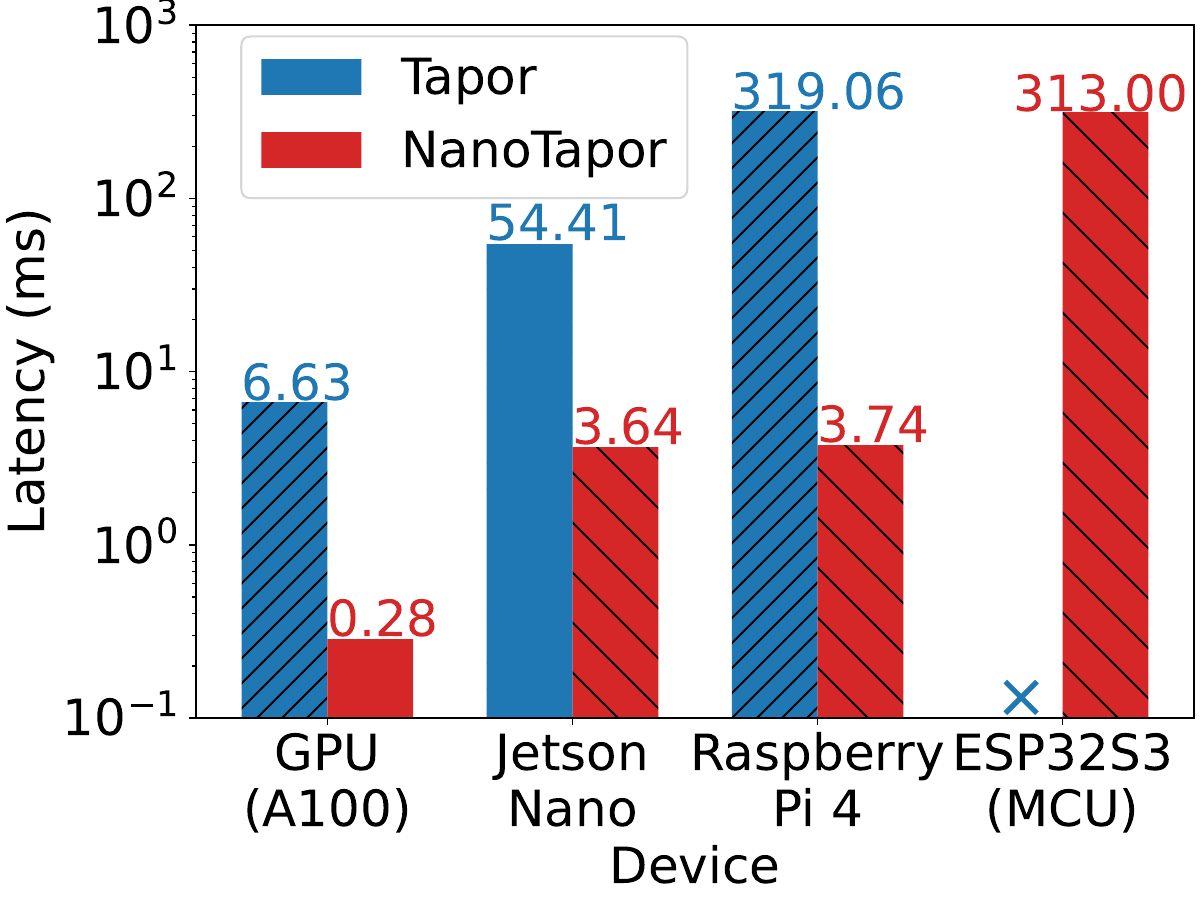}
	  }
    \subfloat[Gesture recognition f1 score.]{%
    \label{subfig:gesture_reco_compare}
    \raisebox{0.45\baselineskip}{\includegraphics[width=0.47\textwidth]{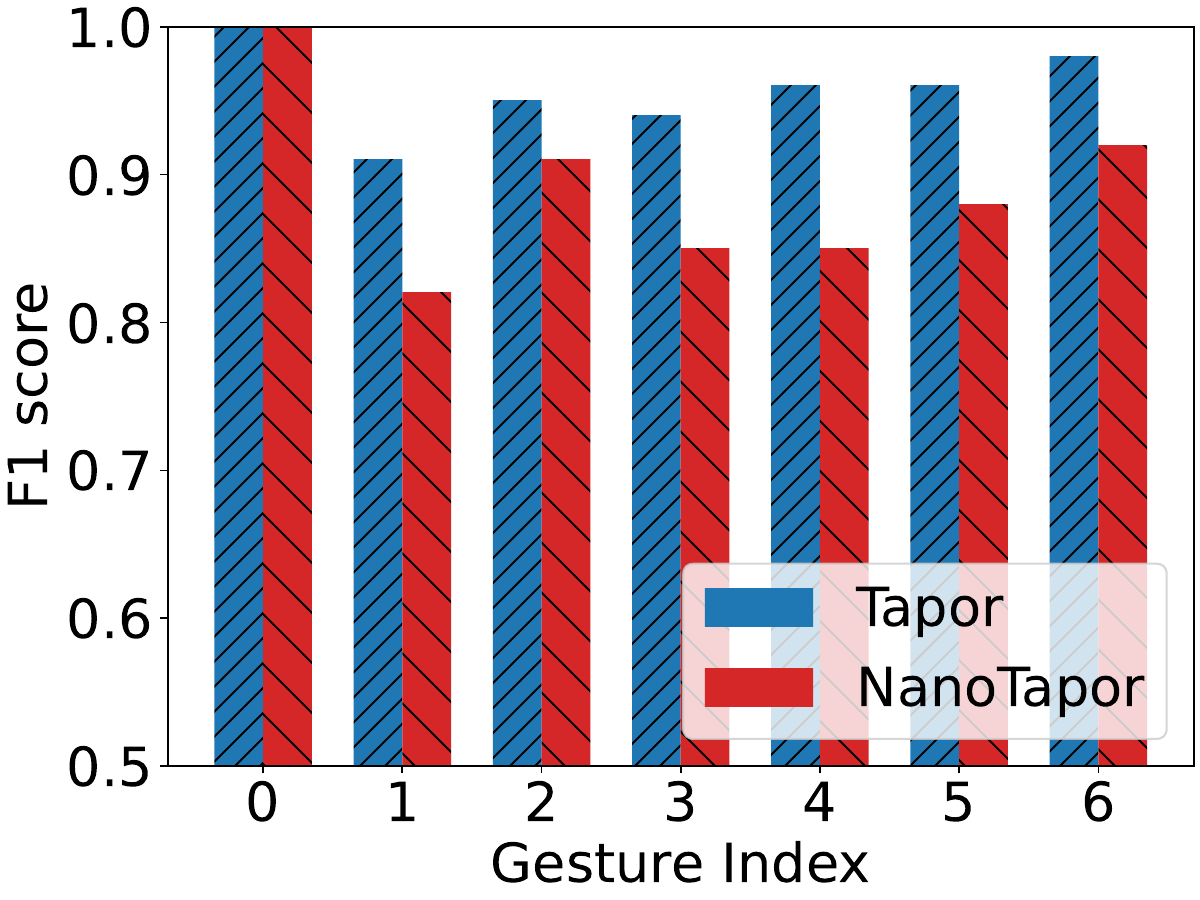}}
	  }
   \vspace{-0.15in}
    \caption{Tapor vs. NanoTapor.}
	\label{fig:tapor_vs_NanoTapor}
  \end{minipage}
    \hfill
  \begin{minipage}{0.4\textwidth}
\begingroup
\renewcommand{\arraystretch}{1.3} 
    \centering
    \footnotesize
    \begin{tabular}{cccc}
\hline
\textbf{\begin{tabular}[c]{@{}l@{}}ESP32S3 \\ w/ UART\end{tabular}} & \textbf{\begin{tabular}[c]{@{}l@{}}MLX90640 \\ w/ I2C\end{tabular}} & \textbf{\begin{tabular}[c]{@{}l@{}}NanoTapor\\ Operating\end{tabular}} & \textbf{\begin{tabular}[c]{@{}l@{}}Power\\ (mW)\end{tabular}} \\ \hline
\ding{51}                                                          & \ding{55}                                                                               & \ding{55}                                                             & 559                                                  \\
\ding{51}                                                          & \ding{51}                                                                               & \ding{55}                                                             & 643                                                  \\
\ding{51}                                                          & \ding{51}                                                                               & \ding{51}                                                             & 767                                                  \\ \hline
\end{tabular}
\endgroup
\vspace{2pt}
\captionof{table}{Power consumption. \rm{The thermal array sensor with $I^2C$ increases power by 84 mW, and the NanoTapor adds an additional 124 mW.}}
\label{tab:power_compare}
  \end{minipage}
\end{figure}

\begin{figure}[t]
 \begin{minipage}{0.35\textwidth}
    \centering
    \raisebox{-3\baselineskip}{\includegraphics[width=1.0\textwidth]{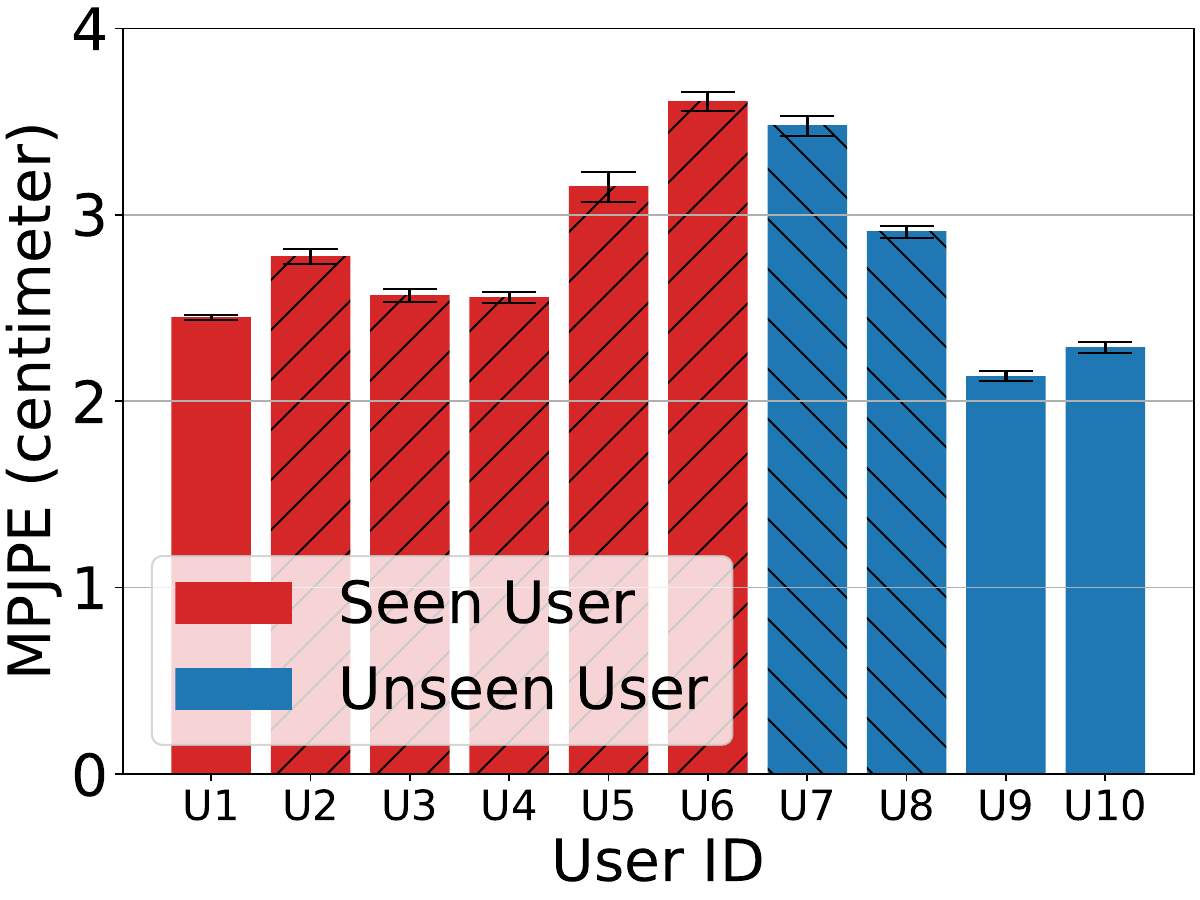}}
    \vspace{-0.2in}
    \caption{Tapor performance on different users.}
    \label{fig:cross_user}
  \end{minipage}
  \hfill
  \begin{minipage}{0.35\textwidth}
    \centering
    \includegraphics[width=1.0\textwidth]{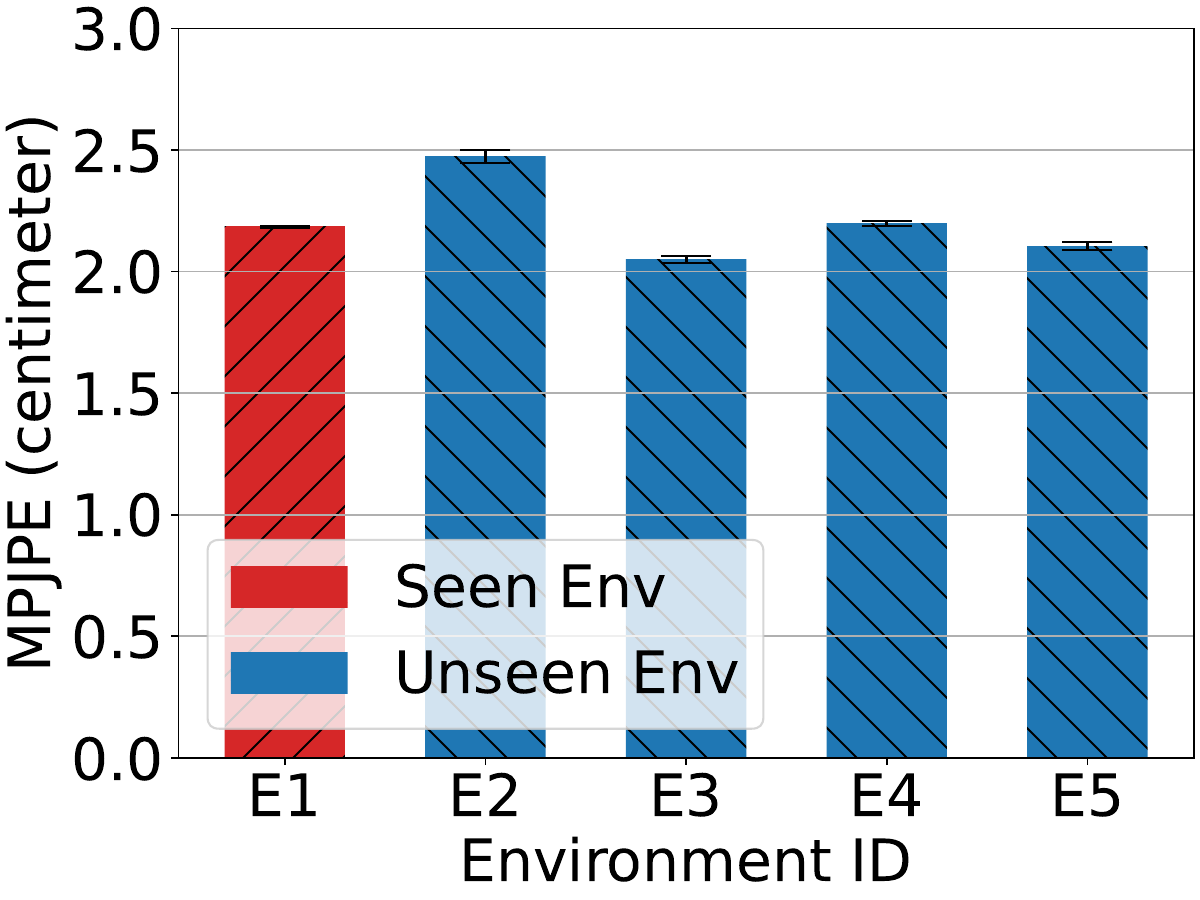}
    \vspace{-0.2in}
    \caption{Impact of different environments.}
    \label{fig:cross_env}
  \end{minipage}
  \hfill
  \begin{minipage}{0.26\textwidth}
   \centering
    \includegraphics[width=1.0\textwidth]{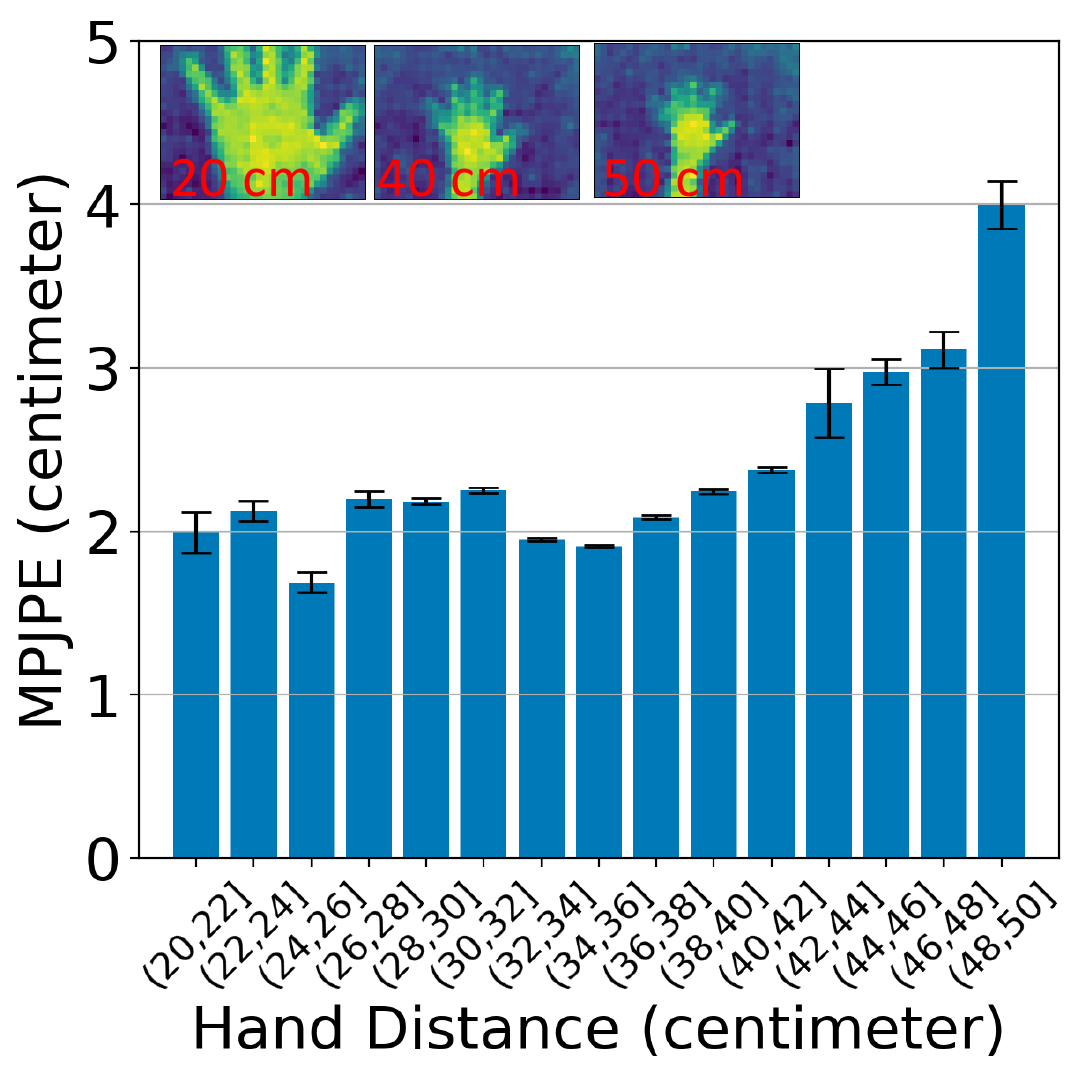}
    \vspace{-0.3in}
    \caption{Impact of hand distances.}
    \label{fig:dist_performance}
  \end{minipage}
\end{figure}

\head{System complexity} 
NanoTapor requires only 1.93M of RAM and 0.502M of Flash for operation, well within the capacity of most IoT devices, such as the ESP32S3-devkitc-1 board which boasts 3.28M of RAM and 3.34M of Flash.
As demonstrated in \tab\ref{tab:power_compare}, activating the MLX90640 thermal array sensor and enabling the $I^2C$ interface increases power consumption by 84 mW. 
Additionally, running the NanoTapor system adds a further 124 mW, resulting in a total power consumption of 767 mW. 
This is significantly lower than other 3D hand pose reconstruction systems, such as the Leap Motion, which consumes over 1495 mW.
It is worth noting that the majority of the power consumption is attributed to the ESP32S3 microcontroller unit (MCU) board. By transitioning to a lower-power MCU, the overall power consumption of the \sysname system could be further reduced.

\begin{figure*}[t]
\begin{minipage}{0.3\textwidth}
    \centering
    \raisebox{3.4\baselineskip}{
    \includegraphics[width=1.0\textwidth]{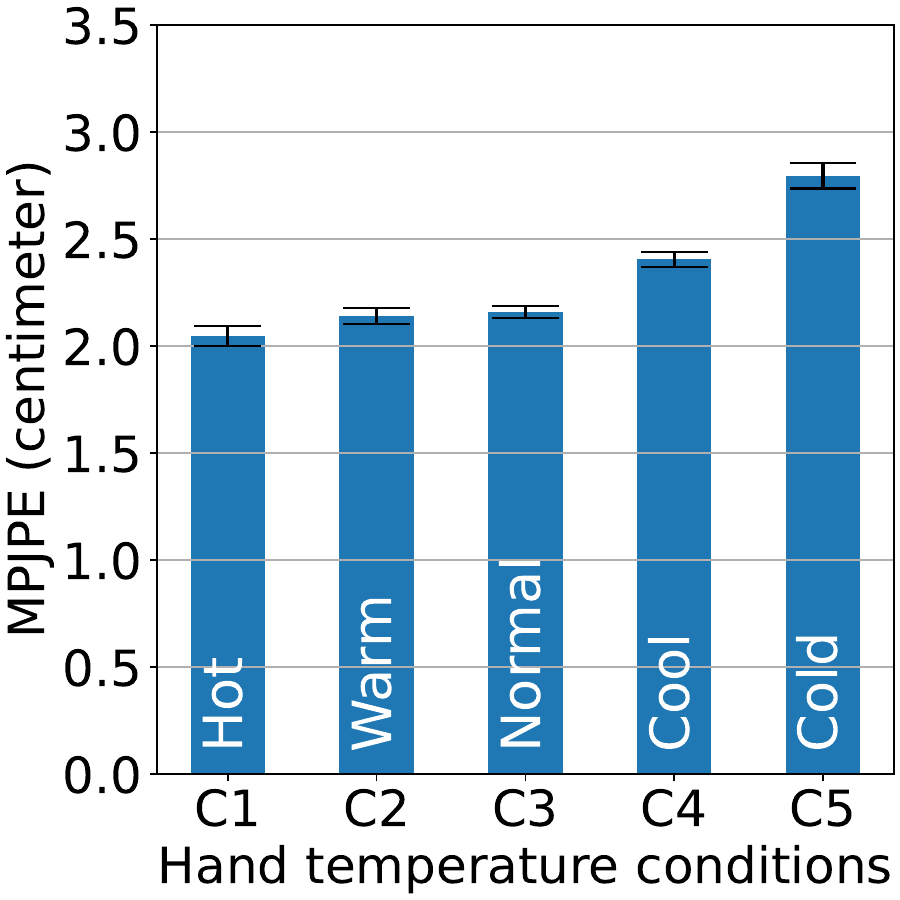}
    }
    \vspace{-0.7in}
    \caption{Impact of hand temperature.}
    \label{fig:hand_temperature}
  \end{minipage}
  \hfill
  \begin{minipage}{0.31\textwidth}
    \centering
    \includegraphics[width=1.0\textwidth]{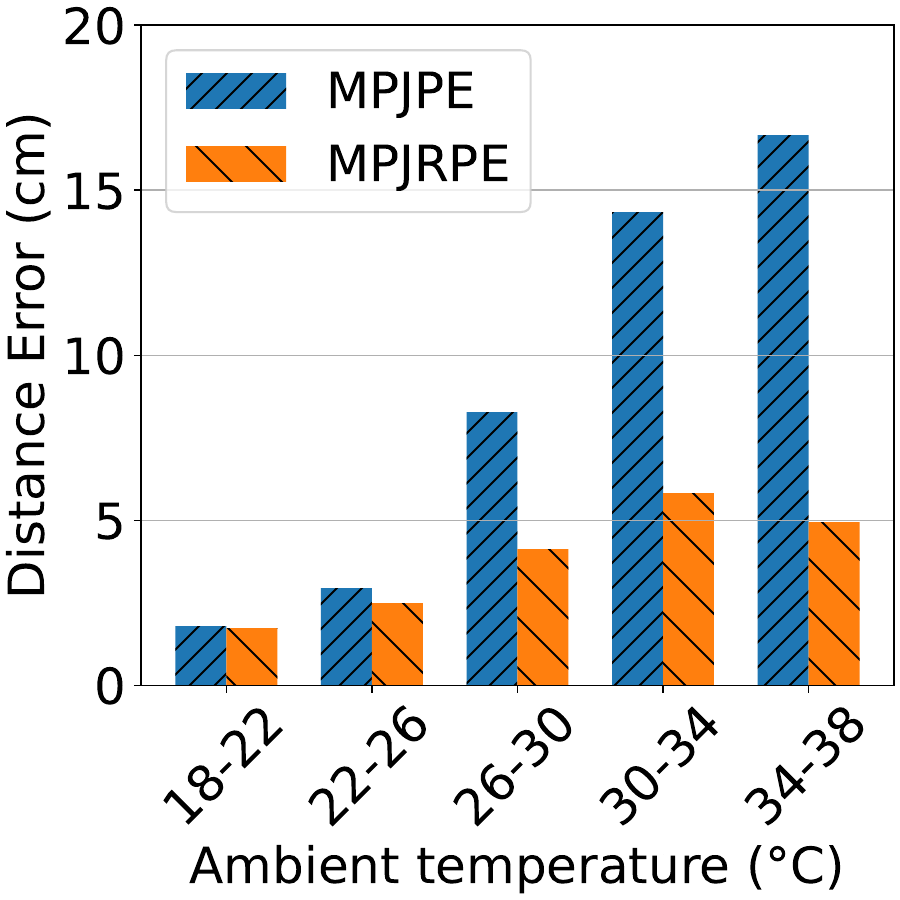}
    \vspace{-0.3in}
    \caption{Performance w.r.t. ambient temperature.}
    \label{fig:cross_temperature}
  \end{minipage}
  \hfill
  \begin{minipage}{0.31\textwidth}
    \centering
    \includegraphics[width=1.0\textwidth]{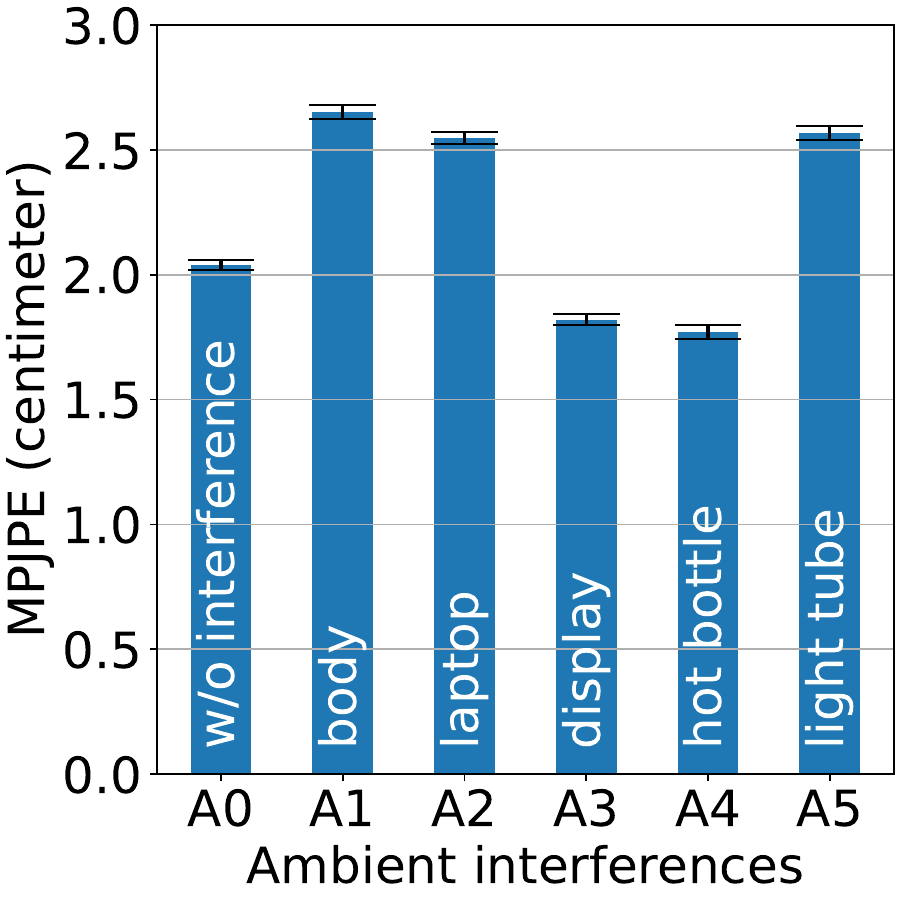}
    \vspace{-0.3in}
    \caption{Impact of ambient interference.}
    \label{fig:cross_interference}
  \end{minipage}
\end{figure*}

\begin{figure}[t]
  \begin{minipage}{0.3\textwidth}
    \centering
    \includegraphics[width=1.0\textwidth]{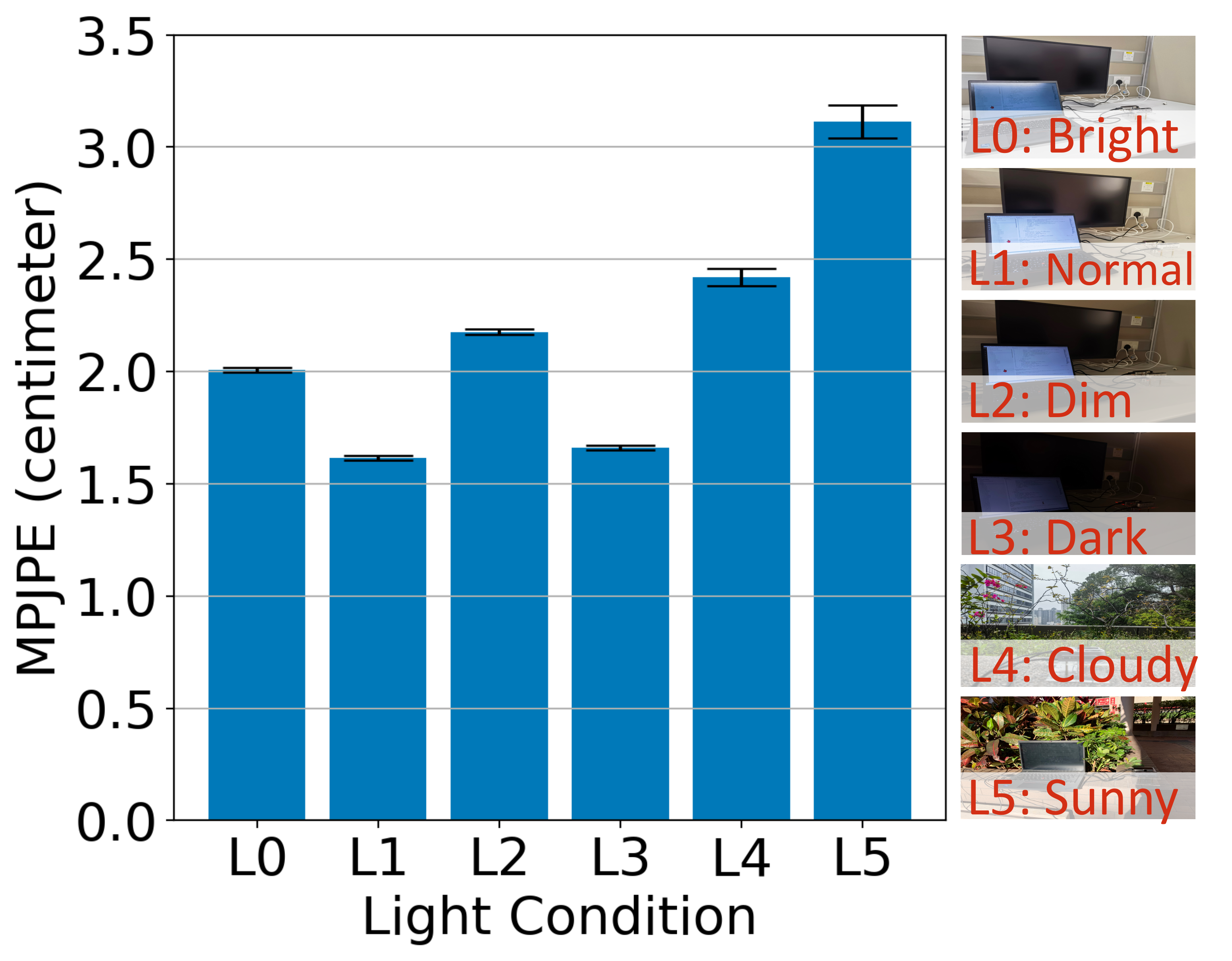}
   \vspace{-0.3in}
    \caption{Performance across different light conditions.}
	\label{fig:cross_light_cond}
  \end{minipage}
  \hfill
  \begin{minipage}{0.68\textwidth}
   \centering
   \subfloat[Different hand coverings.]{%
    \label{subfig:hand_cover}
      \raisebox{0.9\baselineskip}{\includegraphics[width=0.5\textwidth]{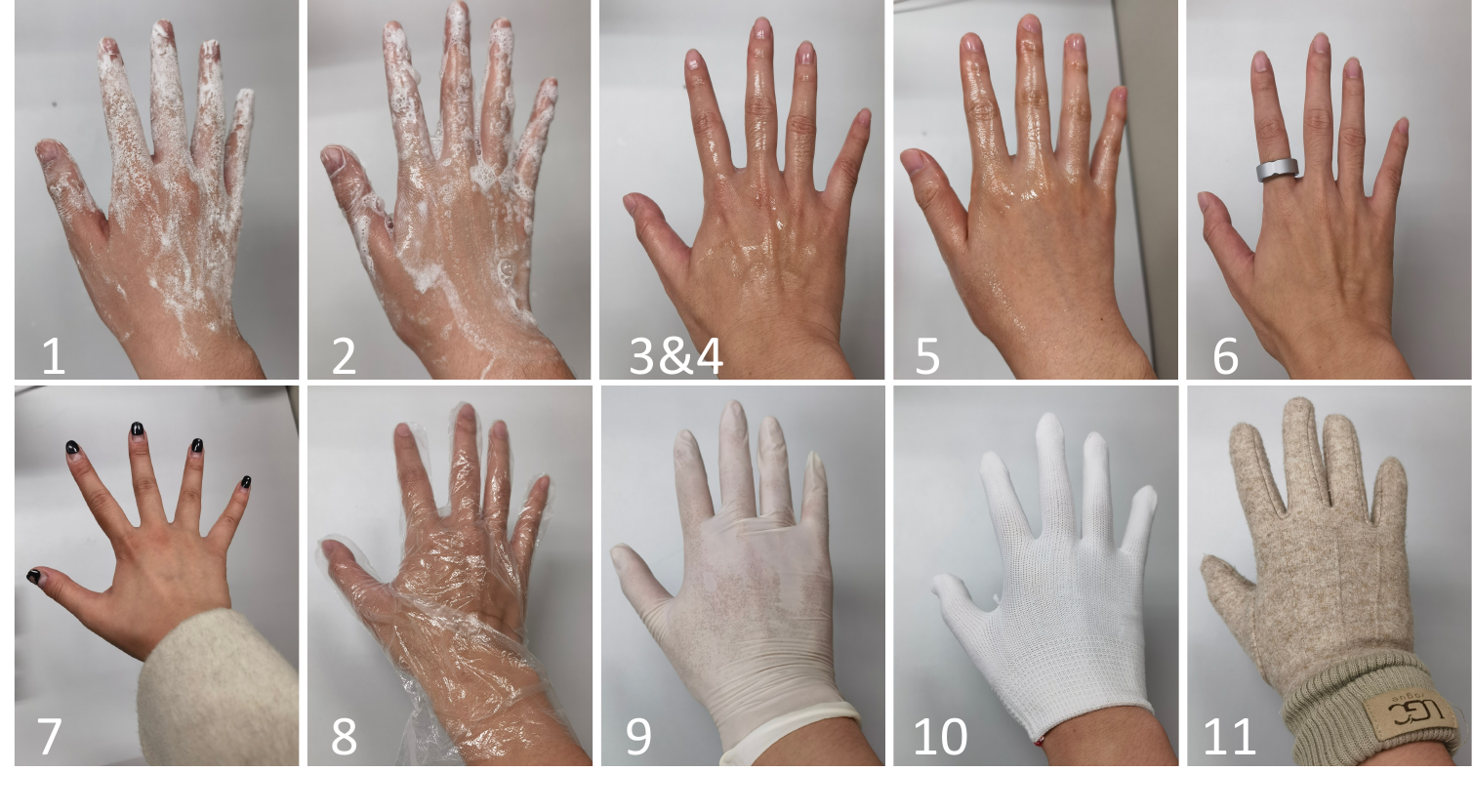}}
	  }
    \subfloat[\sysname performances.]{%
    \label{subfig:cross_cover}
      \includegraphics[width=0.5\textwidth]{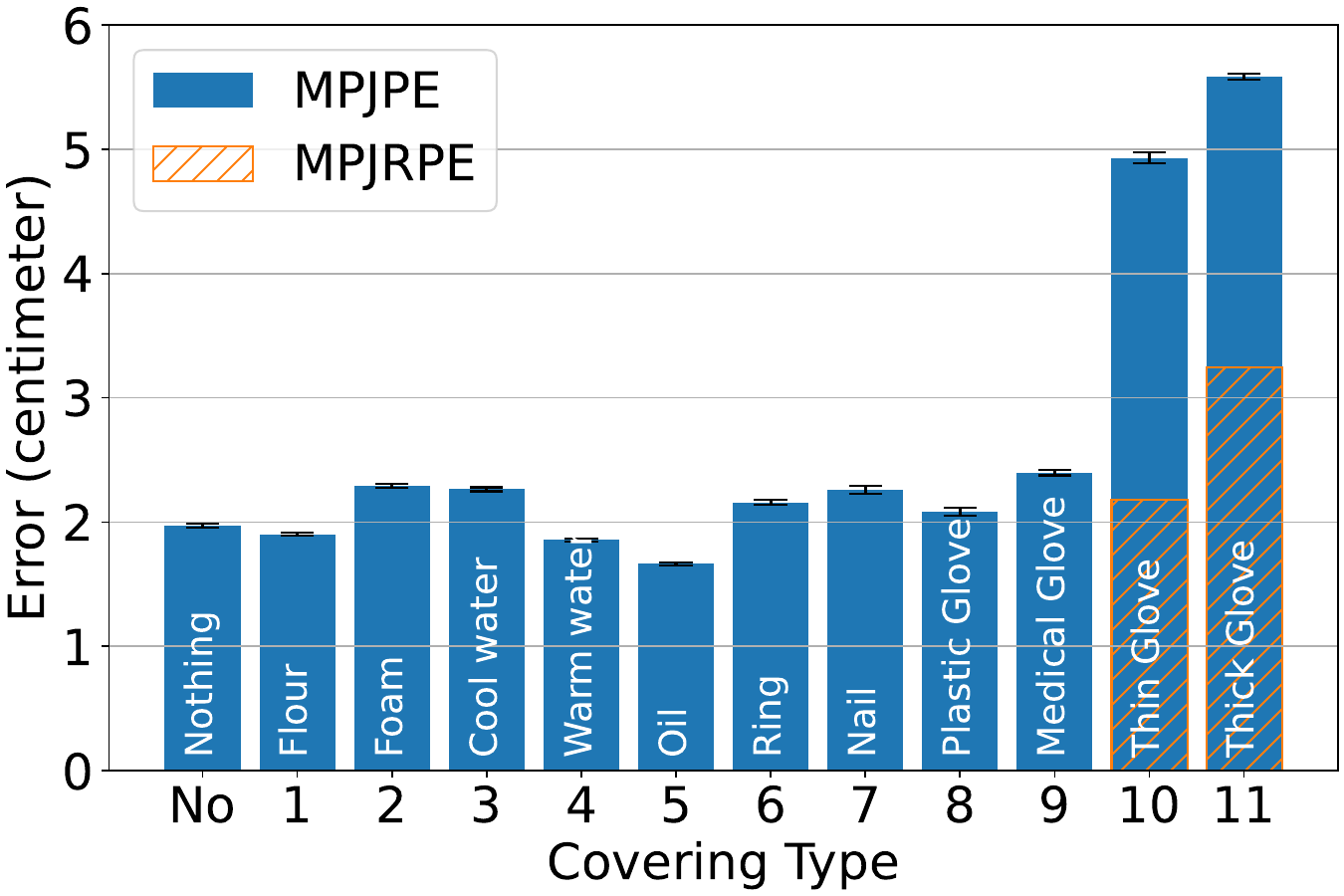}
	  }
   \vspace{-0.1in}
    \caption{Impact of the hand covering.}
	\label{fig:cross_hand_cover}
  \end{minipage}
\end{figure}

\subsection{Robustness Evaluation}
\label{ssec:robust}

\head{Cross user} 
For a practical 3D hand pose reconstruction system, robust performance across users is crucial. Hence, we assess Tapor's performance using data from ten participants, labeled U1-U10, by training on a subset of the data from U1-U6 and testing on the rest. As illustrated in \fig\ref{fig:cross_user}, the performance discrepancies among users are within 1.5 cm MPJPE, and there is no bias towards the users seen during training, demonstrating Tapor's strong generalization ability across different users.

\head{Cross environment} \sysname demonstrates robust performance across various environments, as shown in \fig\ref{fig:cross_env}. The Tapor model is trained with data from an office environment (E1 in \fig\ref{fig:data_env}) and then tested in all five environments. \sysname maintains consistent performance, with MPJPE variations of less than 0.5 cm across all tested environments.

\head{Impact of distance} We also evaluate the impact of the distance between the user's hand and the sensor, \ie, depth. As depicted in \fig\ref{fig:dist_performance}, \sysname maintains stable performance with MPJPE of approximately 2 cm when the distance is within 40 cm. The system performance slightly degrades, with an increase of at most 2 cm in MPJPE, when the distance ranges from 40 cm to 50 cm.
This is attributed to the limited resolution of the thermal map, which hinders the separation and reconstruction of individual joints for distant hands.
Despite this, the sensing range remains comparable to that of commercial hand pose estimation devices like Leap Motion, which operates up to 60 cm using stereo vision. 
Overall, the MPJPE remains below 3 cm for distances up to 48 cm, sufficient for hand pose-based interaction applications primarily within the near field.


\head{Impact of hand temperature} 
While human body temperatures are relatively stable, our hand surface temperatures vary depending on the environment. 
To evaluate the impact of hand temperature, participants immersed their hands in hot, warm, cold, and iced water for over 20 seconds before each respective experiment. 
The results, as \fig\ref{fig:hand_temperature}, show robust performance with slightly increased MPJPE when the hand temperature gets lower.

\head{Impact of ambient temperature} 
We evaluate performance under various ambient temperatures to study the effect of environmental thermal radiation. To isolate this factor, participants remained in rooms with different ambient temperatures for over five minutes prior to data collection, allowing the thermal background to stabilize.
As shown in \fig\ref{fig:cross_temperature}, MPJPE increases significantly beyond 26°C, while MPJRPE remains largely unaffected. This suggests that ambient temperature impacts depth estimation more than relative pose reconstruction, similar to the effect of thick gloves on thermal radiation. 
Despite this, with the sufficient accuracy of relative hand pose reconstruction, \sysname can still support important applications like gesture control.

\begin{figure*}[t]
    \centering
    \subfloat[Training schemes]{
    \includegraphics[width=0.19\textwidth]{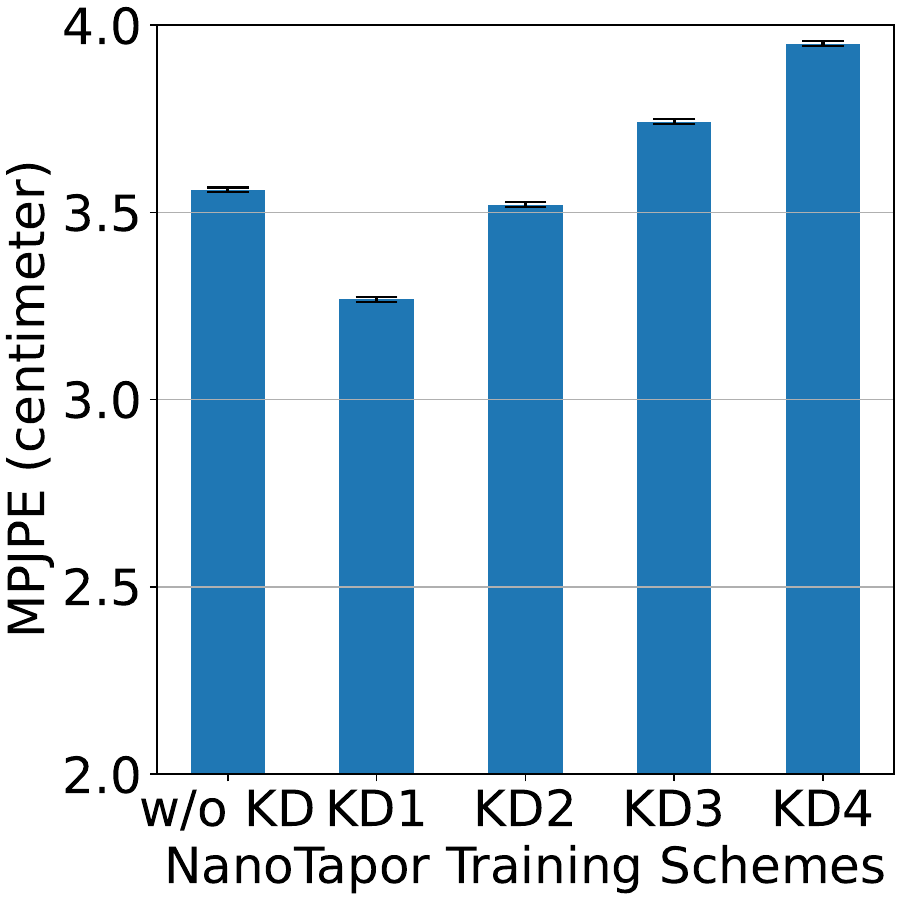}
    \label{fig:kd_NanoTapor}
    }
    \hfill
        \subfloat[Skeleton init.]{\includegraphics[width=0.19\textwidth]{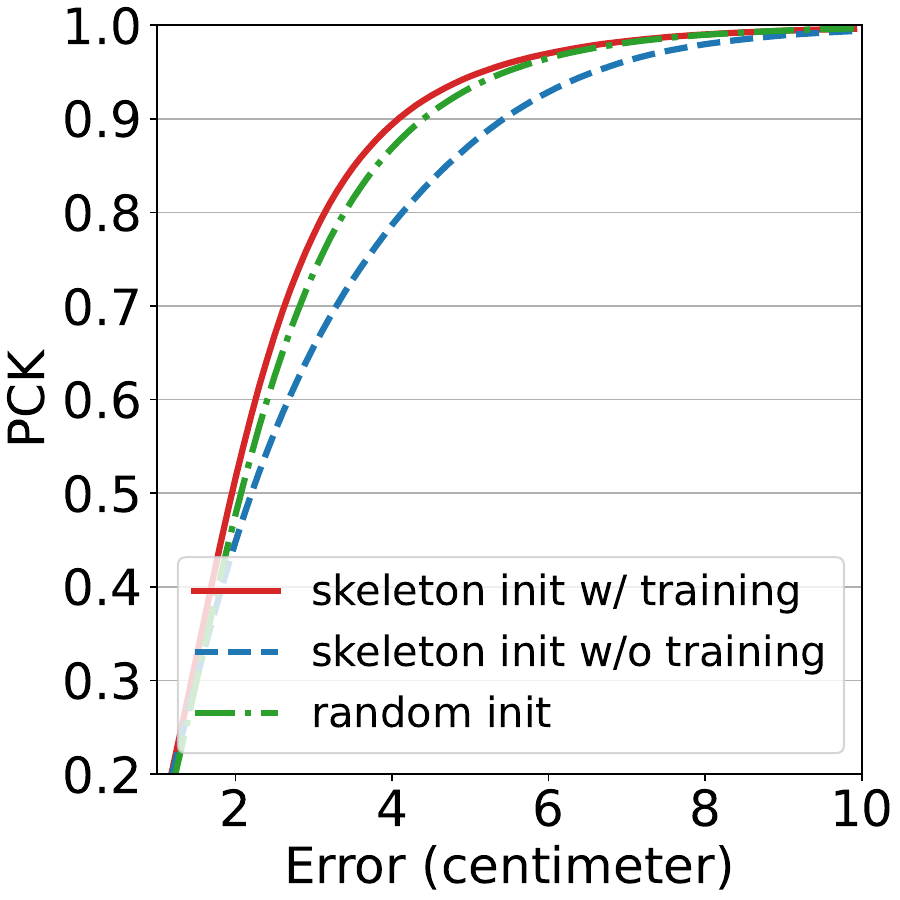}\label{fig:sub1}}
    \hfill
        \subfloat[ Training loss. ]{\includegraphics[width=0.19\textwidth]{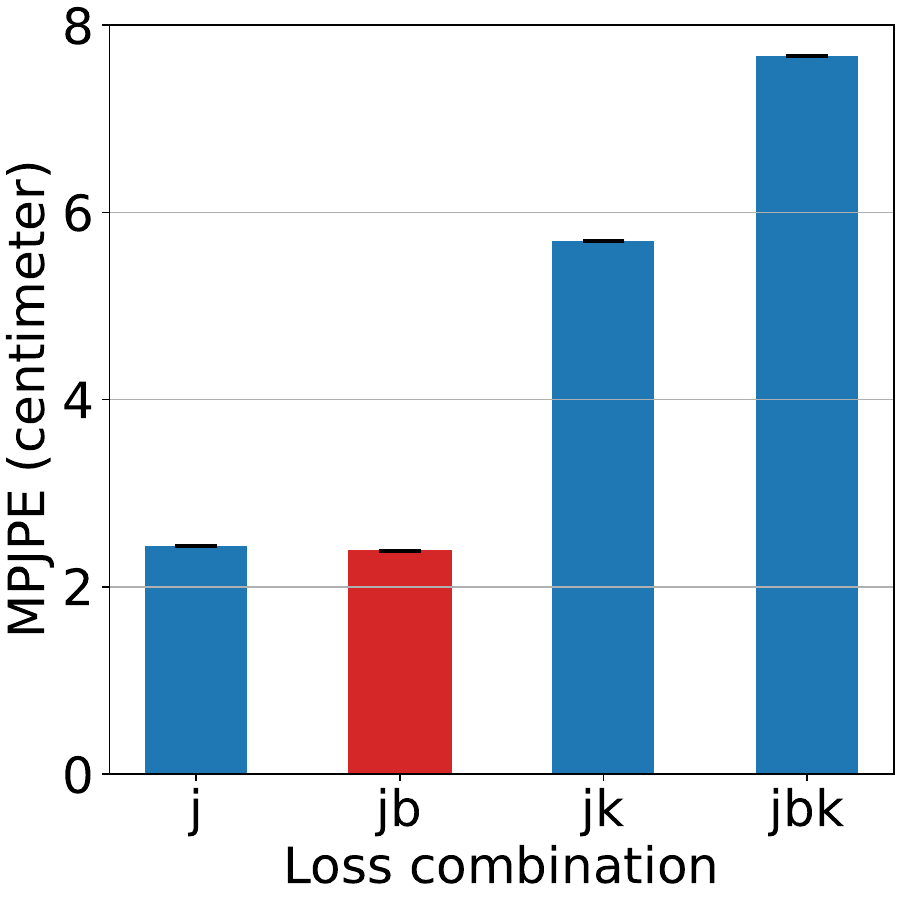}\label{fig:sub3}
        }
        \hfill
        \subfloat[Temporal length.]{\includegraphics[width=0.19\textwidth]{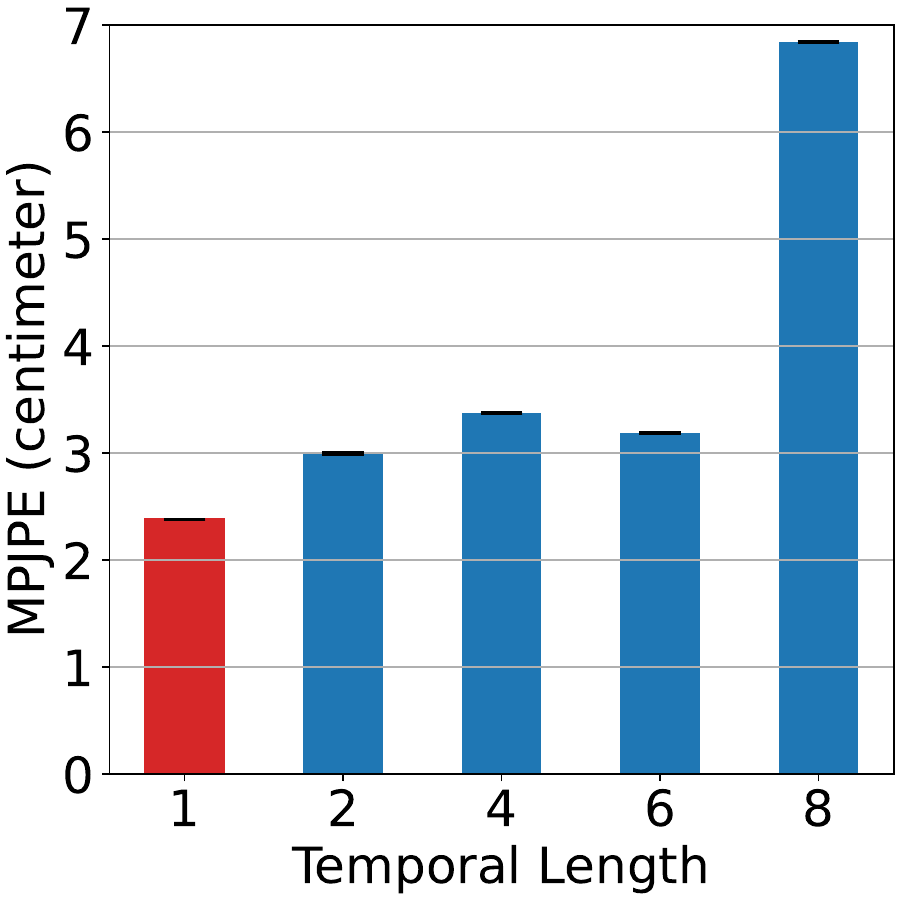}\label{fig:sub2}}
\hfill
        \subfloat[Upsample scales $\alpha$.]{\includegraphics[width=0.19\textwidth]{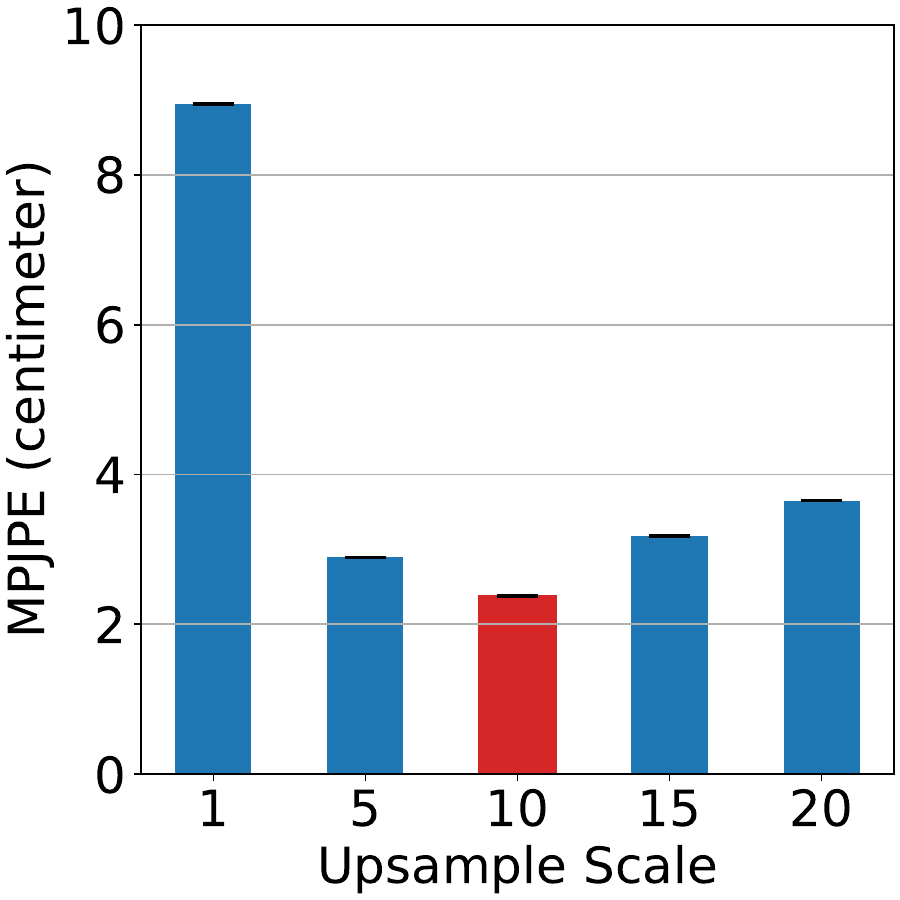}\label{fig:sub4}}
     \vspace{-0.1in}
    \caption{Ablation study results. {\rm The ablation study is conducted on the Tapor model except for (a) on NanoTapor.}}
  \label{fig:ablation_result}
\end{figure*}

\head{Impact of ambient interference} 
\rev{Certain everyday objects, such as electronic devices or heated items, may exhibit higher temperatures than their surroundings, potentially introducing interference to \sysname. However, since \sysname is designed for around-device interaction, the user's hand typically dominates the sensor's FoV during operation, effectively minimizing the influence of surrounding objects. 
To systematically evaluate \sysname's robustness against ambient interference, we conducted experiments involving various common heat-emitting objects.
These included a running laptop, an active display, a hot water bottle placed near the device, as well as the presence of the user's body and light tubes within the sensor's FoV. As illustrated in \fig\ref{fig:cross_interference}, the results demonstrate that \sysname remains robust and maintains reliable performance even in the presence of such environmental interferences.}

\head{Impact of light condition} We evaluate the effect of lighting conditions on \sysname's performance, as illustrated in \fig\ref{fig:cross_light_cond}. The results demonstrate consistent performance and robustness, even in darkness or under sunlight, with negligible variations across different light conditions.

\head{Impact of hand coverings}
As mentioned in \S\ref{sec:primer}, thermal array readings depend on thermal radiation emitted by the target. Hand coverings, particularly thick gloves, can partially block the radiation and affect the performance of \sysname. Therefore, assessing the impact of different hand coverings is essential to gauge Tapor's real-world effectiveness. As in \fig\ref{fig:cross_hand_cover}, Tapor performs well when the hand is covered with liquids (covering types $1-5$), rings, and nail decorations, as these coverings do not significantly alter the thermal radiation of the hand. With gloves made from materials like plastic or latex, Tapor remains effective, as these materials allow thermal radiation to pass through. However, thick cotton gloves, which block most thermal radiation, lead to a notable increase in MPJPE. 
Nevertheless, the MPJRPE remains low, suggesting that the performance decline is mainly due to an underestimation of the hand's depth. Therefore, \sysname can still accurately reconstruct the relative hand pose for various applications, including gesture recognition.

\head{Left vs. Right Hands} We evaluate \sysname for left and right hands separately. The MPJPE for left-hand pose reconstruction is 2.40 cm, and for right-hand pose reconstruction, it is 2.22 cm, showing no bias towards either hand.

\subsection{Ablation Study}
\label{ssec:ablation_study}
\head{NanoTapor Training Schemes} We evaluate the performance of our heterogeneous knowledge distillation (HKD) technique compared to other methods. As shown in \fig\ref{fig:kd_NanoTapor}, KD1 and KD2 use HKD with pose feature $f_{pose}$ and location feature $f_{loc}$, respectively. KD3 and KD4, on the other hand, use PCA for dimension reduction with $f_{pose}$ and $f_{loc}$ respectively. NanoTapor performs best with HKD, outperforming PCA-based distillation and direct training.

\head{Skeleton Init} As illustrated in \fig\ref{fig:sub1}, Tapor with skeleton init shows the best performance, which verifies the effectiveness of injecting hand structure knowledge for training. 

\head{Loss Combination} We use joint loss (j) and bone loss (b) for Tapor optimization, as shown in \fig\ref{fig:sub3}. Kinematic loss (k) transforms key points' positions into kinematic representation \cite{dragulescu3DActiveWorkspace2007} and calculates MSE with ground truth. Due to poor performance, kinematic loss is not used in \sysname.

\begin{figure}[t]
    \subfloat[Tapor.]{%
    \label{subfig:gesture_cm}
      \includegraphics[width=0.32\textwidth]{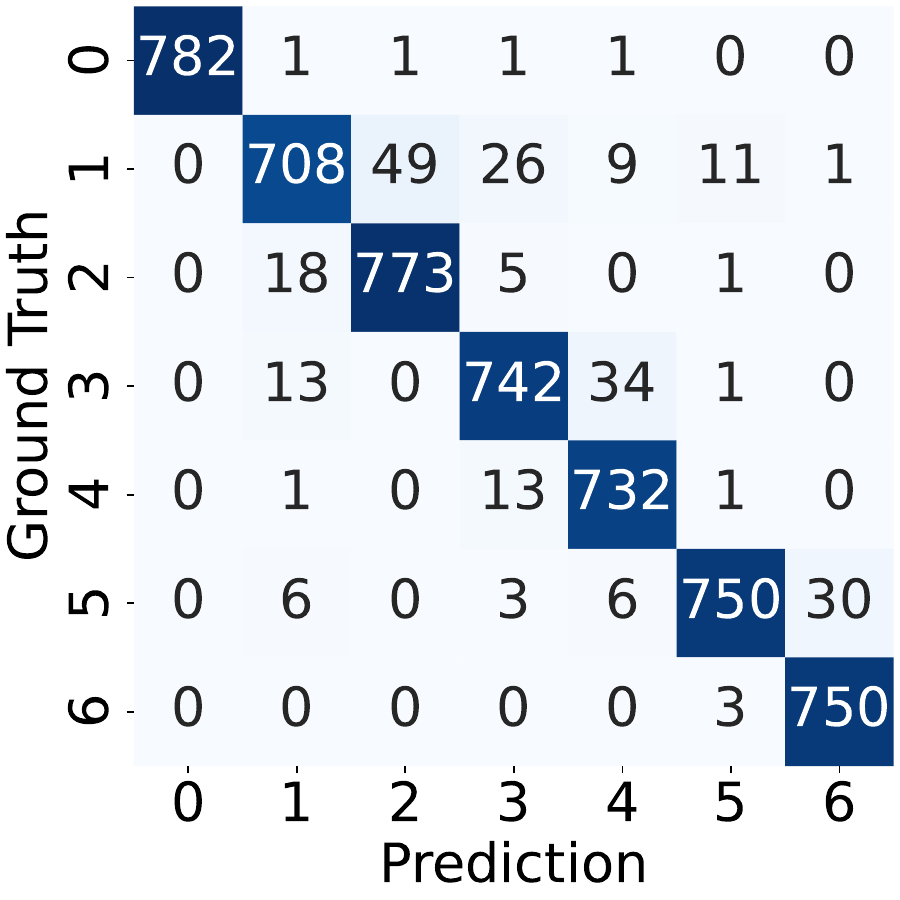}
	  }
    \hspace{2.6\baselineskip}
    \subfloat[NanoTapor.]{%
    \label{subfig:gesture_f1}
      \includegraphics[width=0.32\textwidth]{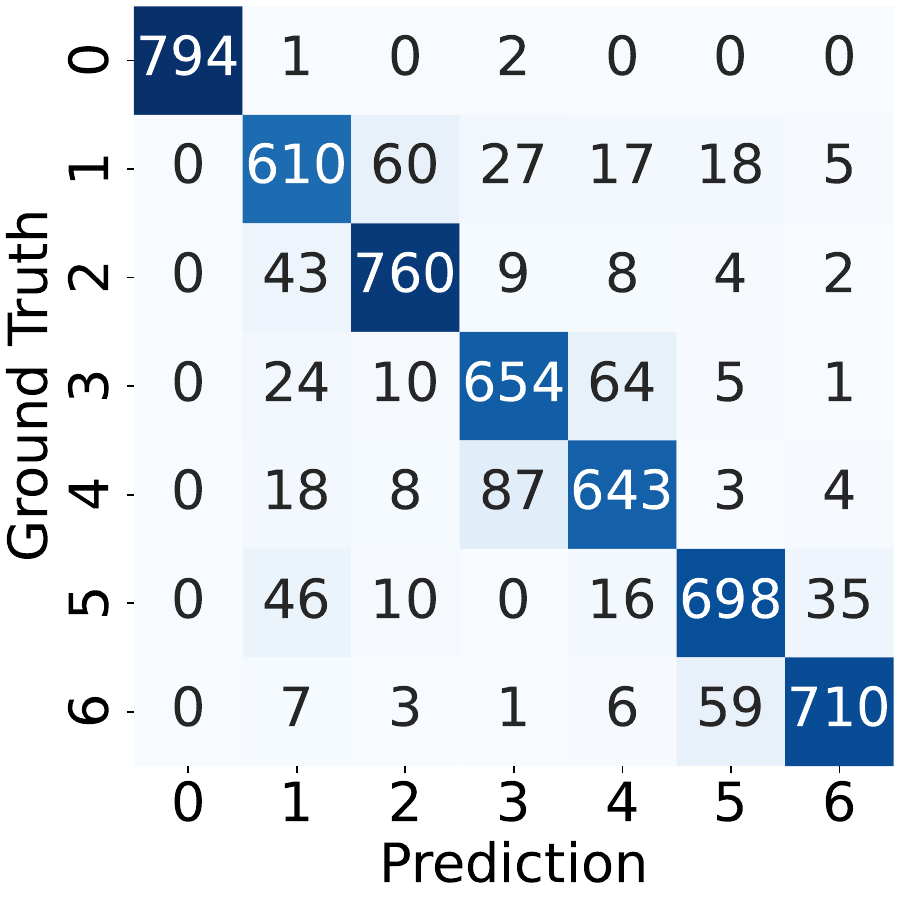}
	  }
   \vspace{-0.6\baselineskip}
    \caption{Gesture recognition performance. \rm{We consider six gestures: 1:thumbs up, 2:raised fist, 3:crossed fingers, 4:victory hand, 5:OK hand, 6:raised hand, plus 0: No gesture.}}
	\label{fig:gesture_reco_case}
\end{figure}

\begin{figure}[t]
    \includegraphics[width=0.7\textwidth]{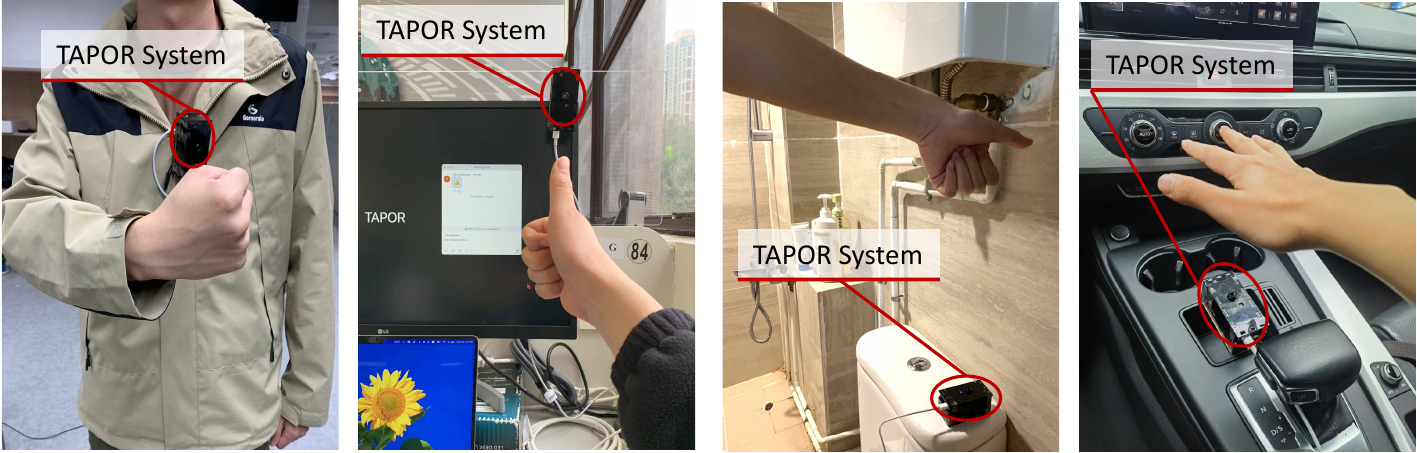}
   \vspace{-0.3\baselineskip}
    \caption{Gesture control applications with \sysname.}
\label{fig:gest_control_vis}
\end{figure}

\head{Temporal Length} Temporal length $l$ indicates the number of KP features used in KP fusion, where $l=1$ uses only the current feature. As in \fig\ref{fig:sub2}, previous features do not improve performance due to pose changes and thermal sensor noise. We use only the current feature for optimal performance and efficiency in our prototype system.

\head{Upsample Scale} Tapor upsamples the temperature map with a scale $\alpha = 10$, which is shown to be optimal in \fig\ref{fig:sub4}.

\subsection{Case Studies}
\label{ssec:case_study}

We develop two applications, gesture control, and finger tracking, on top of \sysname to showcase its applications.

\head{Gesture control}
To develop a gesture control system based on \sysname, we trained a linear layer to map the reconstructed hand pose from Tapor and NanoTapor to six gestures.
As in \fig\ref{fig:gesture_reco_case}, both Tapor and NanoTapor achieve high accuracy for all gestures and nearly zero false alarms (Gesture \#0). 

With the promising performance, we envision \sysname can enable applications in various scenarios. To demonstrate this, we conduct four case studies as shown in \fig\ref{fig:gest_control_vis}. In each case, we collect 300 samples for each of the six gestures (as in \fig\ref{fig:gesture_reco_case}) and evaluate the recognition performance. 

\begin{figure}[t]
    \includegraphics[width=0.8\textwidth]{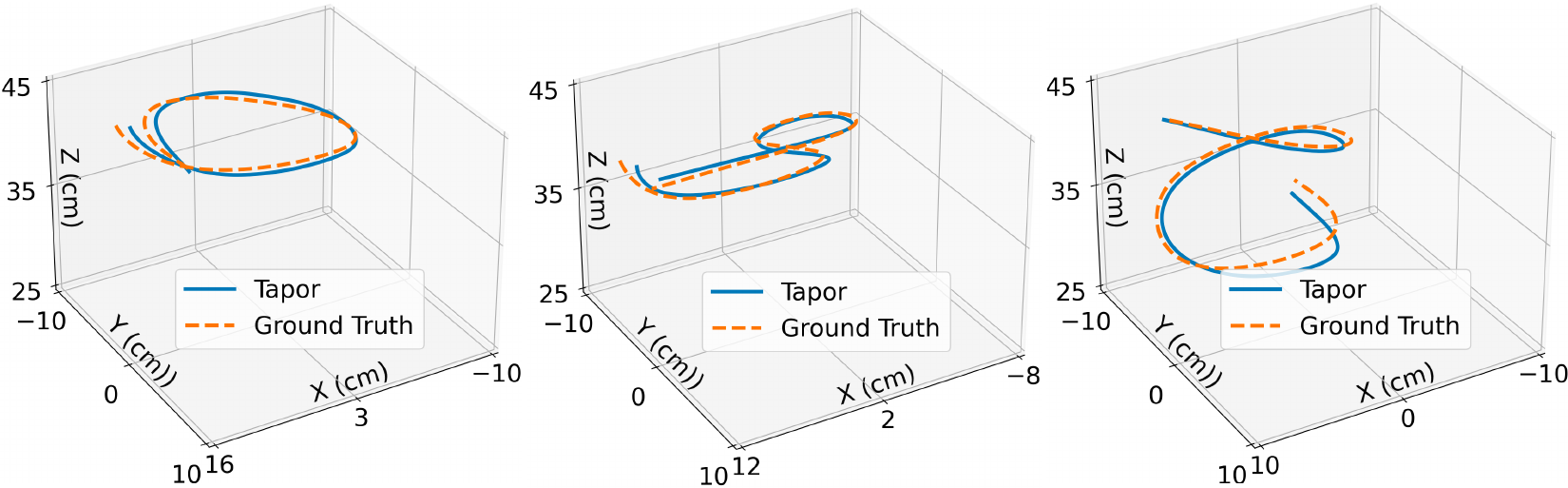}
    \caption{Finger tracking with \sysname. {\rm }}
\label{fig:track_examples}
\end{figure}

{\noindent\textit{1) Wearable AI pin control}}: 
Recently, AI pins like Humane and Rabbit R1\footnote{Humane AI Pin: https://humane.com/; Rabbit R1: https://www.rabbit.tech/.} have attracted attention for integrating AI into daily life. Though equipped with a camera and microphone, these units remain mostly inactive to save power and protect privacy, requiring a button press for activation. \sysname, with its lightweight design, can be integrated into wearable AI assistants to enable privacy-conscious, touch-free control with more command options than a single button.

{\noindent\textit{2) Conference calls}}: 
Similarly, \sysname functions as a gesture-based controller for devices like TVs and AC units. It can integrate with computers to facilitate gesture interactions in online meetings like Zoom. Users can respond with emojis, such as thumbs-up, through gestures, eliminating the need to activate the microphone or camera, or use the mouse.

{\noindent\textit{3) Toilet device control}}: 
We also deploy \sysname in bathrooms for operating appliances. In bathrooms, users' hands are often wet, making it risky to press electrical buttons. 

{\noindent\textit{4) In-Car gesture control}}: 
Similarly, we also showcase how \sysname can be deployed for in-car gesture control. 

Overall, Tapor achieves 97\% recognition accuracy in both bathroom and car applications and maintains over 90\% accuracy in the first two scenarios of wearable AI pin control and online meeting integration, while NanoTapor yields 92\% and 83\%, respectively. 
In practice, many applications often do not need six gestures (\eg, only audio on/off control for AI pins). 
Therefore, we further evaluate the accuracy by varying the gesture vocabulary. 
Our results show that NanoTapor achieves an accuracy of 93\% for 4 gestures and 98\% for 2 gestures, with random combinations from the six gestures.

\head{Finger Tracking} 
\rev{Finger tracking is a highly sought-after capability with numerous practical applications, garnering significant attention in both research and industry. 
Leveraging its robust 3D hand pose reconstruction functionality, \sysname demonstrates exceptional precision in tracking intricate finger movements within three-dimensional physical space.
As depicted in \fig\ref{fig:track_examples}, \sysname effectively tracks the random motion of an index finger, achieving an average position error of just 2.69 cm across 40 trials when combined with a Kalman filter for post-processing.
This outstanding performance underscores \sysname's potential as a powerful tool for advanced finger-tracking applications.}

\section{Discussions and Future Works}
\label{sec:discussion}
\head{System cost} The \sysname prototype costs around US\$20 for a thermal sensor and US\$5 for the ESP32-S3 DevKit. The cost can be further cut down when building in bulk. 

\head{Thermal depth} \sysname's performance, especially depth accuracy, can be impacted by ambient temperatures and target radiation, a well-known issue in thermal sensing \cite{davis2004robust,trofimova2017indoor}. 
In its current form, \sysname maintains high precision at typical indoor temperatures (18°C-26°C) and low pose error at higher temperatures. Future work will explore techniques like background thermal cancellation for improvement.

\head{Better accuracy} It remains an open question whether thermal array-based methods can achieve hand pose estimation accuracy comparable to RGB camera-based systems. 
Exploring approaches such as stereo thermal arrays or multi-sensor fusion may help close this gap and is an important direction for future work.

\head{Hand orientation failure} Cases in which the palm faces upward (\ie, away from the sensor) are rare but represent a potential failure mode due to limited training data for such scenarios. Since our work focuses on around-device interactions, we primarily consider common postures where the palm naturally faces downward toward the sensor, as is typical in interaction scenarios.

\head{Multi-hand pose} 
\rev{Currently, \sysname supports the reconstruction of single-hand poses, which is sufficient for the majority of interaction scenarios. Notably, many existing gesture control systems, such as those in the Apple Watch, Google Soli \cite{lienSoliUbiquitousGesture2016}, and Vtouch \cite{Holo_Button} commercial systems, are also designed for single-hand use. However, extending \sysname to support multi-hand and multi-user interactions presents an exciting avenue for future research, enabling more complex and collaborative applications.}

\head{Close-range interactions} 
\rev{\sysname is designed for close-range interactions, similar to systems like Leap Motion \cite{LeapMotionController} and Google Soli \cite{lienSoliUbiquitousGesture2016}. \sysname's around-device interaction capabilities hold significant potential for a wide range of applications. For instance, users can employ customized or preferred gesture sets to control TVs, in-car media systems in low-light conditions, or navigate maps on smartwatches. Additionally, \sysname can be used for AR/VR headset interaction, controlling smart appliances in the kitchen with a wet hand, and gesture control in the absolute dark. Mid- and long-range interactions remain an area for future exploration.}

\head{Hand-Object interactions}
\rev{This work focuses on free-hand pose estimation for around-device interaction. However, hand pose estimation while holding objects is useful for applications in virtual reality (VR) and augmented reality (AR). Investigating this capability will be a key focus of our future research, as it opens up new possibilities for immersive and interactive experiences.}

\section{Related Works}
\label{sec:related_works}

\subsection{3D Hand Pose Estimation} 
3D hand pose reconstruction has been widely studied. We mainly survey the literature using the below modalities. 

\head{Camera-based} These methods predominantly utilize RGB(D) images for hand pose estimation and shape reconstruction. 
Our focus is on monocular 3D hand estimation methods, which can be categorized into three categories:
1) Direct regression methods estimate 3D hand joint positions from RGB(D) images using end-to-end neural networks \cite{li3DHumanPose2015,oberwegerHandsDeepDeep2016,zhangMediaPipeHandsOndevice2020}. 
However, these methods face challenges due to limited labeled 3D hand datasets and insufficient supervision, leading to research on lifting 2D positions to 3D. 
2) 2D lifting approaches detect key points in 2D frames and then convert these to 3D positions  \cite{ramakrishnaReconstructing3DHuman2012,tomeLiftingDeepConvolutional2017}. 
3) Model-based methods leverage prior knowledge of hand structure to enhance estimation accuracy \cite{zhouModelbasedDeepHand2016,spurrWeaklySupervised3D2020, chenModelBased3DHand2021}, with MANO model \cite{romeroEmbodiedHandsModeling2017} being a foundational model in this category with many followers like HTML \cite{qianHTMLParametricHand2020}, NIMBLE \cite{liNIMBLENonrigidHand2022}, DART \cite{gaoDARTArticulatedHand2022}.
Despite their accuracy, camera-based approaches face challenges like privacy concerns and sensitivity to lighting conditions.

\head{Signal-based}
Wireless, acoustic, and light signals are exploited as more privacy-friendly alternatives for hand pose reconstruction, mainly focusing on gesture recognition \cite{sunVSkinSensingTouch2018,patraMmWaveRadarBased2018,puWholehomeGestureRecognition2013,chi2018ear,li2022room} or hand/finger tracking \cite{maoRNNBasedRoomScale2019,liWiFingerTalkYour2016,lienSoliUbiquitousGesture2016,nandakumarFingerIOUsingActive2016,wangDevicefreeGestureTracking2016,sunWiDrawEnablingHandsfree2015}. 
Among these, Aili \cite{liReconstructingHandPoses2017} reconstructs hand skeletons by capturing the blockage of light from a custom LED-panel-and-photodiode-array setup, while HandFi \cite{jiConstruct3DHand2023} leverages WiFi signals to infer 3D hand poses using the existing communication infrastructure.
Despite their creativity, these systems face challenges: Aili relies on bulky, non-integrable hardware not suited for compact IoT environments, while HandFi suffers from weak boundary localization and generalization limitations, which may cause unintended activations in real-world settings.

\head{Wearable-based}
Involving various sensors attached to the wrist, fingers, or integrated into gloves, these methods are used for gesture recognition \cite{Kato_Takemura_2016, laputSensingFineGrainedHand2019, xuEnablingHandGesture2022, Lu_Ding_Wang_Xue_2024} and hand pose estimation \cite{wuBackHandPose3DHand2020, zhouLearningRingsSelfSupervised2022}. 
For instance, MagX \cite{chenMagXWearableUntethered2021} enables untethered hand tracking by detecting magnetic fields from passive magnet rings worn on fingers, while RoFin \cite{zhangRoFin3DHand2023} employs custom gloves with embedded LED nodes to achieve fine-grained, real-time 3D pose reconstruction.
In a similar vein, Pyro \cite{gongPyroThumbTipGesture2017} introduces a compact, low-power thumb-tip gesture recognition system using pyroelectric sensors. Its design demonstrates robustness across lighting conditions and motion scenarios, and its small form factor makes it attractive for wearable and mobile use.

However, despite their performance, all these wearable-based systems require users to wear specific devices, which can be intrusive or uncomfortable over extended periods. This makes them less suitable for ambient or large-scale IoT deployments, where ease of use, unobtrusiveness, and seamless integration are critical.

\subsection{Thermal Array-based Sensing}
Thermal array sensors are gaining prominence due to their versatility in a variety of sensing applications. 
These applications include human detection \cite{ceruttiConvolutionalNeuralNetwork2019,tanakaLowPowerWireless2014}, occupancy estimation \cite{sahooOccupancyCountingDense2023,chiduralaOccupancyEstimationUsing2021,tyndallOccupancyEstimationUsing2016}, daily activity recognition \cite{muthukumarInfraredArraySensorBased2022,rezaeiUnobtrusiveHumanActivity2022,pollaActionRecognitionLowResolution2019}, fall detection \cite{heNonContactFallDetection2023,naserMultipleThermalSensor2022,zhongMultiOccupancyFallDetection2021,rezaeiUnobtrusiveFallDetection2021}, and indoor localization \cite{bouaziziLowResolutionInfraredArray2022,faulknerDeviceFreeLocalizationUsing2021,hevesiMonitoringHouseholdActivities2014}.
For gesture recognition, although there has been extensive work using thermal cameras, such as Birkeland \etal \cite{birkelandVideoBasedHand2024}, who provided a large thermal hand gesture dataset captured with the FLIR Lepton thermal camera, and ThermoHands \cite{dingThermoHandsBenchmark3D2024a}, which introduced an egocentric benchmark dataset and a transformer-based baseline for robust 3D hand pose estimation under diverse real-world conditions, research on thermal array sensors remains limited.

As a new and emerging sensing modality, thermal arrays offer unique advantages such as extremely low resolution, low signal-to-noise ratio, low power consumption, and significantly lower cost compared to thermal cameras. Despite these promising features, only a few studies have explored their potential for gesture recognition or hand pose estimation.
Notable efforts in this direction include Tateno \etal \cite{tatenoHandGestureRecognition2019}, who developed an in-car interface system using a thermal array sensor and a CNN model, and Vandersteegen \etal \cite{vandersteegenLowLatencyHandGesture2020}, who proposed a low-latency gesture recognition system based on temporal convolutional networks applied to thermal array map sequences.
A particularly relevant study, FingerTrak \cite{huFingerTrakContinuous3D2020}, employs multiple thermal array sensors embedded in a wristband to continuously track hand poses. While this design effectively addresses occlusion issues, it introduces significant hardware complexity and imposes wearability constraints.
In contrast, \sysname is tailored for ambient, non-wearable scenarios. It leverages a single low-cost thermal array sensor to enable seamless around-device interaction, significantly simplifying deployment and enhancing usability.

\sysname distinguishes itself by enabling 3D hand pose reconstruction solely from a thermal array sensor, offering a scalable, privacy-preserving, and cost-effective solution well-suited for widespread IoT applications.

\section{Conclusions}
\label{sec:conclusion}
\rev{We present \sysname, an accurate and efficient 3D hand pose reconstruction system for around-device interaction applications using a single low-cost thermal array sensor. 
\sysname achieves this by contributing a physics-inspired neural network that learns from thermo-depth physics, 2D thermo-pose context, and hand structure knowledge. 
\sysname also presents a novel heterogeneous knowledge distillation technique that enables NanoTapor, an ultra-lightweight model suitable for real-time IoT adoption. 
We conduct extensive real-world experiments to demonstrate \sysname's high accuracy with low complexity, and carry out case studies on gesture control and finger tracking to showcase its potential applications. \sysname sheds light on a ubiquitous around-device interaction interface for the Internet of Thermal Things.}
\begin{acks}
This work is supported by the NSFC under grant No. 62222216 and Hong Kong RGC ECS under grant No. 27204522. 
\end{acks}

\bibliographystyle{ACM-Reference-Format}
\bibliography{reference}


\begin{thebibliography}{92}


\ifx \showCODEN    \undefined \def \showCODEN     #1{\unskip}     \fi
\ifx \showISBNx    \undefined \def \showISBNx     #1{\unskip}     \fi
\ifx \showISBNxiii \undefined \def \showISBNxiii  #1{\unskip}     \fi
\ifx \showISSN     \undefined \def \showISSN      #1{\unskip}     \fi
\ifx \showLCCN     \undefined \def \showLCCN      #1{\unskip}     \fi
\ifx \shownote     \undefined \def \shownote      #1{#1}          \fi
\ifx \showarticletitle \undefined \def \showarticletitle #1{#1}   \fi
\ifx \showURL      \undefined \def \showURL       {\relax}        \fi
\providecommand\bibfield[2]{#2}
\providecommand\bibinfo[2]{#2}
\providecommand\natexlab[1]{#1}
\providecommand\showeprint[2][]{arXiv:#2}

\bibitem[Akter et~al\mbox{.}(2020)]%
        {akterPrivacyConsiderationsVisually2020}
\bibfield{author}{\bibinfo{person}{Taslima Akter}, \bibinfo{person}{Tousif Ahmed}, \bibinfo{person}{Apu Kapadia}, {and} \bibinfo{person}{Swami~Manohar Swaminathan}.} \bibinfo{year}{2020}\natexlab{}.
\newblock \showarticletitle{Privacy {{Considerations}} of the {{Visually Impaired}} with {{Camera Based Assistive Technologies}}: {{Misrepresentation}}, {{Impropriety}}, and {{Fairness}}}. In \bibinfo{booktitle}{\emph{The 22nd {{International ACM SIGACCESS Conference}} on {{Computers}} and {{Accessibility}}}}. \bibinfo{address}{{Virtual Event Greece}}.
\newblock
\showISBNx{978-1-4503-7103-2}


\bibitem[Bai et~al\mbox{.}(2019)]%
        {bai2019onnx}
\bibfield{author}{\bibinfo{person}{Junjie Bai}, \bibinfo{person}{Fang Lu}, \bibinfo{person}{Ke Zhang}, {et~al\mbox{.}}} \bibinfo{year}{2019}\natexlab{}.
\newblock \bibinfo{title}{ONNX: Open Neural Network Exchange}.
\newblock \bibinfo{howpublished}{\url{https://github.com/onnx/onnx}}.
\newblock


\bibitem[Birkeland et~al\mbox{.}(2024)]%
        {birkelandVideoBasedHand2024}
\bibfield{author}{\bibinfo{person}{Simen Birkeland}, \bibinfo{person}{Lin~Julie Fjeldvik}, \bibinfo{person}{Nadia Noori}, \bibinfo{person}{Sreenivasa~Reddy Yeduri}, {and} \bibinfo{person}{Linga~Reddy Cenkeramaddi}.} \bibinfo{year}{2024}\natexlab{}.
\newblock \showarticletitle{Video Based Hand Gesture Recognition Dataset Using Thermal Camera}.
\newblock \bibinfo{journal}{\emph{Data in Brief}}  \bibinfo{volume}{54} (\bibinfo{date}{June} \bibinfo{year}{2024}), \bibinfo{pages}{110299}.
\newblock
\showISSN{2352-3409}
\href{https://doi.org/10.1016/j.dib.2024.110299}{doi:\nolinkurl{10.1016/j.dib.2024.110299}}


\bibitem[Bouazizi et~al\mbox{.}(2022)]%
        {bouaziziLowResolutionInfraredArray2022}
\bibfield{author}{\bibinfo{person}{Mondher Bouazizi}, \bibinfo{person}{Chen Ye}, {and} \bibinfo{person}{Tomoaki Ohtsuki}.} \bibinfo{year}{2022}\natexlab{}.
\newblock \showarticletitle{Low-{{Resolution Infrared Array Sensor}} for {{Counting}} and {{Localizing People Indoors}}: {{When Low End Technology Meets Cutting Edge Deep Learning Techniques}}}.
\newblock \bibinfo{journal}{\emph{Information}} \bibinfo{volume}{13}, \bibinfo{number}{3} (\bibinfo{date}{March} \bibinfo{year}{2022}), \bibinfo{pages}{132}.
\newblock
\showISSN{2078-2489}


\bibitem[Budzier and Gerlach(2011)]%
        {budzierThermalInfraredSensors2011a}
\bibfield{author}{\bibinfo{person}{Helmut Budzier} {and} \bibinfo{person}{Gerald Gerlach}.} \bibinfo{year}{2011}\natexlab{}.
\newblock \bibinfo{booktitle}{\emph{Thermal {{Infrared Sensors}}: {{Theory}}, {{Optimisation}} and {{Practice}}} (\bibinfo{edition}{1} ed.)}.
\newblock \bibinfo{publisher}{{Wiley}}.
\newblock
\showISBNx{978-0-470-87192-8 978-0-470-97691-3}


\bibitem[Cerutti et~al\mbox{.}(2019)]%
        {ceruttiConvolutionalNeuralNetwork2019}
\bibfield{author}{\bibinfo{person}{Gianmarco Cerutti}, \bibinfo{person}{Rahul Prasad}, {and} \bibinfo{person}{Elisabetta Farella}.} \bibinfo{year}{2019}\natexlab{}.
\newblock \showarticletitle{Convolutional {{Neural Network}} on {{Embedded Platform}} for {{People Presence Detection}} in {{Low Resolution Thermal Images}}}. In \bibinfo{booktitle}{\emph{Proceedings of IEEE {{ICASSP}}}}. \bibinfo{pages}{7610--7614}.
\newblock
\showISSN{2379-190X}


\bibitem[Chen et~al\mbox{.}(2021b)]%
        {chenMagXWearableUntethered2021}
\bibfield{author}{\bibinfo{person}{Dongyao Chen}, \bibinfo{person}{Mingke Wang}, \bibinfo{person}{Chenxi He}, \bibinfo{person}{Qing Luo}, \bibinfo{person}{Yasha Iravantchi}, \bibinfo{person}{Alanson Sample}, \bibinfo{person}{Kang~G. Shin}, {and} \bibinfo{person}{Xinbing Wang}.} \bibinfo{year}{2021}\natexlab{b}.
\newblock \showarticletitle{{{MagX}}: Wearable, Untethered Hands Tracking with Passive Magnets}. In \bibinfo{booktitle}{\emph{Proceedings of ACM MobiCom}}. \bibinfo{pages}{269--282}.
\newblock
\showISBNx{978-1-4503-8342-4}


\bibitem[Chen et~al\mbox{.}(2021a)]%
        {chenModelBased3DHand2021}
\bibfield{author}{\bibinfo{person}{Yujin Chen}, \bibinfo{person}{Zhigang Tu}, \bibinfo{person}{Di Kang}, \bibinfo{person}{Linchao Bao}, \bibinfo{person}{Ying Zhang}, \bibinfo{person}{Xuefei Zhe}, \bibinfo{person}{Ruizhi Chen}, {and} \bibinfo{person}{Junsong Yuan}.} \bibinfo{year}{2021}\natexlab{a}.
\newblock \showarticletitle{Model-{{Based 3D Hand Reconstruction}} via {{Self-Supervised Learning}}}. In \bibinfo{booktitle}{\emph{Proceedings of the {{IEEE}}/{{CVF Conference}} on {{Computer Vision}} and {{Pattern Recognition}}}}. \bibinfo{pages}{10451--10460}.
\newblock


\bibitem[Chi et~al\mbox{.}(2018)]%
        {chi2018ear}
\bibfield{author}{\bibinfo{person}{Zicheng Chi}, \bibinfo{person}{Yao Yao}, \bibinfo{person}{Tiantian Xie}, \bibinfo{person}{Xin Liu}, \bibinfo{person}{Zhichuan Huang}, \bibinfo{person}{Wei Wang}, {and} \bibinfo{person}{Ting Zhu}.} \bibinfo{year}{2018}\natexlab{}.
\newblock \showarticletitle{EAR: Exploiting uncontrollable ambient RF signals in heterogeneous networks for gesture recognition}. In \bibinfo{booktitle}{\emph{Proceedings of the 16th ACM conference on embedded networked sensor systems}}. \bibinfo{pages}{237--249}.
\newblock


\bibitem[Chidurala and Li(2021)]%
        {chiduralaOccupancyEstimationUsing2021}
\bibfield{author}{\bibinfo{person}{Veena Chidurala} {and} \bibinfo{person}{Xinrong Li}.} \bibinfo{year}{2021}\natexlab{}.
\newblock \showarticletitle{Occupancy {{Estimation Using Thermal Imaging Sensors}} and {{Machine Learning Algorithms}}}.
\newblock \bibinfo{journal}{\emph{IEEE Sensors Journal}} \bibinfo{volume}{21}, \bibinfo{number}{6} (\bibinfo{date}{March} \bibinfo{year}{2021}), \bibinfo{pages}{8627--8638}.
\newblock
\showISSN{1558-1748}


\bibitem[Cordts et~al\mbox{.}(2016)]%
        {cordts2016cityscapes}
\bibfield{author}{\bibinfo{person}{Marius Cordts}, \bibinfo{person}{Mohamed Omran}, \bibinfo{person}{Sebastian Ramos}, \bibinfo{person}{Timo Rehfeld}, \bibinfo{person}{Markus Enzweiler}, \bibinfo{person}{Rodrigo Benenson}, \bibinfo{person}{Uwe Franke}, \bibinfo{person}{Stefan Roth}, {and} \bibinfo{person}{Bernt Schiele}.} \bibinfo{year}{2016}\natexlab{}.
\newblock \showarticletitle{The cityscapes dataset for semantic urban scene understanding}. In \bibinfo{booktitle}{\emph{Proceedings of the IEEE conference on computer vision and pattern recognition}}. \bibinfo{pages}{3213--3223}.
\newblock


\bibitem[David et~al\mbox{.}(2021)]%
        {david2021tensorflow}
\bibfield{author}{\bibinfo{person}{Robert David}, \bibinfo{person}{Jared Duke}, \bibinfo{person}{Advait Jain}, \bibinfo{person}{Vijay Janapa~Reddi}, \bibinfo{person}{Nat Jeffries}, \bibinfo{person}{Jian Li}, \bibinfo{person}{Nick Kreeger}, \bibinfo{person}{Ian Nappier}, \bibinfo{person}{Meghna Natraj}, \bibinfo{person}{Tiezhen Wang}, {et~al\mbox{.}}} \bibinfo{year}{2021}\natexlab{}.
\newblock \showarticletitle{Tensorflow lite micro: Embedded machine learning for tinyml systems}.
\newblock \bibinfo{journal}{\emph{Proceedings of Machine Learning and Systems}}  \bibinfo{volume}{3} (\bibinfo{year}{2021}), \bibinfo{pages}{800--811}.
\newblock


\bibitem[Davis and Sharma(2004)]%
        {davis2004robust}
\bibfield{author}{\bibinfo{person}{James~W Davis} {and} \bibinfo{person}{Vinay Sharma}.} \bibinfo{year}{2004}\natexlab{}.
\newblock \showarticletitle{Robust Background-Subtraction for Person Detection in Thermal Imagery}. In \bibinfo{booktitle}{\emph{CVPR Workshops}}. \bibinfo{pages}{128}.
\newblock


\bibitem[Deng et~al\mbox{.}(2009)]%
        {deng2009imagenet}
\bibfield{author}{\bibinfo{person}{Jia Deng}, \bibinfo{person}{Wei Dong}, \bibinfo{person}{Richard Socher}, \bibinfo{person}{Li-Jia Li}, \bibinfo{person}{Kai Li}, {and} \bibinfo{person}{Li Fei-Fei}.} \bibinfo{year}{2009}\natexlab{}.
\newblock \showarticletitle{Imagenet: A large-scale hierarchical image database}. In \bibinfo{booktitle}{\emph{2009 IEEE conference on computer vision and pattern recognition}}. Ieee, \bibinfo{pages}{248--255}.
\newblock


\bibitem[Ding et~al\mbox{.}(2024)]%
        {dingThermoHandsBenchmark3D2024a}
\bibfield{author}{\bibinfo{person}{Fangqiang Ding}, \bibinfo{person}{Yunzhou Zhu}, \bibinfo{person}{Xiangyu Wen}, \bibinfo{person}{Gaowen Liu}, {and} \bibinfo{person}{Chris~Xiaoxuan Lu}.} \bibinfo{year}{2024}\natexlab{}.
\newblock \bibinfo{title}{{{ThermoHands}}: {{A Benchmark}} for {{3D Hand Pose Estimation}} from {{Egocentric Thermal Images}}}.
\newblock
\href{https://doi.org/10.48550/arXiv.2403.09871}{doi:\nolinkurl{10.48550/arXiv.2403.09871}}
\showeprint[arxiv]{2403.09871}~[cs]


\bibitem[Dragulescu et~al\mbox{.}(2007)]%
        {dragulescu3DActiveWorkspace2007}
\bibfield{author}{\bibinfo{person}{Doina Dragulescu}, \bibinfo{person}{V{\'e}ronique Perdereau}, \bibinfo{person}{Michel Drouin}, \bibinfo{person}{Loredana Ungureanu}, {and} \bibinfo{person}{Karoly Menyhardt}.} \bibinfo{year}{2007}\natexlab{}.
\newblock \showarticletitle{{{3D}} Active Workspace of Human Hand Anatomical Model}.
\newblock \bibinfo{journal}{\emph{BioMedical Engineering OnLine}} \bibinfo{volume}{6}, \bibinfo{number}{1} (\bibinfo{date}{May} \bibinfo{year}{2007}), \bibinfo{pages}{15}.
\newblock
\showISSN{1475-925X}


\bibitem[Faulkner et~al\mbox{.}(2021)]%
        {faulknerDeviceFreeLocalizationUsing2021}
\bibfield{author}{\bibinfo{person}{Nathaniel Faulkner}, \bibinfo{person}{Fakhrul Alam}, \bibinfo{person}{Mathew Legg}, {and} \bibinfo{person}{Serge Demidenko}.} \bibinfo{year}{2021}\natexlab{}.
\newblock \showarticletitle{Device-{{Free Localization Using Privacy-Preserving Infrared Signatures Acquired From Thermopiles}} and {{Machine Learning}}}.
\newblock \bibinfo{journal}{\emph{IEEE Access}}  \bibinfo{volume}{9} (\bibinfo{year}{2021}), \bibinfo{pages}{81786--81797}.
\newblock
\showISSN{2169-3536}


\bibitem[Gao et~al\mbox{.}(2022b)]%
        {gaoDARTArticulatedHand2022}
\bibfield{author}{\bibinfo{person}{Daiheng Gao}, \bibinfo{person}{Yuliang Xiu}, \bibinfo{person}{Kailin Li}, \bibinfo{person}{Lixin Yang}, \bibinfo{person}{Feng Wang}, \bibinfo{person}{Peng Zhang}, \bibinfo{person}{Bang Zhang}, \bibinfo{person}{Cewu Lu}, {and} \bibinfo{person}{Ping Tan}.} \bibinfo{year}{2022}\natexlab{b}.
\newblock \showarticletitle{{{DART}}: {{Articulated Hand Model}} with {{Diverse Accessories}} and {{Rich Textures}}}. In \bibinfo{booktitle}{\emph{Advances in {{Neural Information Processing Systems}}}}, Vol.~\bibinfo{volume}{35}. \bibinfo{pages}{37055--37067}.
\newblock


\bibitem[Gao et~al\mbox{.}(2022a)]%
        {Gao_Li_Xie_Yi_Wang_Wu_Zhang_2022}
\bibfield{author}{\bibinfo{person}{Ruiyang Gao}, \bibinfo{person}{Wenwei Li}, \bibinfo{person}{Yaxiong Xie}, \bibinfo{person}{Enze Yi}, \bibinfo{person}{Leye Wang}, \bibinfo{person}{Dan Wu}, {and} \bibinfo{person}{Daqing Zhang}.} \bibinfo{year}{2022}\natexlab{a}.
\newblock \showarticletitle{Towards Robust Gesture Recognition by Characterizing the Sensing Quality of WiFi Signals}.
\newblock \bibinfo{journal}{\emph{Proc. ACM Interact. Mob. Wearable Ubiquitous Technol.}} \bibinfo{volume}{6}, \bibinfo{number}{1} (\bibinfo{year}{2022}), \bibinfo{pages}{11:1--11:26}.
\newblock
\href{https://doi.org/10.1145/3517241}{doi:\nolinkurl{10.1145/3517241}}


\bibitem[Gong et~al\mbox{.}(2017)]%
        {gongPyroThumbTipGesture2017}
\bibfield{author}{\bibinfo{person}{Jun Gong}, \bibinfo{person}{Yang Zhang}, \bibinfo{person}{Xia Zhou}, {and} \bibinfo{person}{Xing-Dong Yang}.} \bibinfo{year}{2017}\natexlab{}.
\newblock \showarticletitle{Pyro: {{Thumb-Tip Gesture Recognition Using Pyroelectric Infrared Sensing}}}. In \bibinfo{booktitle}{\emph{Proceedings of the 30th {{Annual ACM Symposium}} on {{User Interface Software}} and {{Technology}}}} \emph{(\bibinfo{series}{{{UIST}} '17})}. \bibinfo{publisher}{Association for Computing Machinery}, \bibinfo{address}{New York, NY, USA}, \bibinfo{pages}{553--563}.
\newblock
\showISBNx{978-1-4503-4981-9}
\href{https://doi.org/10.1145/3126594.3126615}{doi:\nolinkurl{10.1145/3126594.3126615}}


\bibitem[He et~al\mbox{.}(2023)]%
        {heNonContactFallDetection2023}
\bibfield{author}{\bibinfo{person}{Chunhua He}, \bibinfo{person}{Shuibin Liu}, \bibinfo{person}{Guangxiong Zhong}, \bibinfo{person}{Heng Wu}, \bibinfo{person}{Lianglun Cheng}, \bibinfo{person}{Juze Lin}, {and} \bibinfo{person}{Qinwen Huang}.} \bibinfo{year}{2023}\natexlab{}.
\newblock \showarticletitle{A {{Non-Contact Fall Detection Method}} for {{Bathroom Application Based}} on {{MEMS Infrared Sensors}}}.
\newblock \bibinfo{journal}{\emph{Micromachines}} \bibinfo{volume}{14}, \bibinfo{number}{1} (\bibinfo{date}{Jan.} \bibinfo{year}{2023}), \bibinfo{pages}{130}.
\newblock
\showISSN{2072-666X}


\bibitem[Hevesi et~al\mbox{.}(2014)]%
        {hevesiMonitoringHouseholdActivities2014}
\bibfield{author}{\bibinfo{person}{Peter Hevesi}, \bibinfo{person}{Sebastian Wille}, \bibinfo{person}{Gerald Pirkl}, \bibinfo{person}{Norbert Wehn}, {and} \bibinfo{person}{Paul Lukowicz}.} \bibinfo{year}{2014}\natexlab{}.
\newblock \showarticletitle{Monitoring Household Activities and User Location with a Cheap, Unobtrusive Thermal Sensor Array}. In \bibinfo{booktitle}{\emph{Proceedings of the 2014 {{ACM International Joint Conference}} on {{Pervasive}} and {{Ubiquitous Computing}}}} \emph{(\bibinfo{series}{{{UbiComp}} '14})}. \bibinfo{address}{{New York, NY, USA}}, \bibinfo{pages}{141--145}.
\newblock
\showISBNx{978-1-4503-2968-2}


\bibitem[Hinton et~al\mbox{.}(2015)]%
        {hinton2015distilling}
\bibfield{author}{\bibinfo{person}{Geoffrey Hinton}, \bibinfo{person}{Oriol Vinyals}, {and} \bibinfo{person}{Jeff Dean}.} \bibinfo{year}{2015}\natexlab{}.
\newblock \showarticletitle{Distilling the knowledge in a neural network}.
\newblock \bibinfo{journal}{\emph{arXiv preprint arXiv:1503.02531}} (\bibinfo{year}{2015}).
\newblock


\bibitem[Hu et~al\mbox{.}(2020)]%
        {huFingerTrakContinuous3D2020}
\bibfield{author}{\bibinfo{person}{Fang Hu}, \bibinfo{person}{Peng He}, \bibinfo{person}{Songlin Xu}, \bibinfo{person}{Yin Li}, {and} \bibinfo{person}{Cheng Zhang}.} \bibinfo{year}{2020}\natexlab{}.
\newblock \showarticletitle{{{FingerTrak}}: {{Continuous 3D Hand Pose Tracking}} by {{Deep Learning Hand Silhouettes Captured}} by {{Miniature Thermal Cameras}} on {{Wrist}}}.
\newblock \bibinfo{journal}{\emph{Proceedings of the ACM on Interactive, Mobile, Wearable and Ubiquitous Technologies}} \bibinfo{volume}{4}, \bibinfo{number}{2} (\bibinfo{date}{June} \bibinfo{year}{2020}), \bibinfo{pages}{71:1--71:24}.
\newblock


\bibitem[Insights(2024)]%
        {futuremarketinsightsGestureControlMarket}
\bibfield{author}{\bibinfo{person}{Future~Market Insights}.} \bibinfo{year}{2024}\natexlab{}.
\newblock \bibinfo{title}{Gesture Control Market}.
\newblock
\urldef\tempurl%
\url{https://www.futuremarketinsights.com/reports/gesture-control-market}
\showURL{%
\tempurl}


\bibitem[Ionescu et~al\mbox{.}(2013)]%
        {ionescu2013human3}
\bibfield{author}{\bibinfo{person}{Catalin Ionescu}, \bibinfo{person}{Dragos Papava}, \bibinfo{person}{Vlad Olaru}, {and} \bibinfo{person}{Cristian Sminchisescu}.} \bibinfo{year}{2013}\natexlab{}.
\newblock \showarticletitle{Human3. 6m: Large scale datasets and predictive methods for 3d human sensing in natural environments}.
\newblock \bibinfo{journal}{\emph{IEEE transactions on pattern analysis and machine intelligence}} \bibinfo{volume}{36}, \bibinfo{number}{7} (\bibinfo{year}{2013}), \bibinfo{pages}{1325--1339}.
\newblock


\bibitem[Ji et~al\mbox{.}(2023)]%
        {jiConstruct3DHand2023}
\bibfield{author}{\bibinfo{person}{Sijie Ji}, \bibinfo{person}{Xuanye Zhang}, \bibinfo{person}{Yuanqing Zheng}, {and} \bibinfo{person}{Mo Li}.} \bibinfo{year}{2023}\natexlab{}.
\newblock \showarticletitle{Construct {{3D Hand Skeleton}} with {{Commercial WiFi}}}. In \bibinfo{booktitle}{\emph{Proceedings of ACM SenSys}}.
\newblock


\bibitem[Kato and Takemura(2016)]%
        {Kato_Takemura_2016}
\bibfield{author}{\bibinfo{person}{Hiroyuki Kato} {and} \bibinfo{person}{Kentaro Takemura}.} \bibinfo{year}{2016}\natexlab{}.
\newblock \showarticletitle{Hand pose estimation based on active bone-conducted sound sensing}. In \bibinfo{booktitle}{\emph{Proceedings of the 2016 ACM International Joint Conference on Pervasive and Ubiquitous Computing: Adjunct}} \emph{(\bibinfo{series}{UbiComp ’16})}. \bibinfo{publisher}{Association for Computing Machinery}, \bibinfo{address}{New York, NY, USA}, \bibinfo{pages}{109–112}.
\newblock
\showISBNx{978-1-4503-4462-3}
\href{https://doi.org/10.1145/2968219.2971403}{doi:\nolinkurl{10.1145/2968219.2971403}}


\bibitem[Laput and Harrison(2019)]%
        {laputSensingFineGrainedHand2019}
\bibfield{author}{\bibinfo{person}{Gierad Laput} {and} \bibinfo{person}{Chris Harrison}.} \bibinfo{year}{2019}\natexlab{}.
\newblock \showarticletitle{Sensing {{Fine-Grained Hand Activity}} with {{Smartwatches}}}. In \bibinfo{booktitle}{\emph{Proceedings of the 2019 {{CHI Conference}} on {{Human Factors}} in {{Computing Systems}}}} \emph{(\bibinfo{series}{{{CHI}} '19})}. \bibinfo{address}{{New York, NY, USA}}, \bibinfo{pages}{1--13}.
\newblock
\showISBNx{978-1-4503-5970-2}


\bibitem[Li et~al\mbox{.}(2022a)]%
        {li2022experience}
\bibfield{author}{\bibinfo{person}{Dong Li}, \bibinfo{person}{Shirui Cao}, \bibinfo{person}{Sunghoon~Ivan Lee}, {and} \bibinfo{person}{Jie Xiong}.} \bibinfo{year}{2022}\natexlab{a}.
\newblock \showarticletitle{Experience: practical problems for acoustic sensing}. In \bibinfo{booktitle}{\emph{Proceedings of the 28th Annual International Conference on Mobile Computing And Networking}}. \bibinfo{pages}{381--390}.
\newblock


\bibitem[Li et~al\mbox{.}(2022b)]%
        {li2022room}
\bibfield{author}{\bibinfo{person}{Dong Li}, \bibinfo{person}{Jialin Liu}, \bibinfo{person}{Sunghoon~Ivan Lee}, {and} \bibinfo{person}{Jie Xiong}.} \bibinfo{year}{2022}\natexlab{b}.
\newblock \showarticletitle{Room-scale hand gesture recognition using smart speakers}. In \bibinfo{booktitle}{\emph{Proceedings of the 20th ACM Conference on Embedded Networked Sensor Systems}}. \bibinfo{pages}{462--475}.
\newblock


\bibitem[Li et~al\mbox{.}(2016)]%
        {liWiFingerTalkYour2016}
\bibfield{author}{\bibinfo{person}{Hong Li}, \bibinfo{person}{Wei Yang}, \bibinfo{person}{Jianxin Wang}, \bibinfo{person}{Yang Xu}, {and} \bibinfo{person}{Liusheng Huang}.} \bibinfo{year}{2016}\natexlab{}.
\newblock \showarticletitle{{{WiFinger}}: Talk to Your Smart Devices with Finger-Grained Gesture}. In \bibinfo{booktitle}{\emph{Proceedings of the 2016 {{ACM International Joint Conference}} on {{Pervasive}} and {{Ubiquitous Computing}}}} \emph{(\bibinfo{series}{{{UbiComp}} '16})}. \bibinfo{address}{{New York, NY, USA}}, \bibinfo{pages}{250--261}.
\newblock
\showISBNx{978-1-4503-4461-6}


\bibitem[Li and Chan(2015)]%
        {li3DHumanPose2015}
\bibfield{author}{\bibinfo{person}{Sijin Li} {and} \bibinfo{person}{Antoni~B. Chan}.} \bibinfo{year}{2015}\natexlab{}.
\newblock \showarticletitle{{{3D Human Pose Estimation}} from {{Monocular Images}} with {{Deep Convolutional Neural Network}}}. In \bibinfo{booktitle}{\emph{Computer {{Vision}} -- {{ACCV}} 2014}} \emph{(\bibinfo{series}{Lecture {{Notes}} in {{Computer Science}}})}, \bibfield{editor}{\bibinfo{person}{Daniel Cremers}, \bibinfo{person}{Ian Reid}, \bibinfo{person}{Hideo Saito}, {and} \bibinfo{person}{Ming-Hsuan Yang}} (Eds.). \bibinfo{publisher}{{Springer International Publishing}}, \bibinfo{address}{{Cham}}, \bibinfo{pages}{332--347}.
\newblock
\showISBNx{978-3-319-16808-1}


\bibitem[Li et~al\mbox{.}(2017)]%
        {liReconstructingHandPoses2017}
\bibfield{author}{\bibinfo{person}{Tianxing Li}, \bibinfo{person}{Xi Xiong}, \bibinfo{person}{Yifei Xie}, \bibinfo{person}{George Hito}, \bibinfo{person}{Xing-Dong Yang}, {and} \bibinfo{person}{Xia Zhou}.} \bibinfo{year}{2017}\natexlab{}.
\newblock \showarticletitle{Reconstructing {{Hand Poses Using Visible Light}}}. In \bibinfo{booktitle}{\emph{Proceedings of the {{ACM}} on {{Interactive}}, {{Mobile}}, {{Wearable}} and {{Ubiquitous Technologies}}}}, Vol.~\bibinfo{volume}{1}. \bibinfo{pages}{1--20}.
\newblock


\bibitem[Li et~al\mbox{.}(2023)]%
        {Li_Zhang_Chen_Wan_Zhang_Hu_Sun_Chen_2023}
\bibfield{author}{\bibinfo{person}{Yadong Li}, \bibinfo{person}{Dongheng Zhang}, \bibinfo{person}{Jinbo Chen}, \bibinfo{person}{Jinwei Wan}, \bibinfo{person}{Dong Zhang}, \bibinfo{person}{Yang Hu}, \bibinfo{person}{Qibin Sun}, {and} \bibinfo{person}{Yan Chen}.} \bibinfo{year}{2023}\natexlab{}.
\newblock \showarticletitle{Towards Domain-Independent and Real-Time Gesture Recognition Using mmWave Signal}.
\newblock \bibinfo{journal}{\emph{IEEE Transactions on Mobile Computing}} \bibinfo{volume}{22}, \bibinfo{number}{12} (\bibinfo{date}{Dec.} \bibinfo{year}{2023}), \bibinfo{pages}{7355–7369}.
\newblock
\showISSN{1558-0660}
\href{https://doi.org/10.1109/TMC.2022.3207570}{doi:\nolinkurl{10.1109/TMC.2022.3207570}}


\bibitem[Li et~al\mbox{.}(2022c)]%
        {liNIMBLENonrigidHand2022}
\bibfield{author}{\bibinfo{person}{Yuwei Li}, \bibinfo{person}{Longwen Zhang}, \bibinfo{person}{Zesong Qiu}, \bibinfo{person}{Yingwenqi Jiang}, \bibinfo{person}{Nianyi Li}, \bibinfo{person}{Yuexin Ma}, \bibinfo{person}{Yuyao Zhang}, \bibinfo{person}{Lan Xu}, {and} \bibinfo{person}{Jingyi Yu}.} \bibinfo{year}{2022}\natexlab{c}.
\newblock \showarticletitle{{{NIMBLE}}: A Non-Rigid Hand Model with Bones and Muscles}.
\newblock \bibinfo{journal}{\emph{ACM Transactions on Graphics}} \bibinfo{volume}{41}, \bibinfo{number}{4} (\bibinfo{date}{July} \bibinfo{year}{2022}), \bibinfo{pages}{120:1--120:16}.
\newblock
\showISSN{0730-0301}


\bibitem[Lien et~al\mbox{.}(2016)]%
        {lienSoliUbiquitousGesture2016}
\bibfield{author}{\bibinfo{person}{Jaime Lien}, \bibinfo{person}{Nicholas Gillian}, \bibinfo{person}{M.~Emre Karagozler}, \bibinfo{person}{Patrick Amihood}, \bibinfo{person}{Carsten Schwesig}, \bibinfo{person}{Erik Olson}, \bibinfo{person}{Hakim Raja}, {and} \bibinfo{person}{Ivan Poupyrev}.} \bibinfo{year}{2016}\natexlab{}.
\newblock \showarticletitle{Soli: Ubiquitous Gesture Sensing with Millimeter Wave Radar}.
\newblock \bibinfo{journal}{\emph{ACM Transactions on Graphics}} \bibinfo{volume}{35}, \bibinfo{number}{4} (\bibinfo{date}{July} \bibinfo{year}{2016}), \bibinfo{pages}{142:1--142:19}.
\newblock
\showISSN{0730-0301}


\bibitem[Lin et~al\mbox{.}(2014)]%
        {lin2014microsoft}
\bibfield{author}{\bibinfo{person}{Tsung-Yi Lin}, \bibinfo{person}{Michael Maire}, \bibinfo{person}{Serge Belongie}, \bibinfo{person}{James Hays}, \bibinfo{person}{Pietro Perona}, \bibinfo{person}{Deva Ramanan}, \bibinfo{person}{Piotr Doll{\'a}r}, {and} \bibinfo{person}{C~Lawrence Zitnick}.} \bibinfo{year}{2014}\natexlab{}.
\newblock \showarticletitle{Microsoft coco: Common objects in context}. In \bibinfo{booktitle}{\emph{Computer Vision--ECCV 2014: 13th European Conference, Zurich, Switzerland, September 6-12, 2014, Proceedings, Part V 13}}. Springer, \bibinfo{pages}{740--755}.
\newblock


\bibitem[Liu et~al\mbox{.}(2024)]%
        {Liu_Yu_Wang_Guo_Li_Yi_Zhang_2024}
\bibfield{author}{\bibinfo{person}{Yan Liu}, \bibinfo{person}{Anlan Yu}, \bibinfo{person}{Leye Wang}, \bibinfo{person}{Bin Guo}, \bibinfo{person}{Yang Li}, \bibinfo{person}{Enze Yi}, {and} \bibinfo{person}{Daqing Zhang}.} \bibinfo{year}{2024}\natexlab{}.
\newblock \showarticletitle{UniFi: A Unified Framework for Generalizable Gesture Recognition with Wi-Fi Signals Using Consistency-guided Multi-View Networks}.
\newblock \bibinfo{journal}{\emph{Proc. ACM Interact. Mob. Wearable Ubiquitous Technol.}} \bibinfo{volume}{7}, \bibinfo{number}{4} (\bibinfo{year}{2024}), \bibinfo{pages}{168:1--168:29}.
\newblock
\href{https://doi.org/10.1145/3631429}{doi:\nolinkurl{10.1145/3631429}}


\bibitem[Lu et~al\mbox{.}(2024)]%
        {Lu_Ding_Wang_Xue_2024}
\bibfield{author}{\bibinfo{person}{Yu Lu}, \bibinfo{person}{Dian Ding}, \bibinfo{person}{Ran Wang}, {and} \bibinfo{person}{Guangtao Xue}.} \bibinfo{year}{2024}\natexlab{}.
\newblock \showarticletitle{HCMG: Human-Capacitance based Micro Gesture for VR/AR}. In \bibinfo{booktitle}{\emph{Companion of the 2024 on ACM International Joint Conference on Pervasive and Ubiquitous Computing}} \emph{(\bibinfo{series}{UbiComp ’24})}. \bibinfo{publisher}{Association for Computing Machinery}, \bibinfo{address}{New York, NY, USA}, \bibinfo{pages}{766–770}.
\newblock
\showISBNx{9798400710582}
\href{https://doi.org/10.1145/3675094.3678386}{doi:\nolinkurl{10.1145/3675094.3678386}}


\bibitem[Mao et~al\mbox{.}(2019)]%
        {maoRNNBasedRoomScale2019}
\bibfield{author}{\bibinfo{person}{Wenguang Mao}, \bibinfo{person}{Mei Wang}, \bibinfo{person}{Wei Sun}, \bibinfo{person}{Lili Qiu}, \bibinfo{person}{Swadhin Pradhan}, {and} \bibinfo{person}{Yi-Chao Chen}.} \bibinfo{year}{2019}\natexlab{}.
\newblock \showarticletitle{{{RNN-Based Room Scale Hand Motion Tracking}}}. In \bibinfo{booktitle}{\emph{The 25th {{Annual International Conference}} on {{Mobile Computing}} and {{Networking}}}}. \bibinfo{address}{{New York, NY, USA}}, \bibinfo{pages}{1--16}.
\newblock
\showISBNx{978-1-4503-6169-9}


\bibitem[M{\~o}llmann and Vollmer(2018)]%
        {mollmannInfraredThermalImaging2018}
\bibfield{author}{\bibinfo{person}{Klaus-Peter M{\~o}llmann} {and} \bibinfo{person}{Michael Vollmer}.} \bibinfo{year}{2018}\natexlab{}.
\newblock \bibinfo{booktitle}{\emph{Infrared {{Thermal Imaging}}: {{Fundamentals}}, {{Research}} and {{Applications}}} (\bibinfo{edition}{2nd edition} ed.)}.
\newblock \bibinfo{publisher}{{Wiley-VCH}}, \bibinfo{address}{{Weinheim, Germany}}.
\newblock
\showISBNx{978-3-527-41351-5}


\bibitem[Muthukumar et~al\mbox{.}(2022)]%
        {muthukumarInfraredArraySensorBased2022}
\bibfield{author}{\bibinfo{person}{Krishnan~Arumugasamy Muthukumar}, \bibinfo{person}{Mondher Bouazizi}, {and} \bibinfo{person}{Tomoaki Ohtsuki}.} \bibinfo{year}{2022}\natexlab{}.
\newblock \showarticletitle{An {{Infrared Array Sensor-Based Approach}} for {{Activity Detection}}, {{Combining Low-Cost Technology}} with {{Advanced Deep Learning Techniques}}}.
\newblock \bibinfo{journal}{\emph{Sensors}} \bibinfo{volume}{22}, \bibinfo{number}{10} (\bibinfo{date}{Jan.} \bibinfo{year}{2022}), \bibinfo{pages}{3898}.
\newblock
\showISSN{1424-8220}


\bibitem[Nandakumar et~al\mbox{.}(2016)]%
        {nandakumarFingerIOUsingActive2016}
\bibfield{author}{\bibinfo{person}{Rajalakshmi Nandakumar}, \bibinfo{person}{Vikram Iyer}, \bibinfo{person}{Desney Tan}, {and} \bibinfo{person}{Shyamnath Gollakota}.} \bibinfo{year}{2016}\natexlab{}.
\newblock \showarticletitle{{{FingerIO}}: {{Using Active Sonar}} for {{Fine-Grained Finger Tracking}}}. In \bibinfo{booktitle}{\emph{Proceedings of the 2016 {{CHI Conference}} on {{Human Factors}} in {{Computing Systems}}}} \emph{(\bibinfo{series}{{{CHI}} '16})}. \bibinfo{address}{{New York, NY, USA}}, \bibinfo{pages}{1515--1525}.
\newblock
\showISBNx{978-1-4503-3362-7}


\bibitem[Naser et~al\mbox{.}(2021)]%
        {naserHumanDistanceEstimation2021}
\bibfield{author}{\bibinfo{person}{Abdallah Naser}, \bibinfo{person}{Ahmad Lotfi}, {and} \bibinfo{person}{Joni Zhong}.} \bibinfo{year}{2021}\natexlab{}.
\newblock \showarticletitle{Towards Human Distance Estimation Using a Thermal Sensor Array}.
\newblock \bibinfo{journal}{\emph{Neural Computing and Applications}} (\bibinfo{date}{June} \bibinfo{year}{2021}).
\newblock
\showISSN{1433-3058}


\bibitem[Naser et~al\mbox{.}(2022)]%
        {naserMultipleThermalSensor2022}
\bibfield{author}{\bibinfo{person}{Abdallah Naser}, \bibinfo{person}{Ahmad Lotfi}, {and} \bibinfo{person}{Junpei Zhong}.} \bibinfo{year}{2022}\natexlab{}.
\newblock \showarticletitle{Multiple {{Thermal Sensor Array Fusion Toward Enabling Privacy-Preserving Human Monitoring Applications}}}.
\newblock \bibinfo{journal}{\emph{IEEE Internet of Things Journal}} \bibinfo{volume}{9}, \bibinfo{number}{17} (\bibinfo{date}{Sept.} \bibinfo{year}{2022}), \bibinfo{pages}{16677--16688}.
\newblock
\showISSN{2327-4662}


\bibitem[Nguyen et~al\mbox{.}(2019)]%
        {nguyen2019handsense}
\bibfield{author}{\bibinfo{person}{Viet Nguyen}, \bibinfo{person}{Siddharth Rupavatharam}, \bibinfo{person}{Luyang Liu}, \bibinfo{person}{Richard Howard}, {and} \bibinfo{person}{Marco Gruteser}.} \bibinfo{year}{2019}\natexlab{}.
\newblock \showarticletitle{HandSense: capacitive coupling-based dynamic, micro finger gesture recognition}. In \bibinfo{booktitle}{\emph{Proceedings of the 17th Conference on Embedded Networked Sensor Systems}}. \bibinfo{pages}{285--297}.
\newblock


\bibitem[Oberweger et~al\mbox{.}(2016)]%
        {oberwegerHandsDeepDeep2016}
\bibfield{author}{\bibinfo{person}{Markus Oberweger}, \bibinfo{person}{Paul Wohlhart}, {and} \bibinfo{person}{Vincent Lepetit}.} \bibinfo{year}{2016}\natexlab{}.
\newblock \bibinfo{title}{Hands {{Deep}} in {{Deep Learning}} for {{Hand Pose Estimation}}}.
\newblock
\showeprint[arxiv]{1502.06807}~[cs]


\bibitem[Paszke et~al\mbox{.}(2019)]%
        {paszke2019pytorch}
\bibfield{author}{\bibinfo{person}{Adam Paszke}, \bibinfo{person}{Sam Gross}, \bibinfo{person}{Francisco Massa}, \bibinfo{person}{Adam Lerer}, \bibinfo{person}{James Bradbury}, \bibinfo{person}{Gregory Chanan}, \bibinfo{person}{Trevor Killeen}, \bibinfo{person}{Zeming Lin}, \bibinfo{person}{Natalia Gimelshein}, \bibinfo{person}{Luca Antiga}, {et~al\mbox{.}}} \bibinfo{year}{2019}\natexlab{}.
\newblock \showarticletitle{Pytorch: An imperative style, high-performance deep learning library}.
\newblock \bibinfo{journal}{\emph{Advances in neural information processing systems}}  \bibinfo{volume}{32} (\bibinfo{year}{2019}).
\newblock


\bibitem[Patra et~al\mbox{.}(2018)]%
        {patraMmWaveRadarBased2018}
\bibfield{author}{\bibinfo{person}{Avishek Patra}, \bibinfo{person}{Philipp Geuer}, \bibinfo{person}{Andrea Munari}, {and} \bibinfo{person}{Petri M{\"a}h{\"o}nen}.} \bibinfo{year}{2018}\natexlab{}.
\newblock \showarticletitle{Mm-{{Wave Radar Based Gesture Recognition}}: {{Development}} and {{Evaluation}} of a {{Low-Power}}, {{Low-Complexity System}}}. In \bibinfo{booktitle}{\emph{Proceedings of the 2nd {{ACM Workshop}} on {{Millimeter Wave Networks}} and {{Sensing Systems}}}} \emph{(\bibinfo{series}{{{mmNets}} '18})}. \bibinfo{address}{{New York, NY, USA}}, \bibinfo{pages}{51--56}.
\newblock
\showISBNx{978-1-4503-5928-3}


\bibitem[Polla et~al\mbox{.}(2019)]%
        {pollaActionRecognitionLowResolution2019}
\bibfield{author}{\bibinfo{person}{F{\'e}lix Polla}, \bibinfo{person}{H{\'e}l{\`e}ne Laurent}, {and} \bibinfo{person}{Bruno Emile}.} \bibinfo{year}{2019}\natexlab{}.
\newblock \showarticletitle{Action {{Recognition}} from {{Low-Resolution Infrared Sensor}} for {{Indoor}} Use: {{A Comparative Study}} between {{Deep Learning}} and {{Classical Approaches}}}. In \bibinfo{booktitle}{\emph{2019 20th {{IEEE International Conference}} on {{Mobile Data Management}} ({{MDM}})}}. \bibinfo{pages}{409--414}.
\newblock
\showISSN{2375-0324}


\bibitem[Pu et~al\mbox{.}(2013)]%
        {puWholehomeGestureRecognition2013}
\bibfield{author}{\bibinfo{person}{Qifan Pu}, \bibinfo{person}{Sidhant Gupta}, \bibinfo{person}{Shyamnath Gollakota}, {and} \bibinfo{person}{Shwetak Patel}.} \bibinfo{year}{2013}\natexlab{}.
\newblock \showarticletitle{Whole-Home Gesture Recognition Using Wireless Signals}. In \bibinfo{booktitle}{\emph{Proceedings of the 19th Annual International Conference on {{Mobile}} Computing \& Networking}}. \bibinfo{pages}{27--38}.
\newblock


\bibitem[Qian et~al\mbox{.}(2020)]%
        {qianHTMLParametricHand2020}
\bibfield{author}{\bibinfo{person}{Neng Qian}, \bibinfo{person}{Jiayi Wang}, \bibinfo{person}{Franziska Mueller}, \bibinfo{person}{Florian Bernard}, \bibinfo{person}{Vladislav Golyanik}, {and} \bibinfo{person}{Christian Theobalt}.} \bibinfo{year}{2020}\natexlab{}.
\newblock \showarticletitle{{{HTML}}: {{A Parametric Hand Texture Model}} for {{3D Hand Reconstruction}} and {{Personalization}}}. In \bibinfo{booktitle}{\emph{Computer {{Vision}} {\textendash} {{ECCV}} 2020}} \emph{(\bibinfo{series}{Lecture {{Notes}} in {{Computer Science}}})}, \bibfield{editor}{\bibinfo{person}{Andrea Vedaldi}, \bibinfo{person}{Horst Bischof}, \bibinfo{person}{Thomas Brox}, {and} \bibinfo{person}{Jan-Michael Frahm}} (Eds.). \bibinfo{publisher}{{Springer International Publishing}}, \bibinfo{address}{{Cham}}, \bibinfo{pages}{54--71}.
\newblock
\showISBNx{978-3-030-58621-8}


\bibitem[Ramakrishna et~al\mbox{.}(2012)]%
        {ramakrishnaReconstructing3DHuman2012}
\bibfield{author}{\bibinfo{person}{Varun Ramakrishna}, \bibinfo{person}{Takeo Kanade}, {and} \bibinfo{person}{Yaser Sheikh}.} \bibinfo{year}{2012}\natexlab{}.
\newblock \showarticletitle{Reconstructing {{3D Human Pose}} from {{2D Image Landmarks}}}. In \bibinfo{booktitle}{\emph{Computer {{Vision}} {\textendash} {{ECCV}} 2012}} \emph{(\bibinfo{series}{Lecture {{Notes}} in {{Computer Science}}})}, \bibfield{editor}{\bibinfo{person}{Andrew Fitzgibbon}, \bibinfo{person}{Svetlana Lazebnik}, \bibinfo{person}{Pietro Perona}, \bibinfo{person}{Yoichi Sato}, {and} \bibinfo{person}{Cordelia Schmid}} (Eds.). \bibinfo{publisher}{{Springer}}, \bibinfo{address}{{Berlin, Heidelberg}}, \bibinfo{pages}{573--586}.
\newblock
\showISBNx{978-3-642-33765-9}


\bibitem[Rezaei et~al\mbox{.}(2021)]%
        {rezaeiUnobtrusiveFallDetection2021}
\bibfield{author}{\bibinfo{person}{Ariyamehr~Mohsen Rezaei}, \bibinfo{person}{Michael~C. Stevens}, \bibinfo{person}{Ahmadreza Argha}, \bibinfo{person}{Alessandro Mascheroni}, \bibinfo{person}{Alessandro Puiatti}, {and} \bibinfo{person}{Nigel~H. Lovell}.} \bibinfo{year}{2021}\natexlab{}.
\newblock \showarticletitle{An {{Unobtrusive Fall Detection System Using Low Resolution Thermal Sensors}} and {{Convolutional Neural Networks}}}. In \bibinfo{booktitle}{\emph{2021 43rd {{Annual International Conference}} of the {{IEEE Engineering}} in {{Medicine}} \& {{Biology Society}} ({{EMBC}})}}. \bibinfo{pages}{6949--6952}.
\newblock
\showISSN{2694-0604}


\bibitem[Rezaei et~al\mbox{.}(2023)]%
        {rezaei2023trihorn}
\bibfield{author}{\bibinfo{person}{Mohammad Rezaei}, \bibinfo{person}{Razieh Rastgoo}, {and} \bibinfo{person}{Vassilis Athitsos}.} \bibinfo{year}{2023}\natexlab{}.
\newblock \showarticletitle{TriHorn-net: a model for accurate depth-based 3D hand pose estimation}.
\newblock \bibinfo{journal}{\emph{Expert Systems with Applications}}  \bibinfo{volume}{223} (\bibinfo{year}{2023}), \bibinfo{pages}{119922}.
\newblock


\bibitem[Rezaei et~al\mbox{.}(2022)]%
        {rezaeiUnobtrusiveHumanActivity2022}
\bibfield{author}{\bibinfo{person}{Mohsen Rezaei}, \bibinfo{person}{Michael~C. Stevens}, \bibinfo{person}{Ahmadreza Argha}, \bibinfo{person}{Alessandro Mascheroni}, \bibinfo{person}{Alessandro Puiatti}, {and} \bibinfo{person}{Nigel~H. Lovell}.} \bibinfo{year}{2022}\natexlab{}.
\newblock \showarticletitle{An {{Unobtrusive Human Activity Recognition System Using Low Resolution Thermal Sensors}}, {{Machine}} and {{Deep Learning}}}.
\newblock \bibinfo{journal}{\emph{IEEE Transactions on Biomedical Engineering}} (\bibinfo{year}{2022}), \bibinfo{pages}{1--9}.
\newblock
\showISSN{1558-2531}


\bibitem[Romero et~al\mbox{.}(2017)]%
        {romeroEmbodiedHandsModeling2017}
\bibfield{author}{\bibinfo{person}{Javier Romero}, \bibinfo{person}{Dimitrios Tzionas}, {and} \bibinfo{person}{Michael~J. Black}.} \bibinfo{year}{2017}\natexlab{}.
\newblock \showarticletitle{Embodied Hands: Modeling and Capturing Hands and Bodies Together}.
\newblock \bibinfo{journal}{\emph{ACM Transactions on Graphics}} \bibinfo{volume}{36}, \bibinfo{number}{6} (\bibinfo{date}{Nov.} \bibinfo{year}{2017}), \bibinfo{pages}{245:1--245:17}.
\newblock
\showISSN{0730-0301}


\bibitem[Sahoo and Lone(2023)]%
        {sahooOccupancyCountingDense2023}
\bibfield{author}{\bibinfo{person}{Soumya~R. Sahoo} {and} \bibinfo{person}{Haroon~R. Lone}.} \bibinfo{year}{2023}\natexlab{}.
\newblock \showarticletitle{Occupancy Counting in Dense and Sparse Settings with a Low-Cost Thermal Camera}. In \bibinfo{booktitle}{\emph{2023 15th {{International Conference}} on {{COMmunication Systems}} \& {{NETworkS}} ({{COMSNETS}})}}. \bibinfo{pages}{537--544}.
\newblock
\showISSN{2155-2509}


\bibitem[Sandler et~al\mbox{.}(2018)]%
        {sandlerMobileNetV2InvertedResiduals2018}
\bibfield{author}{\bibinfo{person}{Mark Sandler}, \bibinfo{person}{Andrew Howard}, \bibinfo{person}{Menglong Zhu}, \bibinfo{person}{Andrey Zhmoginov}, {and} \bibinfo{person}{Liang-Chieh Chen}.} \bibinfo{year}{2018}\natexlab{}.
\newblock \showarticletitle{{{MobileNetV2}}: {{Inverted Residuals}} and {{Linear Bottlenecks}}}. In \bibinfo{booktitle}{\emph{Proceedings of the {{IEEE Conference}} on {{Computer Vision}} and {{Pattern Recognition}}}}. \bibinfo{pages}{4510--4520}.
\newblock


\bibitem[Spurr et~al\mbox{.}(2020)]%
        {spurrWeaklySupervised3D2020}
\bibfield{author}{\bibinfo{person}{Adrian Spurr}, \bibinfo{person}{Umar Iqbal}, \bibinfo{person}{Pavlo Molchanov}, \bibinfo{person}{Otmar Hilliges}, {and} \bibinfo{person}{Jan Kautz}.} \bibinfo{year}{2020}\natexlab{}.
\newblock \showarticletitle{Weakly {{Supervised 3D Hand Pose Estimation}} via {{Biomechanical Constraints}}}. In \bibinfo{booktitle}{\emph{Computer {{Vision}} {\textendash} {{ECCV}} 2020}} \emph{(\bibinfo{series}{Lecture {{Notes}} in {{Computer Science}}})}, \bibfield{editor}{\bibinfo{person}{Andrea Vedaldi}, \bibinfo{person}{Horst Bischof}, \bibinfo{person}{Thomas Brox}, {and} \bibinfo{person}{Jan-Michael Frahm}} (Eds.). \bibinfo{publisher}{{Springer International Publishing}}, \bibinfo{address}{{Cham}}, \bibinfo{pages}{211--228}.
\newblock
\showISBNx{978-3-030-58520-4}


\bibitem[Sun et~al\mbox{.}(2018)]%
        {sunVSkinSensingTouch2018}
\bibfield{author}{\bibinfo{person}{Ke Sun}, \bibinfo{person}{Ting Zhao}, \bibinfo{person}{Wei Wang}, {and} \bibinfo{person}{Lei Xie}.} \bibinfo{year}{2018}\natexlab{}.
\newblock \showarticletitle{{{VSkin}}: {{Sensing Touch Gestures}} on {{Surfaces}} of {{Mobile Devices Using Acoustic Signals}}}. In \bibinfo{booktitle}{\emph{Proceedings of the 24th {{Annual International Conference}} on {{Mobile Computing}} and {{Networking}}}}. \bibinfo{address}{{New York, NY, USA}}, \bibinfo{pages}{591--605}.
\newblock
\showISBNx{978-1-4503-5903-0}


\bibitem[Sun et~al\mbox{.}(2015)]%
        {sunWiDrawEnablingHandsfree2015}
\bibfield{author}{\bibinfo{person}{Li Sun}, \bibinfo{person}{Souvik Sen}, \bibinfo{person}{Dimitrios Koutsonikolas}, {and} \bibinfo{person}{Kyu-Han Kim}.} \bibinfo{year}{2015}\natexlab{}.
\newblock \showarticletitle{{{WiDraw}}: {{Enabling Hands-free Drawing}} in the {{Air}} on {{Commodity WiFi Devices}}}. In \bibinfo{booktitle}{\emph{Proceedings of the 21st {{Annual International Conference}} on {{Mobile Computing}} and {{Networking}}}}. \bibinfo{address}{{New York, NY, USA}}, \bibinfo{pages}{77--89}.
\newblock
\showISBNx{978-1-4503-3619-2}


\bibitem[Sun et~al\mbox{.}(2017)]%
        {sun2017compositional}
\bibfield{author}{\bibinfo{person}{Xiao Sun}, \bibinfo{person}{Jiaxiang Shang}, \bibinfo{person}{Shuang Liang}, {and} \bibinfo{person}{Yichen Wei}.} \bibinfo{year}{2017}\natexlab{}.
\newblock \showarticletitle{Compositional human pose regression}. In \bibinfo{booktitle}{\emph{Proceedings of the IEEE international conference on computer vision}}. \bibinfo{pages}{2602--2611}.
\newblock


\bibitem[Tanaka et~al\mbox{.}(2014)]%
        {tanakaLowPowerWireless2014}
\bibfield{author}{\bibinfo{person}{Junichi Tanaka}, \bibinfo{person}{Hiroshi Imamoto}, \bibinfo{person}{Tomonori Seki}, {and} \bibinfo{person}{Masatoshi Oba}.} \bibinfo{year}{2014}\natexlab{}.
\newblock \showarticletitle{Low Power Wireless Human Detector Utilizing Thermopile Infrared Array Sensor}. In \bibinfo{booktitle}{\emph{2014 {{IEEE SENSORS}}}}. \bibinfo{pages}{462--465}.
\newblock
\showISSN{1930-0395}


\bibitem[Tateno et~al\mbox{.}(2019)]%
        {tatenoHandGestureRecognition2019}
\bibfield{author}{\bibinfo{person}{Shigeyuki Tateno}, \bibinfo{person}{Yiwei Zhu}, {and} \bibinfo{person}{Fanxing Meng}.} \bibinfo{year}{2019}\natexlab{}.
\newblock \showarticletitle{Hand {{Gesture Recognition System}} for {{In-car Device Control Based}} on {{Infrared Array Sensor}}}. In \bibinfo{booktitle}{\emph{2019 58th {{Annual Conference}} of the {{Society}} of {{Instrument}} and {{Control Engineers}} of {{Japan}} ({{SICE}})}}. \bibinfo{pages}{701--706}.
\newblock


\bibitem[Tome et~al\mbox{.}(2017)]%
        {tomeLiftingDeepConvolutional2017}
\bibfield{author}{\bibinfo{person}{Denis Tome}, \bibinfo{person}{Chris Russell}, {and} \bibinfo{person}{Lourdes Agapito}.} \bibinfo{year}{2017}\natexlab{}.
\newblock \showarticletitle{Lifting {{From}} the {{Deep}}: {{Convolutional 3D Pose Estimation From}} a {{Single Image}}}. In \bibinfo{booktitle}{\emph{Proceedings of the {{IEEE Conference}} on {{Computer Vision}} and {{Pattern Recognition}}}}. \bibinfo{pages}{2500--2509}.
\newblock


\bibitem[Trofimova et~al\mbox{.}(2017)]%
        {trofimova2017indoor}
\bibfield{author}{\bibinfo{person}{Anna~A Trofimova}, \bibinfo{person}{Andrea Masciadri}, \bibinfo{person}{Fabio Veronese}, {and} \bibinfo{person}{Fabio Salice}.} \bibinfo{year}{2017}\natexlab{}.
\newblock \showarticletitle{Indoor human detection based on thermal array sensor data and adaptive background estimation}.
\newblock \bibinfo{journal}{\emph{Journal of Computer and Communications}} \bibinfo{volume}{5}, \bibinfo{number}{4} (\bibinfo{year}{2017}), \bibinfo{pages}{16--28}.
\newblock


\bibitem[Truong et~al\mbox{.}(2018)]%
        {truong2018capband}
\bibfield{author}{\bibinfo{person}{Hoang Truong}, \bibinfo{person}{Shuo Zhang}, \bibinfo{person}{Ufuk Muncuk}, \bibinfo{person}{Phuc Nguyen}, \bibinfo{person}{Nam Bui}, \bibinfo{person}{Anh Nguyen}, \bibinfo{person}{Qin Lv}, \bibinfo{person}{Kaushik Chowdhury}, \bibinfo{person}{Thang Dinh}, {and} \bibinfo{person}{Tam Vu}.} \bibinfo{year}{2018}\natexlab{}.
\newblock \showarticletitle{Capband: Battery-free successive capacitance sensing wristband for hand gesture recognition}. In \bibinfo{booktitle}{\emph{Proceedings of the 16th ACM Conference on Embedded Networked Sensor Systems}}. \bibinfo{pages}{54--67}.
\newblock


\bibitem[Tyndall et~al\mbox{.}(2016)]%
        {tyndallOccupancyEstimationUsing2016}
\bibfield{author}{\bibinfo{person}{Ash Tyndall}, \bibinfo{person}{Rachel {Cardell-Oliver}}, {and} \bibinfo{person}{Adrian Keating}.} \bibinfo{year}{2016}\natexlab{}.
\newblock \showarticletitle{Occupancy {{Estimation Using}} a {{Low-Pixel Count Thermal Imager}}}.
\newblock \bibinfo{journal}{\emph{IEEE Sensors Journal}} \bibinfo{volume}{16}, \bibinfo{number}{10} (\bibinfo{year}{2016}), \bibinfo{pages}{3784--3791}.
\newblock
\showISSN{1558-1748}


\bibitem[Ultraleap(2024)]%
        {LeapMotionController}
\bibfield{author}{\bibinfo{person}{Ultraleap}.} \bibinfo{year}{2024}\natexlab{}.
\newblock \bibinfo{title}{Leap Motion Controller Overview}.
\newblock
\urldef\tempurl%
\url{https://www.ultraleap.com/leap-motion-controller-overview/}
\showURL{%
\tempurl}
\newblock
\shownote{Accessed: 2024-11-12}.


\bibitem[Vandersteegen et~al\mbox{.}(2020)]%
        {vandersteegenLowLatencyHandGesture2020}
\bibfield{author}{\bibinfo{person}{Maarten Vandersteegen}, \bibinfo{person}{Wouter Reusen}, \bibinfo{person}{Kristof Van~Beeck}, {and} \bibinfo{person}{Toon Goedeme}.} \bibinfo{year}{2020}\natexlab{}.
\newblock \showarticletitle{Low-{{Latency Hand Gesture Recognition With}} a {{Low-Resolution Thermal Imager}}}. In \bibinfo{booktitle}{\emph{Proceedings of the {{IEEE}}/{{CVF Conference}} on {{Computer Vision}} and {{Pattern Recognition Workshops}}}}. \bibinfo{pages}{98--99}.
\newblock


\bibitem[Vaswani et~al\mbox{.}(2017)]%
        {vaswani2017attention}
\bibfield{author}{\bibinfo{person}{Ashish Vaswani}, \bibinfo{person}{Noam Shazeer}, \bibinfo{person}{Niki Parmar}, \bibinfo{person}{Jakob Uszkoreit}, \bibinfo{person}{Llion Jones}, \bibinfo{person}{Aidan~N Gomez}, \bibinfo{person}{{\L}ukasz Kaiser}, {and} \bibinfo{person}{Illia Polosukhin}.} \bibinfo{year}{2017}\natexlab{}.
\newblock \showarticletitle{Attention is all you need}.
\newblock \bibinfo{journal}{\emph{Advances in neural information processing systems}}  \bibinfo{volume}{30} (\bibinfo{year}{2017}).
\newblock


\bibitem[Vtouch(2024)]%
        {Holo_Button}
\bibfield{author}{\bibinfo{person}{Vtouch}.} \bibinfo{year}{2024}\natexlab{}.
\newblock
\urldef\tempurl%
\url{https://vtouch.io/en-us/products/holo-button}
\showURL{%
\tempurl}


\bibitem[Wang et~al\mbox{.}(2016)]%
        {wangDevicefreeGestureTracking2016}
\bibfield{author}{\bibinfo{person}{Wei Wang}, \bibinfo{person}{Alex~X. Liu}, {and} \bibinfo{person}{Ke Sun}.} \bibinfo{year}{2016}\natexlab{}.
\newblock \showarticletitle{Device-Free Gesture Tracking Using Acoustic Signals}. In \bibinfo{booktitle}{\emph{Proceedings of the 22nd {{Annual International Conference}} on {{Mobile Computing}} and {{Networking}}}}. \bibinfo{address}{{New York, NY, USA}}, \bibinfo{pages}{82--94}.
\newblock
\showISBNx{978-1-4503-4226-1}


\bibitem[Wu et~al\mbox{.}(2020)]%
        {wuBackHandPose3DHand2020}
\bibfield{author}{\bibinfo{person}{Erwin Wu}, \bibinfo{person}{Ye Yuan}, \bibinfo{person}{Hui-Shyong Yeo}, \bibinfo{person}{Aaron Quigley}, \bibinfo{person}{Hideki Koike}, {and} \bibinfo{person}{Kris~M. Kitani}.} \bibinfo{year}{2020}\natexlab{}.
\newblock \showarticletitle{Back-{{Hand-Pose}}: {{3D Hand Pose Estimation}} for a {{Wrist-worn Camera}} via {{Dorsum Deformation Network}}}. In \bibinfo{booktitle}{\emph{Proceedings of the 33rd {{Annual ACM Symposium}} on {{User Interface Software}} and {{Technology}}}} \emph{(\bibinfo{series}{{{UIST}} '20})}. \bibinfo{address}{{New York, NY, USA}}, \bibinfo{pages}{1147--1160}.
\newblock
\showISBNx{978-1-4503-7514-6}


\bibitem[Xiao et~al\mbox{.}(2021)]%
        {xiao2021onefi}
\bibfield{author}{\bibinfo{person}{Rui Xiao}, \bibinfo{person}{Jianwei Liu}, \bibinfo{person}{Jinsong Han}, {and} \bibinfo{person}{Kui Ren}.} \bibinfo{year}{2021}\natexlab{}.
\newblock \showarticletitle{Onefi: One-shot recognition for unseen gesture via cots wifi}. In \bibinfo{booktitle}{\emph{Proceedings of the 19th ACM Conference on Embedded Networked Sensor Systems}}. \bibinfo{pages}{206--219}.
\newblock


\bibitem[Xu et~al\mbox{.}(2022)]%
        {xuEnablingHandGesture2022}
\bibfield{author}{\bibinfo{person}{Xuhai Xu}, \bibinfo{person}{Jun Gong}, \bibinfo{person}{Carolina Brum}, \bibinfo{person}{Lilian Liang}, \bibinfo{person}{Bongsoo Suh}, \bibinfo{person}{Shivam~Kumar Gupta}, \bibinfo{person}{Yash Agarwal}, \bibinfo{person}{Laurence Lindsey}, \bibinfo{person}{Runchang Kang}, \bibinfo{person}{Behrooz Shahsavari}, \bibinfo{person}{Tu Nguyen}, \bibinfo{person}{Heriberto Nieto}, \bibinfo{person}{Scott~E Hudson}, \bibinfo{person}{Charlie Maalouf}, \bibinfo{person}{Jax~Seyed Mousavi}, {and} \bibinfo{person}{Gierad Laput}.} \bibinfo{year}{2022}\natexlab{}.
\newblock \showarticletitle{Enabling {{Hand Gesture Customization}} on {{Wrist-Worn Devices}}}. In \bibinfo{booktitle}{\emph{Proceedings of the 2022 {{CHI Conference}} on {{Human Factors}} in {{Computing Systems}}}} \emph{(\bibinfo{series}{{{CHI}} '22})}. \bibinfo{address}{{New York, NY, USA}}, \bibinfo{pages}{1--19}.
\newblock
\showISBNx{978-1-4503-9157-3}


\bibitem[Yang et~al\mbox{.}(2021)]%
        {yangSemiHandSemisupervisedHand2021}
\bibfield{author}{\bibinfo{person}{Linlin Yang}, \bibinfo{person}{Shicheng Chen}, {and} \bibinfo{person}{Angela Yao}.} \bibinfo{year}{2021}\natexlab{}.
\newblock \showarticletitle{{{SemiHand}}: {{Semi-supervised Hand Pose Estimation}} with {{Consistency}}}. In \bibinfo{booktitle}{\emph{2021 {{IEEE}}/{{CVF International Conference}} on {{Computer Vision}} ({{ICCV}})}}. \bibinfo{publisher}{{IEEE}}, \bibinfo{address}{{Montreal, QC, Canada}}, \bibinfo{pages}{11344--11353}.
\newblock
\showISBNx{978-1-66542-812-5}


\bibitem[Yang et~al\mbox{.}(2024)]%
        {yang2024depth}
\bibfield{author}{\bibinfo{person}{Lihe Yang}, \bibinfo{person}{Bingyi Kang}, \bibinfo{person}{Zilong Huang}, \bibinfo{person}{Xiaogang Xu}, \bibinfo{person}{Jiashi Feng}, {and} \bibinfo{person}{Hengshuang Zhao}.} \bibinfo{year}{2024}\natexlab{}.
\newblock \showarticletitle{Depth anything: Unleashing the power of large-scale unlabeled data}. In \bibinfo{booktitle}{\emph{Proceedings of the IEEE/CVF Conference on Computer Vision and Pattern Recognition}}. \bibinfo{pages}{10371--10381}.
\newblock


\bibitem[Yu et~al\mbox{.}(2019)]%
        {yu2019rfid}
\bibfield{author}{\bibinfo{person}{Yinggang Yu}, \bibinfo{person}{Dong Wang}, \bibinfo{person}{Run Zhao}, {and} \bibinfo{person}{Qian Zhang}.} \bibinfo{year}{2019}\natexlab{}.
\newblock \showarticletitle{RFID based real-time recognition of ongoing gesture with adversarial learning}. In \bibinfo{booktitle}{\emph{Proceedings of the 17th Conference on Embedded Networked Sensor Systems}}. \bibinfo{pages}{298--310}.
\newblock


\bibitem[Zanuttigh et~al\mbox{.}(2016)]%
        {zanuttigh2016time}
\bibfield{author}{\bibinfo{person}{Pietro Zanuttigh}, \bibinfo{person}{Giulio Marin}, \bibinfo{person}{Carlo Dal~Mutto}, \bibinfo{person}{Fabio Dominio}, \bibinfo{person}{Ludovico Minto}, \bibinfo{person}{Guido~Maria Cortelazzo}, {et~al\mbox{.}}} \bibinfo{year}{2016}\natexlab{}.
\newblock \showarticletitle{Time-of-flight and structured light depth cameras}.
\newblock \bibinfo{journal}{\emph{Technology and Applications}} \bibinfo{volume}{978}, \bibinfo{number}{3} (\bibinfo{year}{2016}).
\newblock


\bibitem[Zhang et~al\mbox{.}(2020)]%
        {zhangMediaPipeHandsOndevice2020}
\bibfield{author}{\bibinfo{person}{Fan Zhang}, \bibinfo{person}{Valentin Bazarevsky}, \bibinfo{person}{Andrey Vakunov}, \bibinfo{person}{Andrei Tkachenka}, \bibinfo{person}{George Sung}, \bibinfo{person}{Chuo-Ling Chang}, {and} \bibinfo{person}{Matthias Grundmann}.} \bibinfo{year}{2020}\natexlab{}.
\newblock \bibinfo{title}{{{MediaPipe Hands}}: {{On-device Real-time Hand Tracking}}}.
\newblock
\showeprint[arxiv]{2006.10214}~[cs]


\bibitem[Zhang et~al\mbox{.}(2023)]%
        {zhangRoFin3DHand2023}
\bibfield{author}{\bibinfo{person}{Xiao Zhang}, \bibinfo{person}{Griffin Klevering}, \bibinfo{person}{Juexing Wang}, \bibinfo{person}{Li Xiao}, {and} \bibinfo{person}{Tianxing Li}.} \bibinfo{year}{2023}\natexlab{}.
\newblock \showarticletitle{{{RoFin}}: {{3D Hand Pose Reconstructing}} via {{2D Rolling Fingertips}}}. In \bibinfo{booktitle}{\emph{Proceedings of the 21st {{Annual International Conference}} on {{Mobile Systems}}, {{Applications}} and {{Services}}}} \emph{(\bibinfo{series}{{{MobiSys}} '23})}. \bibinfo{address}{{New York, NY, USA}}, \bibinfo{pages}{330--342}.
\newblock
\showISBNx{9798400701108}


\bibitem[Zhang et~al\mbox{.}(2021)]%
        {zhang2021wifi}
\bibfield{author}{\bibinfo{person}{Xie Zhang}, \bibinfo{person}{Chengpei Tang}, \bibinfo{person}{Kang Yin}, {and} \bibinfo{person}{Qingqian Ni}.} \bibinfo{year}{2021}\natexlab{}.
\newblock \showarticletitle{WiFi-based cross-domain gesture recognition via modified prototypical networks}.
\newblock \bibinfo{journal}{\emph{IEEE Internet of Things Journal}} \bibinfo{volume}{9}, \bibinfo{number}{11} (\bibinfo{year}{2021}), \bibinfo{pages}{8584--8596}.
\newblock


\bibitem[Zhang and Wu(2024)]%
        {zhang2024tadar}
\bibfield{author}{\bibinfo{person}{Xie Zhang} {and} \bibinfo{person}{Chenshu Wu}.} \bibinfo{year}{2024}\natexlab{}.
\newblock \showarticletitle{TADAR: Thermal Array-based Detection and Ranging for Privacy-Preserving Human Sensing}. In \bibinfo{booktitle}{\emph{Proceedings of the Twenty-fifth International Symposium on Theory, Algorithmic Foundations, and Protocol Design for Mobile Networks and Mobile Computing}}. \bibinfo{pages}{11--20}.
\newblock


\bibitem[Zhao et~al\mbox{.}(2023)]%
        {zhaoIfSightedPeople2023}
\bibfield{author}{\bibinfo{person}{Yuhang Zhao}, \bibinfo{person}{Yaxing Yao}, \bibinfo{person}{Jiaru Fu}, {and} \bibinfo{person}{Nihan Zhou}.} \bibinfo{year}{2023}\natexlab{}.
\newblock \showarticletitle{\{``\vphantom\}{{If}}\vphantom\{\} Sighted People Know, {{I}} Should Be Able to \{know:''\} {{Privacy Perceptions}} of {{Bystanders}} with {{Visual Impairments}} around {{Camera-based Technology}}}. In \bibinfo{booktitle}{\emph{32nd {{USENIX Security Symposium}} ({{USENIX Security}} 23)}}. \bibinfo{pages}{4661--4678}.
\newblock
\showISBNx{978-1-939133-37-3}


\bibitem[Zhong et~al\mbox{.}(2021)]%
        {zhongMultiOccupancyFallDetection2021}
\bibfield{author}{\bibinfo{person}{Cankun Zhong}, \bibinfo{person}{Wing W.~Y. Ng}, \bibinfo{person}{Shuai Zhang}, \bibinfo{person}{Chris~D. Nugent}, \bibinfo{person}{Colin Shewell}, {and} \bibinfo{person}{Javier {Medina-Quero}}.} \bibinfo{year}{2021}\natexlab{}.
\newblock \showarticletitle{Multi-{{Occupancy Fall Detection Using Non-Invasive Thermal Vision Sensor}}}.
\newblock \bibinfo{journal}{\emph{IEEE Sensors Journal}} \bibinfo{volume}{21}, \bibinfo{number}{4} (\bibinfo{date}{Feb.} \bibinfo{year}{2021}), \bibinfo{pages}{5377--5388}.
\newblock
\showISSN{1558-1748}


\bibitem[Zhou et~al\mbox{.}(2023)]%
        {zhou2023edgesam}
\bibfield{author}{\bibinfo{person}{Chong Zhou}, \bibinfo{person}{Xiangtai Li}, \bibinfo{person}{Chen~Change Loy}, {and} \bibinfo{person}{Bo Dai}.} \bibinfo{year}{2023}\natexlab{}.
\newblock \showarticletitle{Edgesam: Prompt-in-the-loop distillation for on-device deployment of sam}.
\newblock \bibinfo{journal}{\emph{arXiv preprint arXiv:2312.06660}} (\bibinfo{year}{2023}).
\newblock


\bibitem[Zhou et~al\mbox{.}(2022)]%
        {zhouLearningRingsSelfSupervised2022}
\bibfield{author}{\bibinfo{person}{Hao Zhou}, \bibinfo{person}{Taiting Lu}, \bibinfo{person}{Yilin Liu}, \bibinfo{person}{Shijia Zhang}, {and} \bibinfo{person}{Mahanth Gowda}.} \bibinfo{year}{2022}\natexlab{}.
\newblock \showarticletitle{Learning on the {{Rings}}: {{Self-Supervised 3D Finger Motion Tracking Using Wearable Sensors}}}.
\newblock \bibinfo{journal}{\emph{Proceedings of the ACM on Interactive, Mobile, Wearable and Ubiquitous Technologies}} \bibinfo{volume}{6}, \bibinfo{number}{2} (\bibinfo{date}{July} \bibinfo{year}{2022}), \bibinfo{pages}{90:1--90:31}.
\newblock


\bibitem[Zhou et~al\mbox{.}(2016)]%
        {zhouModelbasedDeepHand2016}
\bibfield{author}{\bibinfo{person}{Xingyi Zhou}, \bibinfo{person}{Qingfu Wan}, \bibinfo{person}{Wei Zhang}, \bibinfo{person}{Xiangyang Xue}, {and} \bibinfo{person}{Yichen Wei}.} \bibinfo{year}{2016}\natexlab{}.
\newblock \bibinfo{title}{Model-Based {{Deep Hand Pose Estimation}}}.
\newblock
\showeprint[arxiv]{1606.06854}~[cs]


\bibitem[Zimmermann and Brox(2017)]%
        {zimmermannLearningEstimate3D2017}
\bibfield{author}{\bibinfo{person}{Christian Zimmermann} {and} \bibinfo{person}{Thomas Brox}.} \bibinfo{year}{2017}\natexlab{}.
\newblock \showarticletitle{Learning to {{Estimate 3D Hand Pose From Single RGB Images}}}. In \bibinfo{booktitle}{\emph{Proceedings of the {{IEEE International Conference}} on {{Computer Vision}}}}. \bibinfo{pages}{4903--4911}.
\newblock


\end{thebibliography}


\end{document}